\documentclass[aip,jcp,amsmath,amssymb,reprint,longbibliography]{revtex4-1}
\usepackage[utf8]{inputenc}
\usepackage{braket}
\usepackage{amsmath}
\usepackage{appendix}
\usepackage{amsfonts}
\usepackage{comment}
\usepackage{cancel}
\usepackage{bbold}
\usepackage{graphicx}
\usepackage{float}
\usepackage{MnSymbol}
\usepackage{mathtools}
\usepackage{tabularx}
\usepackage{array}
\usepackage{bm}
\usepackage{placeins}
\usepackage{MnSymbol}
\usepackage{xcolor}

\newcolumntype{Y}{>{\centering\arraybackslash}X}

\usepackage[colorlinks, linkcolor=blue]{hyperref}
\usepackage[nameinlink, capitalise]{cleveref}
\crefname{section}{Sec.}{Secs.}

\newcommand{\nn}{\nonumber} 
\newcommand\numeq[1]%
{\stackrel{\scriptscriptstyle(\mkern-1.5mu#1\mkern-1.5mu)}{=}}

\newcommand{\eq}{\mathrm{eq}}

\newcommand\commin[1]{\iffalse #1 \fi}%inline comment

\newcommand{\mr}{\mathrm}
\newcommand{\mc}{\mathcal}

\newcommand{\jor}[1]{{\color{red}#1}}
\newcommand{\sy}{\mathrm{s}}
\newcommand{\ba}{\mathrm{b}}
\newcommand{\syba}{\mathrm{sb}}
\newcommand{\Pb}{\mathrm{Pb}}

\newcommand{\tcut}{t_{\mathrm{cut}}}

\newcommand{\rW}{r_{\mathrm{W}}}

\newcommand{\Pop}{\mathrm{P}_+}

\newcommand{\W}{\mathrm{W}}
\newcommand{\E}{\mathrm{E}}

\newcommand{\eu}[1]{\mathrm{e}^{#1}}

\newcommand{\rmd}{\mathrm{d}}

\newcommand{\e}{\ensuremath{\,\mathrm{e}}}

\newcommand{\C}{\ensuremath{\mathcal{C}}}
\newcommand{\K}{\ensuremath{\mathcal{K}}}

\newcommand{\thalf}{\ensuremath{\tfrac{1}{2}}}

\DeclareMathOperator{\tr}{tr}

\usepackage{chngcntr}
\begin{document}

\title{Quasiclassical approaches to the generalized quantum master equation}
\author{Graziano Amati}
\affiliation{Laboratory of Physical Chemistry, ETH Z\"urich, 8093 Z\"urich, Switzerland}
\author{Maximilian A. C. Saller}
\altaffiliation{Present address: Department of Chemistry, University of Michigan, Ann Arbor, 48109 MI, USA}
\affiliation{Laboratory of Physical Chemistry, ETH Z\"urich, 8093 Z\"urich, Switzerland}
\author{Aaron Kelly}
\affiliation{The Hamburg Center for Ultrafast Imaging, Universit\"at Hamburg, Max Planck Institute for the Structure and Dynamics of Matter
and Center for Free-Electron Laser Science, 22761 Hamburg, Germany}
\affiliation{Department of Chemistry, Dalhousie University, Halifax, Canada}
\author{Jeremy O. Richardson}
\email[Author to whom correspondence should be addressed: ]{jeremy.richardson@phys.chem.ethz.ch}
\affiliation{Laboratory of Physical Chemistry, ETH Z\"urich, 8093 Z\"urich, Switzerland}
%$^1$,  $^{1,2}$, Aaron Kelly$^{3,4}$, Jeremy O. Richardson$^1$}
%$^2$Deparment of Chemistry, University of Michigan, Ann Arbor, 48109 MI, USA\\
%$^3$The Hamburg Center for Ultrafast Imaging, Universit\"at Hamburg, Max Planck Institute for the Structure and Dynamics of Matter and Center for Free-Electron Laser Science, 22761 Hamburg, Germany\\
%$^4$Department of Chemistry, Dalhousie University, Halifax, Canada}

\date{\today}

\begin{abstract}
The formalism of the generalized quantum master equation (GQME) is an effective tool to simultaneously increase the accuracy and the efficiency of quasiclassical trajectory methods in the simulation of nonadiabatic quantum dynamics. The GQME expresses correlation functions in terms of a non-Markovian equation of motion, involving memory kernels which are typically fast-decaying and can therefore be computed by short-time quasiclassical trajectories.
In this paper we study the approximate solution of the GQME, obtained by calculating the kernels with two methods, namely Ehrenfest mean-field theory and spin mapping. We test the approaches on a range of spin--boson models with increasing energy bias between the two electronic levels and place a particular focus on the long-time limits of the populations. We find that the accuracy of the predictions of the GQME depends strongly on the specific technique used to calculate the kernels. In particular, spin mapping outperforms Ehrenfest for all systems studied. The problem of unphysical negative electronic populations affecting spin mapping is resolved 
by coupling the method with the master equation. Conversely, Ehrenfest in conjunction with the GQME can predict negative populations, despite the fact that the populations calculated from direct dynamics are positive definite.
\end{abstract}

\maketitle
\section{Introduction}

The dynamical coupling between nuclear and electronic degrees of freedom in molecular systems is a key feature of many important processes, from photosynthesis \cite{lee2016} and light harvesting \cite{cheng2009} to vision.\cite{polli2010}
Formally these can be described by nonadiabatic quantum dynamics based on the explicit treatment of the couplings beyond the Born--Oppenheimer approximation. 
There has been considerable interest in the development of quasiclassical techniques, aimed at approximating nonadiabatic quantum correlation functions with classical analogues.\cite{Stock2005nonadiabatic}

An important aspect of quasiclassical methods is that they require low computational effort compared to exact quantum dynamics; while the cost of exact quantum simulations scales exponentially with simulation time and/or system size, quasiclassical dynamics scale polynomially (or even linearly under favourable circumstances). As a trade-off, an error is inevitably introduced by quasiclassical methods, due to the fact that the true quantum dynamics are approximated by classical trajectories.\cite{Miller2001SCIVR, kapral2015} Given that such quasiclassical dynamics relax to the thermal distributions of approximated Hamiltonians, time correlation functions do not generally decay to the correct equilibrium values. This can lead to poor predictions for the long-time limit of the electronic populations, in particular in cases with a strong bias between the levels of the electronic subsystem.
Furthermore, some quasiclassical methods are even known to predict unphysical negative values for the electronic populations of high-lying states. This issue is well-known within linearized semiclassical versions of Meyer--Miller--Stock--Thoss (MMST) mapping.\cite{Meyer1979nonadiabatic,Stock1997mapping,Mueller1999pyrazine,Stock2005nonadiabatic,Sun1998mapping,Wang1999mapping,Kim2008Liouville,Kelly2012mapping,identity,Liu2020linearized}
Although the recently derived spin-mapping method \cite{spinmap,multispin} often gives improved results over MMST, it has still not completely eliminated this problem.\cite{chimia_mapping2022}
In the case of Ehrenfest mean-field theory, the electronic populations are positive definite, but can still be captured poorly in asymmetric systems.\cite{Parandekar2006Ehrenfest} It becomes then a critical issue to increase the accuracy of the predictions while minimizing the computational effort.

The formalism of the generalized quantum master equation (GQME) has been employed to increase the long-time accuracy of quasiclassical techniques.\cite{Shi2004GQME,Kelly2013GQME,Kelly2015nonadiabatic,Pfalzgraff2015GQME,Kelly2016master,Montoya2016GQME,Mulvihill2019LSCGQME,mulvihill2021b,mulvihill2022,ng2022} The main idea of the approach is to make use of quasiclassical dynamics not to directly calculate correlation functions, but rather to obtain the kernels of the GQME\@. 
The solution of the GQME based on these kernels appears to give far more accurate correlation functions.
In addition, the GQME procedure may be computationally more efficient than direct approaches.
This is because memory kernels tend to decay on much faster timescales than the correlations themselves (and even become delta functions in the limit of Markovian dynamics).\cite{OpenQuantum} 
Thus, only short trajectory simulations are needed to predict long-time dynamics.
For this reason, the GQME approach has  also been used to speed-up numerically exact quantum dynamics calculations.\cite{shi2003,erpenbeck2019, ullah2021}% \red{maybe also say it's used for speeding up pure classical dynamics}

Many examples have been published in which the calculation of the kernels of the GQME with Ehrenfest mean-field theory (which we refer to as GQME/Ehrenfest) leads to excellent agreement with benchmark quantum-mechanical results, whereas the Ehrenfest method itself is not sufficient.\cite{Kelly2015nonadiabatic,Kelly2016master,Pfalzgraff2019GQME,mulvihill2021,Pfalzgraff2015GQME,Montoya2016GQME}
However, to the best of the authors' knowledge, it is still an open question to what extent the predictions of the quasiclassical approximation of the GQME can break down and how they can be improved.

The accuracy of direct quasiclassical predictions of correlation functions can vary greatly, depending on the specific method.\cite{linearized}
For example, spin mapping \cite{spinmap,multispin} has been demonstrated to outperform a number of other quasiclassical approaches (including the Ehrenfest method) in many cases.\cite{FMOclassical,chimia_mapping2022} 
%At long times, however, spin mapping can predict unphysical negative electronic populations, while the populations calculated with Ehrenfest are positive definite.
This leads to the interesting question as to whether solving the kernels with accurate short-time trajectories from spin mapping (GQME/spin-mapping) can lead to more reliable results than GQME/Ehrenfest. 
One could additionally ask whether GQME/spin-mapping can solve the problem of negative populations observed in the direct application of the quasiclassical method.
%On the other hand, one might worry that
A potential source of concern is that solving the GQME using input data which includes unphysical negative populations may lead to unphysical output results. 
We might also worry that the GQME procedure will have little or no effect on spin-mapping results, because it appears that
certain elements of the correlation functions used to build the kernel are identical to those of the trivial closure relation described in Ref.~\onlinecite{Kelly2016master}.

In order to answer these questions, we study the time evolution of the electronic population in spin--boson models with increasing values of the energy bias between electronic states. Interestingly, we find that GQME/Ehrenfest fails to
%give a meaningful plateau in the cutoff time for all systems studied, 
converge as a function of the cutoff time of the memory kernels. This issue
%and it 
becomes more problematic for increasing values of the energy bias between the electronic states. 
Conversely, GQME/spin-mapping yields stable and accurate long-time predictions %for all systems considered, 
in many cases, except for those with strong system--bath coupling. In particular, this approach consistently outperforms direct spin-mapping calculations, which can produce negative populations in strongly biased systems.
We investigate these results by comparing the accuracy of the quasiclassical predictions of the kernels for the two methods and identifying the critical correlation functions.
We conclude that the master equation is not guaranteed to correct the long-time dynamics of quasiclassical approaches, nor does the GQME formalism itself ensure the positive definiteness of the electronic populations when using quasiclassical memory kernels. 
However, the master equation can nonetheless still be a valuable tool to improve the predictions of quasiclassical methods in many important cases.
%in cases where they have a high enough intrinsic accuracy.

\section{Theory}
We will study the dynamics of population transfer and loss of coherence in quantum systems coupled to a classical bath.
We will commonly refer to the quantum subsystem as the electronic degrees of freedom and the classical bath as the nuclear degrees of freedom, although any other quantum--classical problem could be treated analogously.
For simplicity of notation, we will consider only two-level quantum systems, but the generalization to an arbitrary number of states is straightforward.
The two-level subsystem is conveniently described in the basis of the Pauli matrices, as the electronic dynamics are equivalent to the precession of a spin driven by an effective magnetic field, and it is thus natural to write a representation of the GQME in this basis. 
The resulting spin--spin correlation functions will then be approximated by quasiclassical methods in later sections. 
\subsection{GQME in the basis of the Pauli matrices}
Let us consider a two-level quantum nonadiabatic system described by the Hamiltonian
\begin{equation}\label{def_H}
\hat H = \hat H_\sy\otimes \hat{\mc I}_\ba + \hat{\mc I}_\sy\otimes \hat H_\ba + \hat H_\syba,
\end{equation}
where $\hat H_\sy$, $\hat H_\ba$ and $\hat H_\syba$ denote the Hamiltonian of the electronic subsystem, the nuclear bath, and an interaction between the two respectively. 
Note that there is no requirement that the bath be harmonic, nor any limitation on the complexity on the system--bath interaction for what follows.
$\hat{\mc I}_\sy$ and $\hat{\mc I}_\ba$ are the identity operators for the electronic and nuclear subsystems. Partial traces with respect to the system and bath are defined by $\tr_\sy[\cdot]$ and $\tr_\ba\{\cdot\}$, while the total trace is $\tr [\cdot]= \tr_\ba\{\tr_\sy[\cdot]\}$.
We denote the two states of the system in bra-ket notation, $\ket +$ and $\ket -$; their populations are measured by
\begin{equation}\label{eq:def_pm}
\hat{\mr P}_{\pm}=\ket \pm \bra\pm = \tfrac 12 (\hat{\mc I}_\sy\pm \hat \sigma_z).
\end{equation}
%where $ \hat \sigma_k$ for $k=x,y,z$ are the Pauli matrices.
The electronic--nuclear coupling operator can be written as a sum of the tensor products \cite{OpenQuantum}
\begin{equation}\label{eq:H_SB_gen}
\hat H_\syba = \hat \sigma_k\otimes \hat V_{\ba,k}.
\end{equation}
Here and in the following we use Einstein's summation convention; %we reserve Greek indices for sums over the electronic identity operator and the Pauli matrices ($\hat \sigma_\mu \in \{\hat\sigma_0, \hat\sigma_x,\hat\sigma_y,\hat\sigma_z \}$ and $\hat \sigma_0= \hat{\mc I}_\sy$), whereas we use Latin indices to sum over the three Pauli matrices ($\hat \sigma_k \in \{\hat\sigma_x,\hat\sigma_y,\hat\sigma_z \}$).
we reserve Latin indices for sums 
over the three Pauli matrices, $\hat \sigma_k \in \{\hat\sigma_x,\hat\sigma_y,\hat\sigma_z \}$,
whereas we use Greek indices,
$\hat \sigma_\mu \in \{\hat\sigma_0, \hat\sigma_x,\hat\sigma_y,\hat\sigma_z \}$,
to additionally include 
the electronic identity operator, $\hat \sigma_0= \hat{\mc I}_\sy$.
In this paper we will refer to the index of the identity interchangeably with $0$ and $\mc I$.

We initialize our system according to the factorized density
%a density of Feynman--Vernon type \cite{feynman1963}
\begin{equation}\label{eq:rho0}
\hat\rho_0 = \tfrac 12 \hat{\mc I}_\sy\otimes \hat \rho_\ba,
\hspace{10mm}
\hat \rho_\ba = \frac{\e^{-\beta \hat H_\ba}}{\tr_\ba\{\e^{-\beta \hat H_\ba}\}},
\end{equation}
where $\hat \rho_\ba$ denotes the initial density of the bath. The density $\hat\rho_0$ describes an out-of-equilibrium state, in which the two electronic levels are equally populated, while the (uncoupled) bath is prepared in a thermal state at inverse temperature $\beta$. 
The results of the following sections are based on the assumption that the condition
\begin{equation}\label{eq:tr_rV}
\tr_\ba\{\hat\rho_\ba\hat V_{\ba,k}\}=0,\hspace{7mm}k=x,y,z,
\end{equation}
holds. This does not imply any loss of generality; by defining 
\begin{equation}\label{eq:Hsk}
H_{\sy,k} = \tfrac 12\tr_\sy[\hat H_\sy\hat\sigma_k], 
\end{equation}
such that $\hat H_\sy = H_{\sy,k}\hat \sigma_k$, we can always shift 
\begin{subequations}
\begin{align}
\hat V_{\ba,k} &\rightarrow \hat V_{\ba,k} -\tr_\ba\{\hat\rho_\ba\hat V_{\ba,k}\}\hat{\mc I}_\ba,\\
H_{\sy, k}&\rightarrow H_{\sy, k}+\tr_\ba\{\hat\rho_\ba\hat V_{\ba,k}\},
\end{align}
\end{subequations}
such that \cref{eq:tr_rV} is fulfilled.\cite{shi2003} Note that we can fix $H_{\sy,0}\equiv 0$, given that any contribution in the Hamiltonian proportional to $\hat {\mc I}_\sy$ can be included in the term $\hat{\mc I}_\sy\otimes \hat H_\ba$. 

Let us now consider two projection superoperators $\mc P$ and $\mc Q$, which act on the Liouville space of the total (electronic and nuclear) system. We take the two projectors to be complementary to each other, that is $\mc P+\mc Q=\textit 1$, where $\textit 1$ denotes the identity superoperator. The dynamics of the propagator %is decomposed along the directions of projection as 
can be decomposed as
\cite{nakajima1958,zwanzig1960,mori1965}
\begin{equation} \label{eq:eom_prop}
\frac{\mathrm d}{\mathrm d t}\e^{{{\mathcal L}} t}
= \e^{{\mathcal L} t}\mathcal P{\mathcal L}+\mathcal Q \e^{ \mathcal { LQ} t}\mathcal {L} +\int_0^t \mathrm d \tau \; \e^{\mathcal { L} (t-\tau)} \mathcal {PLQ} \e^{\mathcal{LQ}\tau}\mathcal {L},
\end{equation}
where we define the Liouvillian ${\mc L \cdot= i[\hat H, \cdot]}$ and take $\hbar=1$ throughout.
Spin--spin correlation functions can be written in terms of the inner product in Liouville space \cite{gyamfi2020}
\begin{equation}\label{eq:scal_Liou}
\C_{\mu\nu}(t) =  \tr[\hat \rho_0\hat \sigma_\mu  \hat \sigma_\nu(t)] = \llangle\hat\rho_0 \hat\sigma_\mu|\e^{\mathcal L t}|\hat \sigma_\nu\rrangle.
\end{equation}
With the given definition of $\hat \rho_0$ in \cref{eq:rho0}, this is normalized such that $\C_{\mu\nu}(0) = \delta_{\mu\nu}$.

A non-Markovian equation of motion for \cref{eq:scal_Liou} can be derived from \cref{eq:eom_prop} with a Redfield-type projection superoperator \cite{Montoya2016GQME}
\begin{equation}\label{eq:P}
\mc P = | \hat \sigma_\lambda \rrangle \llangle \hat \rho_0 \hat \sigma_\lambda |.
\end{equation}
By using the above expression for $\mathcal P$ in \cref{eq:eom_prop} and multiplying from the left by $\llangle\hat \rho_0 \hat \sigma_\mu |$ and from the right by $| \hat \sigma_\nu \rrangle$, we obtain the GQME (in matrix notation)
\begin{align}\label{eq:GQME}
\frac{\mathrm d}{\mathrm dt}\C(t)&= \C(t)\mathcal X-\int_0^t \mathrm d \tau\; \C(t-\tau) \K(\tau),
\end{align}
where we defined
\begin{equation}\label{eq:Xgen}
\mathcal X_{\mu\nu} = \llangle{\hat \rho_0\hat \sigma_\mu}|\mathcal L| {\hat \sigma_\nu}\rrangle = \dot \C_{\mu\nu}(0),
\end{equation}
and the memory kernel is
\begin{equation}\label{eq:def_K}
\K_{\mu\nu}(t) = -\llangle\hat \rho_0\hat \sigma_\mu|{\mathcal L}\mathcal{Q}\e^{{\mathcal L}\mathcal {Q} t}\mathcal{Q}{\mathcal L} |\hat \sigma_\nu \rrangle.
\end{equation}
Given $\mc X$ and $\mc K(t)$, \cref{eq:GQME} is solved for $\mc C(t)$ using a numerical 
integro--differential equation solver based on the trapezoidal rule. \cite{NumRep}

$\mc X$ can be be evaluated exactly [\cref{eq:X}] %, in principle exactly,%\footnote{For the spin--boson model, this can be done analytically, but in general, one would require a molecular dynamics simulation which could in principle be converged to the exact result.}
and can be shown to be antisymmetric.  Also, $\mathcal X_{\mu 0}=0$, given that $\mc L|\hat\sigma_0\rrangle=0$.
%\tcb{%keep this comment
%\begin{align}
%&\tr[\hat \rho_0\hat\sigma_\mu \mc L\hat\sigma_\nu]=\tr[\hat \rho_0\hat\sigma_\mu i [\hat H_\sy+\hat H_\syba+\hat H_\ba, \hat\sigma_\nu]]\nn\\
%&=\tr[\hat \rho_0\hat\sigma_\mu(i[\hat H_\sy,\hat \sigma_\nu]+i\hat V_{\ba,k}\otimes[\hat\sigma_k,\hat\sigma_\nu])]\nn\\
%&=\tr[\hat \rho_0\hat\sigma_\mu(i\hat H_\sy\hat\sigma_\nu-i\hat \sigma_\nu\hat H_\sy)]\nn\\
%&=\tr[\hat \rho_0\hat\sigma_\nu i\hat\sigma_\mu\hat H_\sy]-\tr_\sy[i\hat\rho_0 \hat H_\sy\hat\sigma_\mu\hat \sigma_\nu]\nn\\
%&=\tr[\hat \rho_0\hat\sigma_\nu i\hat\sigma_\mu\hat H_\sy]-\tr_\sy[i\hat\rho_0\hat \sigma_\nu \hat H_\sy\hat\sigma_\mu]= -\tr[\hat\rho_0\sigma_\nu\mc L\hat\sigma_\mu]\nn
%\end{align}}
The calculation of the kernel is more involved.
In particular, the projected propagator $\e^{{\mathcal L}\mathcal {Q} t}$ in \cref{eq:def_K} generates 
%compressible 
non-Hamiltonian dynamics which cannot be easily approximated by a quasiclassical scheme. To circumvent this issue, we follow the seminal papers of Shi and Geva \cite{shi2003,Shi2004GQME} and 
introduce the auxiliary memory kernels involving only the full propagator $\e^{\mc L t}$,
\begin{subequations}\label{eq:Kaux}
\begin{align}
\K_{\mu\nu}^{(1)}(t) &= -\llangle\hat \rho_0\hat \sigma_\mu|{\mathcal L}\mathcal{Q}\e^{{\mathcal L}t}\mathcal{Q}{\mathcal L}  |\hat \sigma_\nu \rrangle, \label{eq:K1} \\
\K_{\mu\nu}^{(3)}(t) &=-\llangle\hat \rho_0\hat \sigma_\mu|{\mathcal L}\mathcal{Q}\e^{{\mathcal L}t}|\hat \sigma_\nu\rrangle.\label{eq:K3} 
\end{align}
\end{subequations}
These can be conveniently rewritten as
\begin{subequations}\label{eq:Kaux_exp}
\begin{align}
\K^{(1)}(t) &= -\dot\C^{\mc L} (t)+\mc X \dot \C(t)+\C^{{\mathcal L}}(t)\mc X -\mc X\C(t)\mc X, \label{eq:K1_der}\\
\K^{(3)}(t) &= - \C^{\mathcal{L}}(t) +\mc X \C(t)\label{eq:K3_der},
\end{align}
\end{subequations}
where 
\begin{equation}\label{eq:CL}
\C^{\mc L}_{\mu\nu}(t) = -\llangle\mc L\hat \rho_0\hat \sigma_\mu|\e^{\mc Lt}|\hat \sigma_\nu\rrangle.
\end{equation}
Note that the two auxiliary kernels are related by the identity \cite{Montoya2016GQME}
\begin{equation}\label{eq:K1_K3}
\K^{(1)}(t) = \dot \K^{(3)}(t) - \K^{(3)}(t)\mc X.
\end{equation}
The full memory kernel, $\K(t)$, is obtained from\cite{Montoya2016GQME,Kelly2016master}
\begin{align} \label{eq:volt_K}
\K(t) = \K^{(1)}(t)+\int_0^t \mathrm d \tau\;\K^{(3)}(\tau)\K(t-\tau),
\end{align}
which is a Volterra equation of the second kind and can be solved using numerical routines.\cite{NumRep} 

We will use quasiclassical techniques to approximate $\K(t)$ via $\C(t)$ and $\C^{\mc L}(t)$ as detailed in \cref{app:GQME_mapping} and \cref{subsec:quasiclass}. 
As shown in previous work,\cite{Kelly2015nonadiabatic} %Ref.~\onlinecite{Kelly2016master},
the solution of the GQME may be significantly different from (and often more accurate than) the original correlation function $\C(t)$ when using approximated methods.
However, if $\C^{\mc L}(t)=\dot\C(t)$, which is true if the correlation functions are obtained with exact quantum dynamics, 
the solution of the GQME is identical to the original correlation function, $\C(t)$.\cite{Kelly2016master}

Note that our formulation of the GQME in terms of Pauli matrices is simply a basis rotation of that used in other works on the master equation, \cite{shi2003,Kelly2015nonadiabatic,Kelly2016master,Montoya2016GQME,Pfalzgraff2019GQME} and will thus lead to identical results if provided with kernels based on the same approximations.
Our choice of the basis, however, leads to slightly simpler expressions which more straightforwardly lend themselves to the analysis we present.
Also, as the Pauli matrices are Hermitian, all correlation functions and kernels are real valued.
Let us finally remark that this formalism for two-state electronic systems could be easily extended to more states by expanding the projectors on generators of the $\mathrm{SU}(N)$ algebra and the identity operator.\cite{multispin}

\subsection{Quasiclassical trajectory methods}\label{subsec:quasiclass}
In the following section we will summarize the main features of the two quasiclassical techniques used in this work. 
\subsubsection{Spin mapping}
Spin mapping\cite{spinmap,multispin,chimia_mapping2022,runesonPhD} is a linearized semiclassical approach which approximates the correlation function in \cref{eq:scal_Liou} with
\begin{equation}\label{eq:C_mn_SM}
\C^{(\W)}_{\mu\nu}(t)=\frac 12\int\mr d q\, \mr d p\;\rho_\ba(q,p)\int \mr d \bm{u} \;\sigma_\mu^{(\W)}(\bm u)\sigma_\nu^{(\W)}(\bm u_t).
\end{equation}
Here, $\bm u$ is a vector on the Bloch sphere with $|\bm u|=1$ and $\mr d\bm u = \frac 1 {2\pi}\mr d \varphi\,\mr d(\cos\theta)$, where $\varphi$ and $\theta$ are the angles in spherical polar coordinates.
The nuclear degrees of freedom are replaced by their classical analogue ($\hat q_\alpha\mapsto q_\alpha$, $\hat p_\alpha\mapsto p_\alpha$), and 
$\rho_\ba(q,p)$ is a phase-space representation of the bath distribution (e.g., a Wigner function or classical distribution), normalized such that $\int\mr d q\,\mr d p \; \rho_\ba(q,p)=1$.

The spin-mapping representation of Pauli matrices in \cref{eq:C_mn_SM} is given in terms of the  Stratonovich--Weyl kernel in the so-called $\W$-representation\cite{stratonovich1957} 
\begin{equation}\label{eq:SWker}
\hat w_{\W}(\bm u)= \tfrac 12 (\hat\sigma_0+r_\W u_k\hat\sigma_k),
\end{equation}
such that
\begin{align}
\sigma_\mu^{(\W)}(\bm u)&=\tr_\sy[\hat\sigma_\mu\hat w_\W(\bm u)]
=\begin{cases}
1,&\mu= 0\\
\rW u_\mu,& \mu=x,y,z
\end{cases}.
\end{align}
In the spin-mapping formalism, the vector with components $\sigma_k^{(\W)}(\bm u)$, for $k=x,y,z$, is thus fixed to a sphere with radius $r_\W=\sqrt 3$ (i.e., larger than the Bloch sphere). This specific choice of the radius guarantees that the initial value of \cref{eq:C_mn_SM} is equal to the correct quantum--classical result:
\begin{equation}\label{eq:init_SM}
\frac 12\int \mr d \bm{u} \;\sigma_\mu^{(\W)}(\bm u)\sigma_\nu^{(\W)}(\bm u) = \thalf \tr_\sy [\hat \sigma_\mu\hat\sigma_\nu]=\delta_{\mu\nu},
\end{equation}
where we used
\begin{equation}
\int\mr d \bm u \; = 2, \hspace{5mm}\int\mr d \bm u \; u_i=0, \hspace{5mm}\int\mr d \bm u \, u_i u_j =\tfrac 23\delta_{ij}.
\end{equation}
%In this approach the nuclear Hilbert space is replaced by a fully classical analogue ($\hat q_\alpha\mapsto q_\alpha$, $\hat p_\alpha\mapsto p_\alpha$). 

The time evolution in \cref{eq:C_mn_SM} is generated by the coupled equations of motion
\begin{subequations}\label{eq:eom}
\begin{align}
\dot u_i &= 2\epsilon_{ijk}H_j u_k,\label{eq:dot_ui}\\
\dot q_\alpha &=  \frac{\partial}{\partial p_\alpha} H^{(\W)},\\
\dot p_\alpha &= - \frac{\partial}{\partial q_\alpha} H^{(\W)},
\end{align}
\end{subequations}
where the quasiclassical Hamiltonian is given by
\begin{equation}
H^{(\W)}(q,p,\bm u) = H_\ba(q,p)+H_k(q,p) \sigma_k^{(\W)}(\bm u).\label{eq:HW}
\end{equation}
Here, $H_{\ba}$ and $V_{\ba,k}$ are the classical analogues of $\hat H_{\ba}$ and $\hat V_{\ba,k}$
and
\begin{equation}
H_k(q,p) %= \tfrac 12 \tr_\sy[\hat H\hat\sigma_k]
=H_{\sy,k}+V_{\ba,k}(q,p)
\end{equation}
is the classical analogue of $\tfrac 12 \tr_\sy[\hat H\hat\sigma_k]$.
Finally, $\epsilon_{ijk}$ denotes the Levi--Civita tensor.
The dynamics in \cref{eq:eom} conserve both the norm $|\bm u|$ and the energy according to the Hamiltonian $H^{(\W)}$.
Although it may not be obvious in this form, the dynamics are formally symplectic, which is most easily seen by rewriting in a different coordinate system. \cite{weigert1995,Wang1999mapping,spinmap,Hele2016Faraday}

\subsubsection{Ehrenfest}\label{subsec:ehrenfest}
The Ehrenfest correlation functions can also be written in terms of the propagation of electronic mapping variables on a sphere, the key difference being that the Pauli matrices are represented by $\sigma_k^{(\E)}(\bm u)$  with a radius $r_\E=1$.
The dynamics are generated by the same equations of motion as for spin mapping, \cref{eq:eom}, with a Hamiltonian $H^{(\E)}$ defined in an equivalent form to \cref{eq:HW}, except that $\sigma_k^{(\W)}(\bm u)$ is replaced by $\sigma_k^{(\E)}(\bm u)$.

Given that within the Ehrenfest method, the equivalent of \cref{eq:init_SM} is no longer valid, a focusing procedure is required in order to sample the initial state from points on the Bloch sphere. The choice of the focusing method is not unique, in the sense that a number of different procedures recover the correct initial values. \cite{spinmap,Pfalzgraff2019GQME,Sato2018CFBT,Montoya2016GQME} However, in general, they may lead to different results for $t>0$.

In the present work, we follow an approach based on the focusing on opposite sides of the mapping sphere as suggested in Ref.~\onlinecite{spinmap}, that is we calculate for instance %note in GQME doc
\begin{align}\label{eq:MFT_xn}
\C_{x\nu}^{(\E)}(t) = \frac 12 \left(\braket{\sigma^{(\E)}_\nu(\bm{u}_t)}_{(1,0,0)} - \braket{\sigma^{(\E)}_\nu(\bm{u}_t)}_{(-1,0,0)}\right),
\end{align}
where 
\begin{equation} 
\braket{A(\bm{u}_t)}_{\bm u}=\int \rmd q\,\rmd p\; \rho_\mathrm{b}(q,p) A(\bm{u}_t)
\end{equation}
denotes a phase average over the initial nuclear distribution, with $\bm{u}$ initialized by the value in the subscript. 
$\C_{y\nu}^{(\E)}(t)$ and $\C_{z\nu}^{(\E)}(t)$ are defined in a way equivalent to \cref{eq:MFT_xn} except with the initial Bloch vector defined in the $y$ or $z$ directions, while for the identity operator, we use
\begin{equation}
\C_{\mc I\nu}^{(\E)}(t) =\frac 12\left( \braket{\sigma_\nu^{(\E)}(\bm{u}_t)}_{(0,0,1)} + \braket{\sigma_\nu^{(\E)}(\bm{u}_t)}_{(0,0,-1)}\right).
\end{equation}

We compared our choice with a different approach which involves uniform sampling on the equator of the Bloch sphere in order to initialize the state in a coherence (i.e., $\sigma_x$ or $\sigma_y$). \cite{Pfalzgraff2019GQME} Although the two methods are not formally identical, we could not identify any significant difference between the two in the numerical results. 

\section{Results and analysis}\label{sec:results}
In this section we discuss our results obtained from the combination of the GQME with either Ehrenfest or spin mapping. After introducing the model system, we investigate the extent to which the master equation improves the long-time dynamics compared to the direct propagation of correlation functions. Finally, we discuss the reasons behind the improvements offered by this approach.
\subsection{Model}\label{subsec:model}
We study the dynamics of the spin--boson model, given by the Hamiltonian \cite{Leggett1987spinboson}
\begin{subequations}\label{eq:Hsp_bo}
\begin{align}
\hat H_\sy &= \Delta \hat \sigma_x+\varepsilon \hat \sigma_z, \label{eq:H_S} \\
 \hat H_\ba &= \frac 12\sum_{\alpha=1}^F\left(\frac { \hat p_\alpha^2}{m_\alpha}+m_\alpha\omega_\alpha^2\hat q_\alpha^2\right),\label{eq:H_B}\\
\hat H_\syba& = \hat \sigma_z \otimes \sum_{\alpha=1}^F c_\alpha\hat q_\alpha. \label{eq:H_SB}
\end{align}
\end{subequations}
The two electronic states, $\ket{\pm}$, are separated by an energy bias of $2\varepsilon$, and $\Delta$ denotes the coupling between those states.
The system--bath coupling constants $c_\alpha$, the frequencies $\omega_\alpha$, and the masses $m_\alpha$ of $F$ nuclear modes are determined by the spectral density of Ohmic form
\begin{equation}\label{eq:spectral}
J(\omega)=\frac{\pi \xi}{2}\omega \e^{-\omega/\omega_c},
\end{equation}
where $\xi$ and $\omega_c$ denote the Kondo parameter and the cutoff frequency respectively.
\Cref{eq:spectral} is discretized following Ref.~\onlinecite{craig2005}, to give
\begin{equation}\label{eq:JF}
J_F(\omega) = \frac \pi 2 \sum_{\alpha=1}^F \frac{c_\alpha^2}{m_\alpha\omega_\alpha}\delta(\omega-\omega_\alpha).
\end{equation}
From a comparison between \cref{eq:H_SB_gen} and \cref{eq:H_SB}, we identify $\hat V_{\ba,z} = \sum_{\alpha=1}^F c_\alpha\hat q_\alpha$ as the only non-zero system--bath coupling function. 

Unless otherwise stated, in all our simulations we fix $\Delta=1$, $\xi=1$, $\omega_{\mathrm c}=1$, $\beta=0.3$ and $\hbar = 1$, and vary $\varepsilon$ between 1 and 7. 
As discussed in \cref{eq:K_elem}, the GQME kernel is defined in terms of correlation functions involving both nuclear and electronic operators. We found that these tend to require more bath modes to converge than the purely electronic correlation functions. Therefore, to calculate the GQME we considered a bath of $F=400$ modes, while $F=100$ sufficed to calculate the direct dynamics of the spin--spin correlation functions.
The masses of the nuclear modes are arbitrary and were thus set to $m_\alpha=1$ for numerical convenience.
Here we restrict our analysis to high temperature and small nuclear frequencies, such that a classical treatment of the nuclear dynamics is justified. 
Consistent with the assumption of classical nuclear dynamics inherent to both Ehrenfest and spin mapping, we sample the initial bath modes from a classical Boltzmann distribution, rather than from the common choice of a Wigner distribution. This choice simplifies our formal analysis of the long-time limits under the ergodic hypothesis as we do not have to worry about possible zero-point energy leakage in the case of a Wigner distribution.\cite{Habershon2009water}
The difference between the two distributions can be quantified by the error
\begin{align}\label{eq:Wig_class}
\frac{\beta\omega_{\mathrm c}}{2}-\tanh\left(\frac{\beta\omega_{\mathrm c}}{2}\right)&=  \frac 1{24}\beta^3\omega_c^3 \simeq 1.1\times 10^{-3}\ll 1,
\end{align}
which appears to be negligible in our parameter regime; this implies that a classical treatment of the nuclei is a valid approximation.
We performed tests (not shown) which confirm that the results
throughout the paper are very similar when initialized in a Wigner distribution and would not affect any of the conclusions.

\subsection{Long-time population}\label{subsec:LT_pop}
\begin{figure*}
\centering
\includegraphics[width=5.25in]{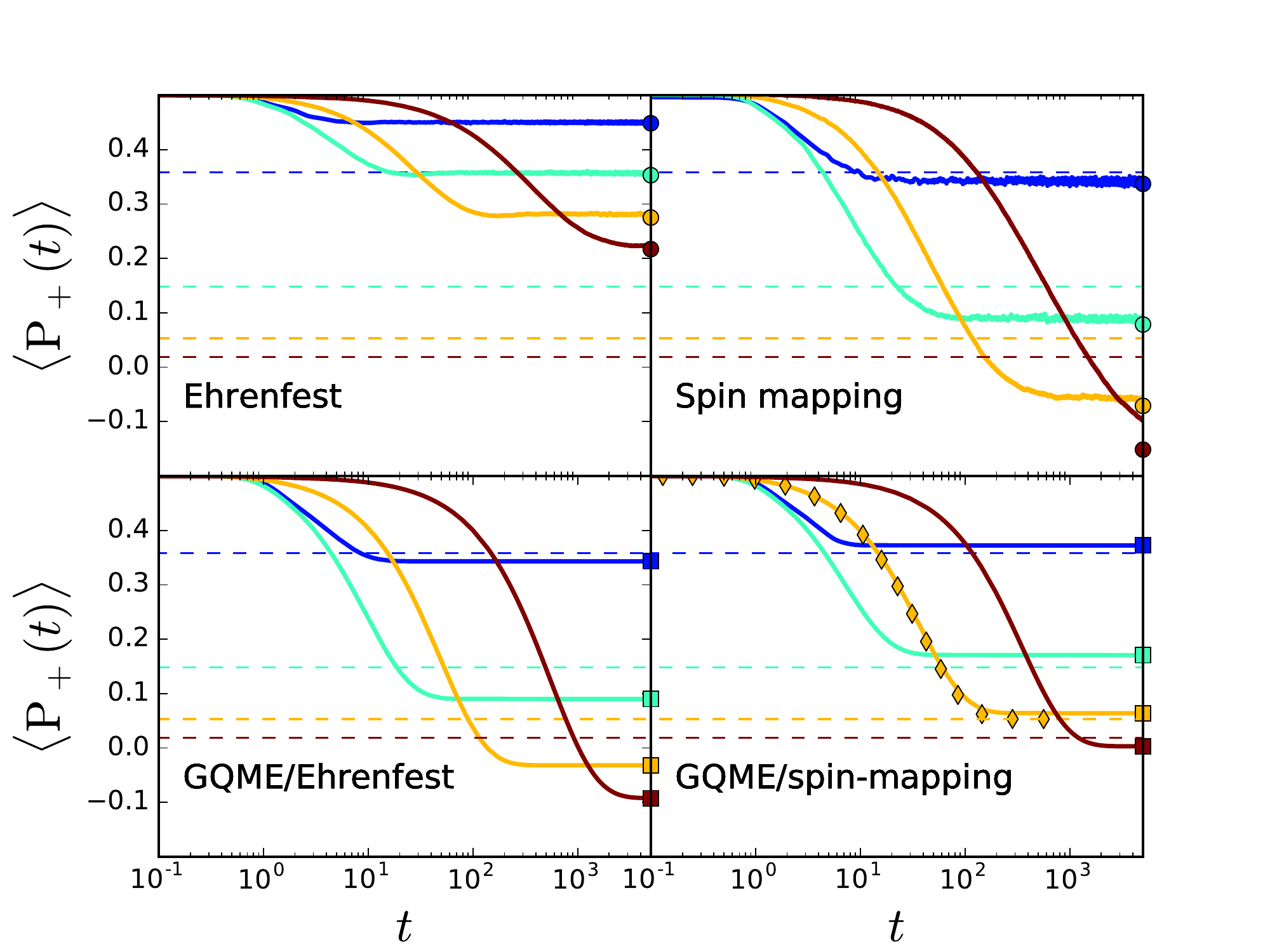}\caption{Average population of state $\ket +$ as a function of time, for four values of the energy bias: $\varepsilon=1$ (blue), $\varepsilon=3$ (turquoise), $\varepsilon=5$ (yellow) and $\varepsilon=7$ (brown). The dashed lines in the four panels indicate the correct quantum--classical equilibrium values. The two upper panels show the direct dynamics of Ehrenfest and spin mapping. The circles at the final time indicate the predictions from the ergodic hypothesis in \cref{eq:Pthermal}. The two lower panels show the solution of the GQME in conjunction with the two quasiclassical techniques. The squares at the final time mark the solution of the long-time population of the GQME, obtained from integrals over the memory kernels [\cref{eq:sigmaz_K}]. Finally, the diamonds in the lower-right panel correspond to the QUAPI result for $\varepsilon=5$ only. }\label{fig:Pexc}
\end{figure*}

We study the long-time dynamics of the population of the higher-energy electronic state, $\hat{\mathrm P}_+ = \ket +\bra +$, for increasing values of the energy bias. %, $\varepsilon$.
In the two upper panels of \cref{fig:Pexc} we show numerical results for 
\begin{equation}\label{eq:P_pm}
\langle\Pop(t)\rangle = \tfrac 12\left[1 + \C_{\mc I z}(t)\right],
\end{equation}
obtained directly from trajectory simulations. 

The quasiclassical Hamiltonian dynamics of both Ehrenfest and spin mapping conserve the energy and the norm of Bloch vector, $|\bm u|=1$.
We thus define the electronic phase space by the surface of the Bloch sphere and note that, in general (i.e., as long as the couplings $\Delta$ and $\xi$ are non-zero), there are no other conserved quantities within this space.
We therefore expect the quasiclassical dynamics of both methods to be ergodic \cite{hawkins2021} on the surface of the Bloch sphere.
Within this assumption, the quasiclassical average of the electronic population is expected to relax at long times to the canonical phase-space average \cite{evans2009,mauri1993}  
\begin{subequations}\label{eq:Pthermal}
\begin{align}\label{eq:ergodic}
\lim_{t\to\infty} \langle \Pop(t)\rangle &=  \langle \Pop\rangle_\eq = \frac 1 {Z}\int \mr d q\,\mr d p\int \mr d\bm u\; \e^{-\beta H^{(\mathrm{m})}}\Pop,\\
Z &= \int \mr d q\,\mr d p\int \mr d\bm u\; \e^{-\beta H^{(\mathrm{m})}},\hspace{6mm}\mathrm{m}=\E,\W.
\end{align}
\end{subequations}

The circles in \cref{fig:Pexc}, calculated from the theoretical prediction of the long-time limit in \cref{eq:Pthermal}, agree satisfactorily with the final plateaus of the correlation functions, implying that the ergodic assumption is valid. Despite such internal consistency, we notice that for both quasiclassical methods the long-time limits deviate significantly from the correct quantum--classical thermal distribution (defined by a quantum trace over the electronic states and a classical phase-space integral over the nuclear variables, following Eq.~(7) from Ref.~\onlinecite{mauri1993}).
These benchmark results are shown in the picture as horizontal dashed lines (and calculated with one-dimensional numerical integration after rewriting the spin--boson Hamiltonian in the reaction coordinate picture \cite{Thoss2001hybrid,wang2017,ellipsoid}).
Note that these are approximately, but not exactly, equal to $1/(1 + \eu{2\beta\varepsilon})$.

We note that the Ehrenfest method overpredicts the long-time population of state $\ket +$ in all cases. Spin mapping is significantly more reliable for small biases, but as we increase $\varepsilon$ this method predicts unphysical negative values. This is a well-known limitation of the spin-mapping approach, \cite{multispin} as well as of several other quasiclassical techniques.\cite{Mueller1998mapping,identity,FMO,chimia_mapping2022,bellonzi2016}

The two lower panels of \cref{fig:Pexc} show the results of the dynamics obtained from the solution of the GQME [\cref{eq:GQME}], using kernels calculated with either Ehrenfest or spin-mapping trajectory simulations. 
%This is carried out numerically using an integro--differential equation solver based on the trapezoidal rule. \cite{NumRep}
For systems with a weak asymmetry, the GQME significantly improves the Ehrenfest result (similarly to what has been shown in previous work \cite{Kelly2015nonadiabatic,Kelly2016master,Montoya2016GQME,mulvihill2019_no_sb}), whereas the spin-mapping result (which is already accurate from direct calculations) is barely changed.
However, for stronger asymmetry, the solution of the GQME exhibits important differences. The results are in both cases shifted in the right direction, but even though the populations of Ehrenfest are guaranteed to be positive if calculated directly, when coupled to the GQME formalism this method predicts negative populations. 
Conversely, the problematic negative populations of spin mapping appear to be resolved by the GQME procedure.

The GQME/spin-mapping approach is also capable of capturing the correct intermediate-time dynamics of these systems. %of the systems studied in \cref{fig:Pexc}.
This can be seen by comparing the results in the lower-right panel of \cref{fig:Pexc} for $\varepsilon=5$ with the numerically exact solution calculated with the quasiadiabatic propagator path integral technique (QUAPI), \cite{makri1995,shi2003} shown for this case only as yellow diamonds. 
We note that QUAPI describes the quantum dynamics of both the electronic and the nuclear subsystems. Thus, the agreement between spin mapping and QUAPI further confirms that the assumption of classical nuclei is valid for the parameter regime considered in this work.

As discussed in detail in \cref{app:zero_modeK}, it is possible to predict the long-time limits of the GQME directly from the time integral of the memory kernels.
In \cref{eq:sigmaz_K} we present a closed-form expression derived from a consideration of the stationary state of the GQME\@. %the Fourier--Laplace transform of the GQME\@.
We include the predictions from this formula as square markers on the lower panels of \cref{fig:Pexc}.
The results are virtually identical to the limits predicted by the time-dependent solution of the GQME, confirming the validity of this useful analysis tool.
%This integral formula will be used in later discussions in the paper.

\begin{figure}
\centering
\includegraphics[width=3.5in]{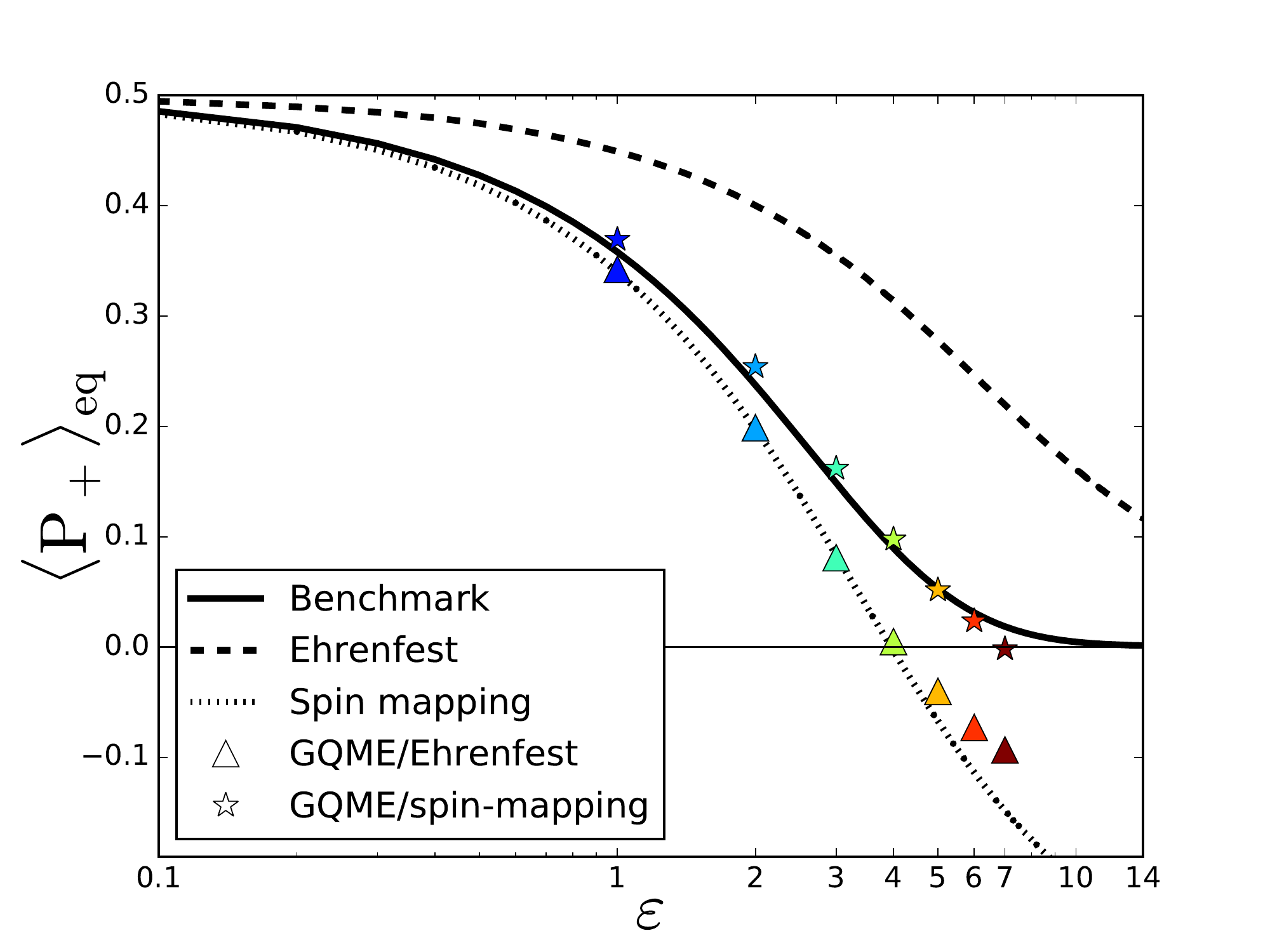}\caption{Predictions for the equilibrium population of state $\ket +$ from different methods. The dashed and dotted black lines denote the theoretical limits computed from the ergodic hypothesis in \cref{eq:Pthermal}, respectively in the case of Ehrenfest and spin mapping. 
The solid line (``Benchmark'') denotes the correct result expected in thermal equilibrium for a mixed quantum--classical system. \cite{mauri1993,ellipsoid}
The colored markers correspond to the solution of the GQME with the kernels calculated with Ehrenfest (triangles) and spin mapping (stars).
The color scheme indicates increasing values of $\varepsilon=1,2,\dots,7$ from blue ($\varepsilon=1$) to brown ($\varepsilon=7$). 
All the long-time limits from the GQME have been calculated by fixing the cutoff time of the kernels to $t_{\mathrm{cut}}=5$. 
We refer to \cref{subsec:cutoff} for a discussion on the issue of determining a well-defined cutoff time for GQME/Ehrenfest.}\label{fig:phase_diag}
\end{figure}

Our findings are summarized in \cref{fig:phase_diag}, which shows the thermal population of state $\ket +$ as a function of the energy bias. 
Here, we include the theoretical predictions from \cref{eq:Pthermal} for spin mapping (black dotted line) and Ehrenfest (black dashed line). The numerically exact quantum-classical benchmark is shown here as a black solid line. The limits calculated from the GQME are shown as stars for spin mapping and as triangles for Ehrenfest, with the same color code as in \cref{fig:Pexc}.
The differences in accuracy between GQME/Ehrenfest and GQME/spin-mapping are evident, and GQME/spin-mapping appears to be the most reliable method overall. Note that it is a coincidence that the predictions of GQME/Ehrenfest lie almost on top of the direct spin-mapping predictions.
In particular, we will show in Sec.~\ref{subsec:cutoff} that these Ehrenfest results are somewhat arbitrary, as a plateau for the cutoff time cannot be uniquely defined.

The fact that GQME/Ehrenfest can predict negative populations demonstrates that the GQME formalism is not guaranteed to be positive definite if coupled with approximate quasiclassical methods. While it is known that the Lindblad master equation is always positive definite, \cite{OpenQuantum,manzano2020} in general even Markovian master equations may not be unless the kernels are obtained in a very careful manner.\cite{rivas2010}
Since the GQME is more general than either of these special cases, our results are in line with what is formally known about master equations.
We could not identify any specific reason leading us to conclude that the populations calculated by GQME/spin-mapping are always guaranteed to be positive; however we did not observe an example of negative populations from this approach.
It is possible that the populations obtained from GQME/spin-mapping will become negative for $\varepsilon\ge 8$. Testing this hypothesis is not trivial, since for such strong asymmetries the approximated memory kernels exhibit large oscillations which do not decay within a short timescale. 
It is therefore not straightforward to define a suitable cutoff time to solve the GQME, as we will discuss further in \cref{subsec:cutoff}.

\subsection{Cutoff times}\label{subsec:cutoff}
\begin{figure}
\includegraphics[width=3.25in]{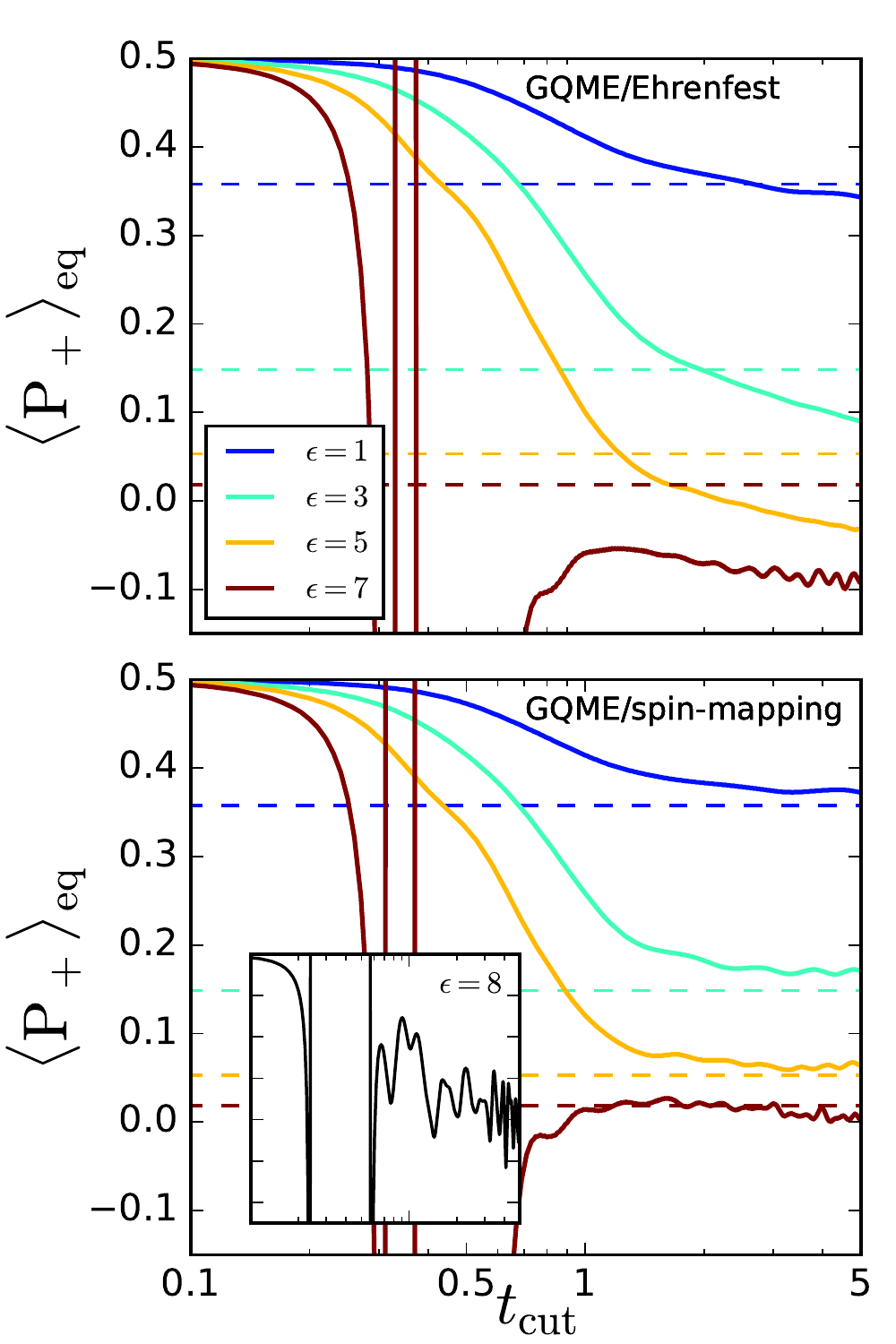}\caption{Equilibrium population of state $\ket +$ obtained by solving the GQME for increasing values of the cutoff time, $\tcut$, with either Ehrenfest (upper panel) or spin mapping (lower panel). Results are shown for increasing values of the energy bias: $\varepsilon=1$ (blue), $\varepsilon=3$ (turquoise), $\varepsilon=5$ (yellow) and $\varepsilon=7$ (brown). The quantum--classical benchmarks are included in both panels as dashed horizontal lines. The inset in the lower panel shows the result of GQME/spin-mapping for the system with $\varepsilon=8$; the axes correspond to the same scale as the main figure. Results for this system (and higher values of $\epsilon$) are not included in the present analysis because of strong oscillations in the memory kernel. }\label{fig:pop_cutoff}
\end{figure}

The choice of the cutoff time, $\tcut$, for the memory kernels can have a significant impact on the accuracy of the results of the GQME. \cite{kidon2015,pfalzgraff2015,Montoya2016GQME}
To examine the dependence of the final electronic population on $\tcut$, we can define
\begin{equation}\label{eq:Kcut}
{\tilde \K}_{\mr{cut}} = \int_0^{\tcut}\mr d \tau \; \K(\tau).
\end{equation}
The limit of $\tcut\to\infty$ in \cref{eq:Kcut} corresponds to the zero-frequency limit of the Fourier--Laplace transform %\cite{beerends2003} 
$\tilde \K(0)$ [\cref{eq:FLT}].
The components of ${\tilde\K}_{\mr{cut}}$ can be inserted into an identity [\cref{eq:sigma_K}] which relates them to the final long-time limits of the GQME\@. %, \tcb{following an approach similar in concept to previous work.\cite{cohen2013}}
We can thus find the relation between the population of the higher-energy state from the GQME and the cutoff time, $\tcut$, shown in \cref{fig:pop_cutoff} for increasing values of the energy bias.
We notice that for all considered systems the population calculated by GQME/Ehrenfest does not plateau to any limit for the considered range of cutoff times. In fact, we could not identify a clear convergence even by increasing the cutoff range up to $\tcut=15$ (not shown). This indicates that a unique solution of the GQME from Ehrenfest is ill-defined in these cases. 

We note that we could in principle artificially extract an optimal cutoff time
%for three of the lines
for $\varepsilon=1$, $\varepsilon=3$ and $\varepsilon=5$, such that the long-time population of GQME/Ehrenfest would be correct by construction.
This is obviously only possible if the exact result is known, and even then the procedure would not be physically justified.
However, in the case of $\varepsilon=7$ (brown line), all reasonable cutoffs for GQME/Ehrenfest result in negative populations. Moreover, an optimal cutoff does not guarantee any improvement in the accuracy of the dynamics at intermediate times.

The same issue in the definition of the cutoff does not occur in the case of spin mapping, shown in the lower panel of \cref{fig:pop_cutoff}. In that case, a plateau for all systems considered allows us to clearly identify the long-time population as determined by the GQME, even without knowing the correct result beforehand.
In all our simulations summarized in \cref{fig:phase_diag} we fixed a cutoff time $\tcut =5$ (corresponding to the last point of \cref{fig:pop_cutoff}). 
This is somewhat arbitrary for Ehrenfest but well justified for spin mapping.

In the inset of the lower panel of \cref{fig:pop_cutoff} we show the solution of the population of GQME/spin-mapping for the system with $\varepsilon=8$. In this case the population clearly fails to converge as a function of the cutoff time for any value $\tcut\le 5$. The issue is due to strong nonvanishing oscillations in some of the components of the memory kernel. Here GQME/spin-mapping is therefore not uniquely defined. Because of this reason, this and any other system with $\varepsilon\ge 8$ are not included in the present analysis. 

\begin{figure}
\centering
\includegraphics[width=3.25in]{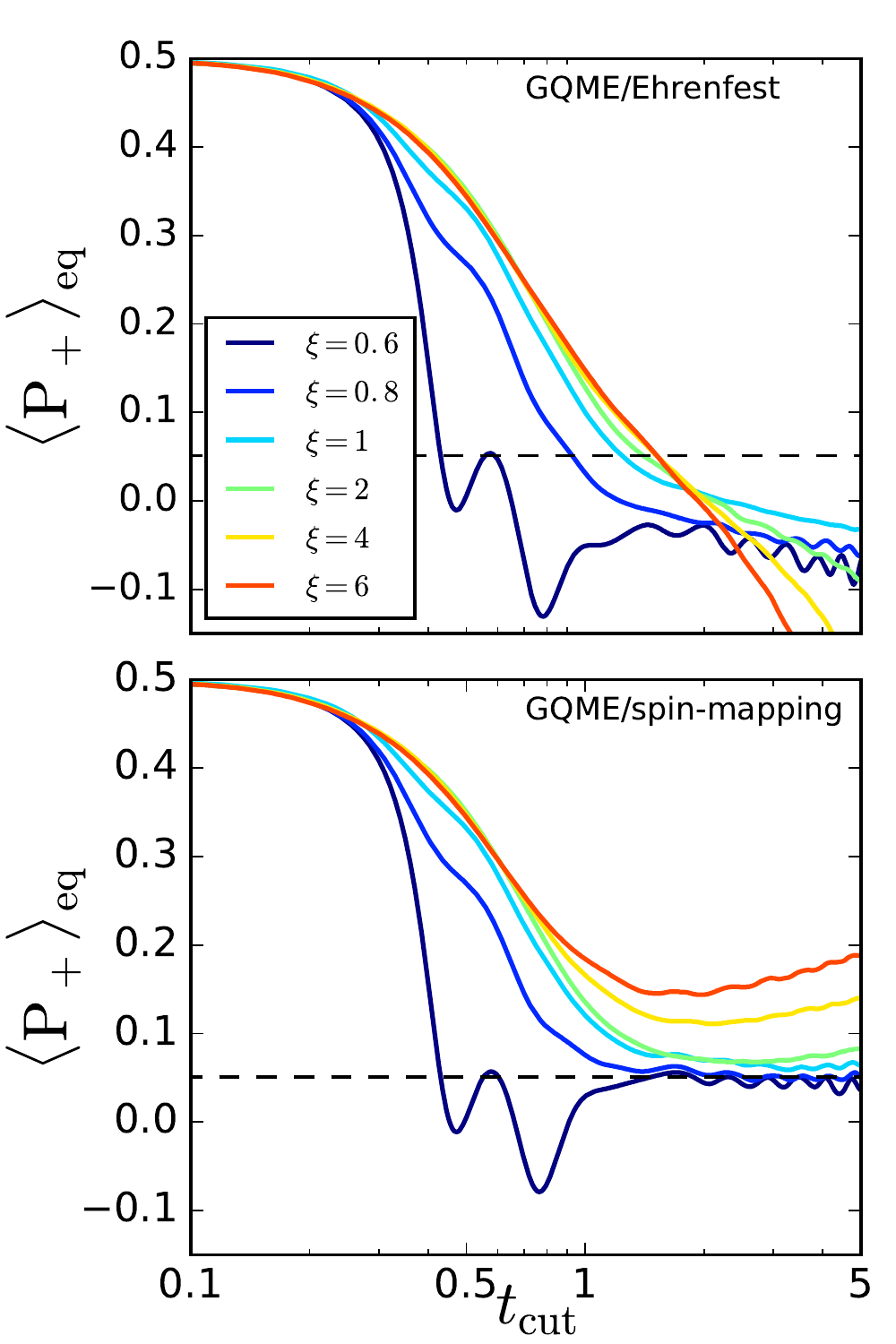}\caption{
Equilibrium population of state $\ket +$ obtained by solving the GQME for increasing values of the cutoff time, $\tcut$, with either Ehrenfest (upper panel) or spin mapping (lower panel).
The different colors correspond to different values of the coupling constant $\xi$, as specified in the legend. The other parameters are fixed to $\beta=0.3$, $\varepsilon=5$, $\omega_{\mathrm{c}}=1$ and $F=400$. The black dashed line indicates the exact quantum--classical value of the thermal population, which is virtually identical for all the considered values of $\xi$. 
Results for $\xi<0.6$ become increasingly oscillatory for large values of $\tcut$.
}\label{fig:pop_cutoff_xi}
\end{figure}

%Nonetheless, GQME/spin-mapping is not guaranteed to return accurate results in every case.  
In addition, we examine how the accuracy of the solution of GQME/Ehrenfest and GQME/spin-mapping is affected by varying the system--bath coupling strength, $\xi$, while keeping $\varepsilon=5$ fixed. 
The results from our numerical analysis are shown in \cref{fig:pop_cutoff_xi}. %, which exhibits the solution of the master equation as a function of the cutoff time $\tcut$ for $\varepsilon=5$ and several values of $\xi$. 
For this range of values of $\xi$, the solution of the thermal populations appears to be virtually constant, and equal to the black dashed line shown on both plots. 
%Results for GQME/Ehrenfest (upper panel) indicate that the method is not able to predict the correct equilibrium population, given that the solution of the master equation as a function of the cutoff decreases for increasing values of $\tcut$, reaching unphysical negative values for $\tcut\gtrsim 1$ for all considered systems. 
Similarly to what was observed in \cref{fig:pop_cutoff}, the populations predicted by GQME/Ehrenfest assume unphysical negative values, and do not plateau with increasing $\tcut$.
This problem is worsened as the system--bath coupling is increased.
Although the GQME/spin-mapping method is able to accurately predict the equilibrium electronic population for small-to-intermediate values of the coupling ($0.6\lesssim\xi\lesssim 1$),
for larger values of $\xi$, the solution does not plateau by increasing the cutoff (even until $\tcut=15$; not shown).
Even though this is a clear breakdown of GQME/spin-mapping, this method does not appear to be worse than GQME/Ehrenfest and interestingly, no negative populations are observed in the GQME/spin-mapping case for any choice of $\tcut$.
Let us remark that in no case do we see an example where the populations plateau with the cutoff time to a \emph{wrong} result. The result is either correct or clearly does not converge. This may thus be a useful diagnostic to assess whether the predictions of the GQME should be trusted or not in cases where the exact value of the population is not known.
%All curves follow the same trend also for higher values of the cutoff till at least $\tcut=15$ (not shown).
%Even in the opposite limit of coupling $\xi =0.5$ (inset of the lower panel) GQME/spin-mapping exhibits limitations. In particular, due to persistent oscillations of the population as a function of the cutoff time we cannot identify a unique value of the plateau, hence we decided not to include the system in the present analysis.

%With further tests, we were able to identify the breakdown of the method in the limit of strong electronic--nuclear coupling ($\xi\gg 1$), as shown in \cref{fig:pop_cutoff_xi} for the system with $\varepsilon=5$ and a range of values of $\xi$.

In order to capture the correct dynamics in these systems with strong asymmetry or strong system--bath coupling,
%In order to simulate these strongly asymmetric systems,
%other quasiclassical trajectory methods may be advantageous.
%In particular, rigorous surface-hopping methods\cite{MASH} may be better suited to the regime of strong system--bath coupling.
%Alternatively, 
it may be necessary to go beyond linearized semiclassical dynamics and couple the GQME with more accurate but expensive partially linearized density-matrix (PLDM) methods, such as spin-PLDM. \cite{spinPLDM1} %,spinPLDM2,Kelly2013GQME} 
By introducing quantum jumps,\cite{spinPLDM2} the dynamics could in principle be systematically converged to the quantum--classical Liouville equation (QCLE) result.\cite{Kapral1999QCLE}
By combining this with GQME, similarly to Ref.~\onlinecite{Kelly2013GQME}, it should be possible to significantly reduce the number of jumps necessary for convergence.
In the following, we will focus our analysis on a case for which GQME/spin-mapping shows a clear plateau in $\tcut$ (namely $\xi=1$, $\varepsilon=5$) in order to better understand what makes GQME/spin-mapping work where GQME/Ehrenfest fails.

\subsection{Approximation of time derivatives}
We now discuss the reasons why the GQME procedure can improve the predictions of thermal population from quasiclassical approaches. 
It has been shown in Ref.~\onlinecite{Kelly2016master} that if one constructs the memory kernels with $\dot\C(t)$ instead of $\C^{\mc L}(t)$, the solution of the master equation will be identical to the direct correlation function $\C(t)$.
Also, if the correlation functions were evaluated with a time-translationally invariant dynamics, $\dot\C(t)$ and $\C^{\mc L}(t)$ would be identical.
We will however show that within both spin mapping and Ehrenfest the two are different.

In spin mapping, the correlation function defined in \cref{eq:CL} is approximated by 
\begin{align}
\C^{\mc L}_{\mu\nu}(t) &= -\frac 12\int\mr d q\, \mr d p\int \mr d \bm u\;\left\{\tr_\sy[(\mc L \hat \rho_\ba\hat \sigma_\mu)\hat w_\W(\bm u)]\right\}(q,p)\nn\\
&\quad\times \sigma_\nu^{(\W)}(\bm u_t).\label{eq:CL_W}
\end{align}
We have introduced the notation, $\{\hat{\mc A}\}(q,p)$, to indicate an operator in the electronic space and a phase-space function of the nuclear variables.  This is formally obtained by first taking the partial Wigner transform with respect to the nuclear degrees of freedom of the operator $\hat{\mc A}$, and then by taking the limit of classical nuclei, according to the prescription derived in Ref.~\onlinecite{leaf1968}.  Roughly speaking, one simply replaces commutators by Poisson brackets and anticommutators by products of phase-space functions. 

We compare \cref{eq:CL_W} with the ``direct'' time derivatives
\begin{subequations}
\begin{align}
\dot\C_{\mu\nu}(t)&=
\frac 12 \int\mr d q\, \mr d p\int \mr d \bm{u} \; \rho_\ba(q,p) \sigma^{(\W)}_\mu(\bm u) 
\mc L_\W\left(\sigma_\nu^{(\W)}(\bm u_t)\right)
\\ &= -\frac 12 \int\mr d q\, \mr d p\int \mr d \bm{u} \;\mc L_\W\left( \rho_\ba(q,p) \sigma^{(\W)}_\mu(\bm u) \right)
\sigma_\nu^{(\W)}(\bm u_t), \label{eq:dC_W}
\end{align}
\end{subequations}
where $\mc L_\W$ denotes the Liouvillian which generates the quasiclassical dynamics in \cref{eq:eom}
and we have used the property of the Liouvillian to transfer the time derivative from one part of the integrand to the other,\cite{Zwanzig} similar to integration by parts.
In general, $\dot\C(t)\ne\C^{\mc L}(t)$ because
\begin{equation}\label{eq:L_compare} \left\{\tr_\sy[(\mc L \hat \rho_\ba\hat \sigma_\mu)\hat w_\W(\bm u)]\right\}(q,p)\neq \mc L_\W\left( \rho_\ba(q,p)\sigma^{(\W)}_\mu(\bm u) \right).
\end{equation}

To prove \cref{eq:L_compare}, we start by expanding the left-hand side using \cref{eq:Lrhosmu1}:
%\tcb{\begin{align*} %useful comment - keep it
%&\mc L(\hat\rho_\ba\hat\sigma_\mu) =\Big\{\tr_\sy\Big[\tfrac i 2H_{\sy,k}[\hat \sigma_k, \hat  \sigma_\mu]\otimes \hat \rho_\ba \\
%&+ \tfrac i 4\Big([\hat \sigma_k, \hat  \sigma_\mu] \otimes [\hat V_{\ba,k},\hat \rho_\ba]_++[\hat \sigma_k, \hat  \sigma_\mu]_+ \otimes [\hat V_{\ba,k},\hat \rho_\ba]\Big)\Big]\Big\}(q,p)\\
%&\to i^2 (H_{\sy,k}+V_{\ba,k})\epsilon_{k\mu l}\sigma_l^{(\W)}(\bm u)  \rho_\ba \\ &+\frac 1 4 2\tr_\sy\left[[\hat\sigma_k,\hat\sigma_\mu]_+\hat w_\W(\bm u)\right](-\{V_{\ba,k},\rho_\ba\}_{\Pb})
%\end{align*}}
\begin{align}%eq A3 notes Jeremy 
&\left\{\tr_\sy[(\mc L \hat \rho_\ba\hat \sigma_\mu)\hat w_\W(\bm u)]\right\}(q,p)\nn\\
&\quad= 2\rho_\ba( H_{\sy,k}+V_{\ba,k})\epsilon_{\mu k l}\sigma^{(\W)}_l(\bm u)\nn\\
&\quad\quad-\tfrac 12\tr_\sy\left[[\hat \sigma_k, \hat  \sigma_\mu]_+\hat w_\W(\bm u)\right] \{ V_{\ba,k}, \rho_\ba\}_{\Pb}.\label{eq:Lqm}
\end{align}
The generalization of the Levi--Civita symbol in \cref{eq:Lqm} allows one or more indices to be zero,
with $\epsilon_{0\mu\nu}=0$. Additionally, we use $\{H_\ba,\rho_\ba\}_\Pb=0$,
where a Poisson bracket for two (classical) nuclear dynamical variables is defined as
\begin{equation}
\{f, g\}_\Pb = \sum_{\alpha=1}^F\left(\frac{\partial f}{\partial q_\alpha}\frac{\partial g}{\partial p_\alpha}-\frac{\partial f}{\partial p_\alpha}\frac{\partial g}{\partial q_\alpha}\right).
\end{equation}
We can now expand the right-hand side of \cref{eq:L_compare} as
\begin{align}
&\mc L_\W\left( \rho_\ba\sigma^{(\W)}_\mu(\bm u) \right)
= 2\rho_\ba\left(H_{\sy,k}+V_{\ba,k}\right)\epsilon_{\mu k l}\sigma^{(\W)}_l(\bm u) \nn\\
&\quad -\sigma_k^{(\W)}(\bm u) \sigma^{(
\W)}_\mu(\bm u) \{V_{\ba,k},\rho_\ba\}_\Pb ,\label{eq:LW}
\end{align}
While the first terms of \cref{eq:Lqm} and \cref{eq:LW} are equivalent, the same does not necessarily hold for the remaining contributions; in particular
\begin{align}
\tfrac 12\tr_\sy\left[[\hat \sigma_k, \hat  \sigma_\mu]_+\hat w_\W(\bm u)\right]&= \tr_\sy\left[(\delta_{\mu 0}\hat\sigma_k+\delta_{k\mu})\hat w_\W(\bm u)\right]\nn\\
&= \delta_{\mu 0}\sigma^{(W)}_k(\bm u)+\delta_{k\mu},\label{eq:CL_diff}
\end{align}
which is clearly not in general equal to $\sigma_k^{(\W)}(\bm u)\sigma_\mu^{(\W)}(\bm u)$.
However, the two expressions are equal for
$\mu=0$, %, $\delta_{k\mu}=0$ for all $k=1,2,3$ in \cref{eq:CL_diff} and
as $\sigma_0^{(\W)}(\bm u)=1$, and hence $\dot\C_{\mc I\nu}(t)=\C_{\mc I\nu}^{\mc L}(t)$ within spin mapping.
Note that this is also the case when using the identity-corrected MMST \cite{identity,FMO,linearized} but not for the standard linearized semiclassical MMST\cite{Sun1997mapping,Kim2008Liouville,Shi2004GQME,Mulvihill2019LSCGQME}
(unless focused initial conditions\cite{Mueller1998mapping,Bonella2003mapping} are used).
Likewise, the arguments used to derive this result do not apply for Ehrenfest, as in this case one needs to account for the different initial distribution %in the correspondent of 
while deriving the expressions %correspondent 
analogous to \cref{eq:Lqm} and \cref{eq:LW}.

\begin{figure*}
\includegraphics[width=7in]{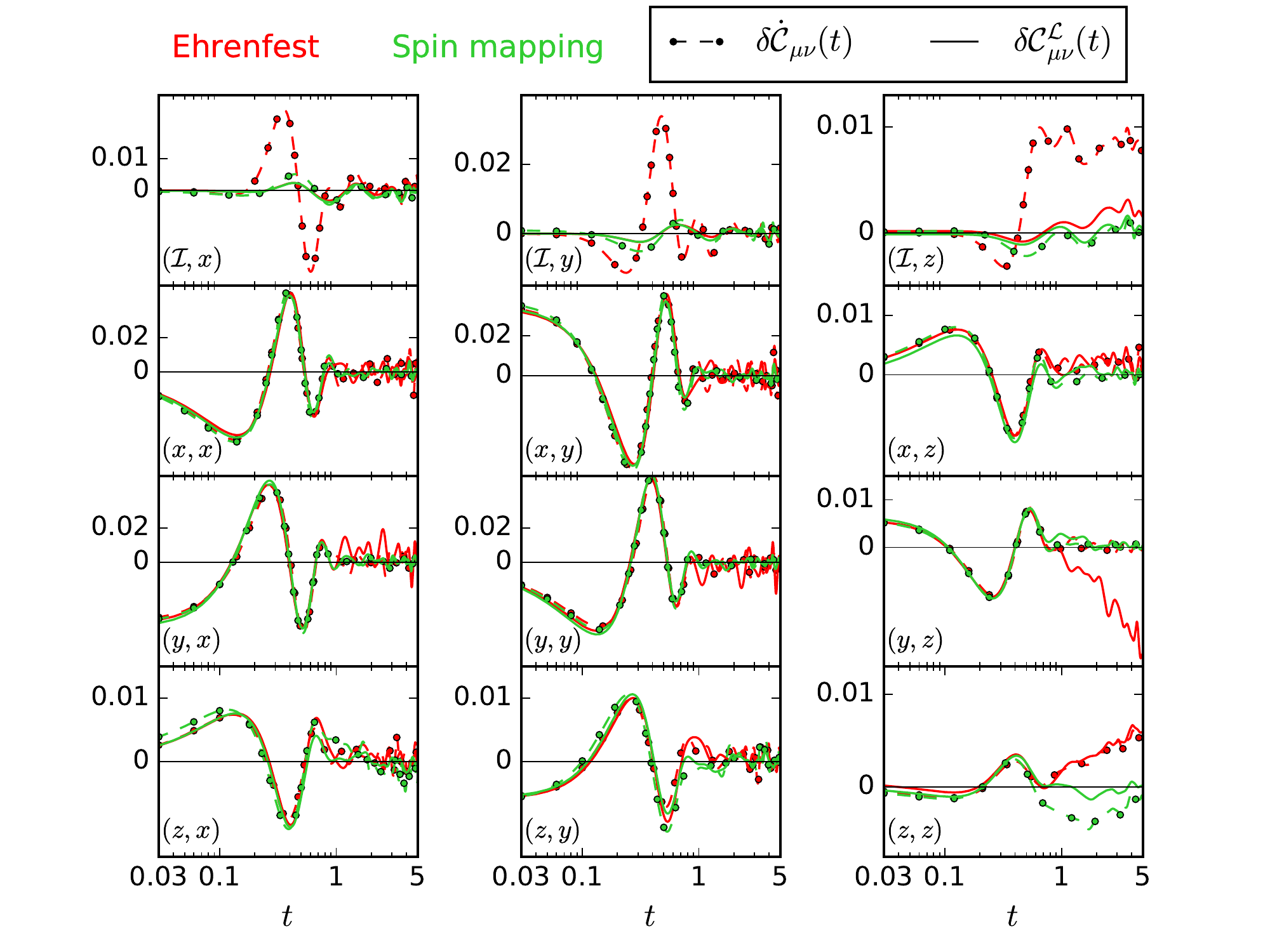}\caption{Error in the first derivatives of the correlation functions according to Ehrenfest (green lines) and spin mapping (red lines), as defined in \cref{eq:delta_der}. We consider here the strongly asymmetric system $\varepsilon=5$. The solid and dashed-marked lines refer respectively to the errors of $\dot \C(t)$  and $\C^{\mc L}(t)$. 
The indices $(\mu,\nu)$ of the components of the correlation functions are shown in the lower left corner of each panel.
We do not include the terms with $\nu=0$, as these are identically zero.
}\label{fig:dC_err}
\end{figure*}

We can study the error of the first derivatives of $\C(t)$ calculated from the two different quasiclassical methods by defining
\begin{subequations}\label{eq:delta_der}
\begin{align}
\delta \dot\C_{\mu\nu}(t) &= \dot\C_{\mu\nu}(t)-\dot\C_{\mu\nu}^{\mathrm{QUAPI}}(t),\label{eq:delta_dC}\\
\delta \C^{\mc L}_{\mu\nu}(t) &= \C^{\mc L}_{\mu\nu}(t)-\dot\C_{\mu\nu}^{\mathrm{QUAPI}}(t),\label{eq:delta_CL}
\end{align}
\end{subequations}
where $\dot\C(t)$ and $\C^{\mc L}(t)$ are calculated with Ehrenfest or spin mapping, while $\dot\C^{\mathrm{QUAPI}}(t)$ denotes the numerically exact time derivative calculated from QUAPI\@. In \cref{fig:dC_err}, we show the errors in \cref{eq:delta_der} for all the non-zero components of the correlation functions, calculated for a representative strongly-biased system ($\varepsilon=5$).

To avoid numerical errors in finite-difference schemes, the direct time derivatives $\dot \C_{\mu\nu}(t)$ are calculated from the expansion of the Liouvillian on the time-evolved operator, according to the exact quantum-mechanical expansion
\begin{subequations}
\begin{align}
\dot \C_{\mu\nu}(t) &= \llangle\hat \rho_0 \hat \sigma_\mu| \e^{\mathcal L t}\mathcal L|\hat \sigma_\nu\rrangle 
\nn\\
&=\llangle\hat \rho_0 \hat \sigma_\mu |\e^{\mathcal L t}|i[\hat H_\sy, \hat \sigma_\nu]+i[\hat H_\syba, \hat \sigma_\nu]\rrangle\nn\\
&=\C_{\mu\lambda}(t)\mathcal X_{\lambda\nu} + \C^{V_{\ba,k}}_{\mu\lambda}(t)\mathcal Y_{\lambda\nu}^{(k)},\label{eq:dC_VB}\\
\C^{V_{\ba,k}}_{\mu\nu}(t) &= \llangle\hat \rho_0 \hat \sigma_\mu |\e^{\mc Lt}|\hat V_{\ba,k} \hat \sigma_\nu\rrangle,
\end{align}
\end{subequations}
where the matrices $\mc X$ and $\mc Y^{(k)}$ are defined in \cref{eq:X} and \cref{eq:Y}, respectively.
The identity \cref{eq:dC_VB} is preserved by both quasiclassical methods. This follows from the fact that the expression for $\mc L \hat \sigma_\nu$ is correctly mapped onto $\mc L_{\mathrm m}\sigma^{(\mathrm m)}(\bm u)$ for $\mathrm m = \W, \E$.
%keep the comment
%\tcb{\begin{align}
%\tr_\sy[ (\mc L \hat \sigma_\nu) \hat w_\W(\bm u)]
%&= \tr_\sy\left[[i [\hat H_\sy\otimes \hat{\mc I}_\ba+\hat\sigma_k\otimes \hat V_{\ba,k}, \hat\sigma_\nu]\hat w_\W(\bm u)\right]\nn\\
%&= \tr_\sy\left[i (H_{\sy,k}\hat{\mc I}_\ba+\hat V_{\ba,k}) 2i\epsilon_{k\nu l}\hat\sigma_l \hat w_\W(\bm u)\right]\nn\\
%&= 2\epsilon_{\nu k l} (H_{\sy,k}+ V_{\ba,k})\sigma_l^{(\W)}(\bm u)
%\end{align}}
%\begin{equation}
%\tr_\sy[ (\mc L \hat \sigma_\nu) \hat w_\W(\bm u)]  = 2\epsilon_{\nu k l} (H_{\sy,k}+ V_{\ba,k})\sigma_l^{(\W)}(\bm u)
%\end{equation}
%(while the same does not apply for $\mc L(\hat\rho_0\hat\sigma_\mu)$, as discussed above). 

By comparing the results from the two derivatives in \cref{fig:dC_err} we note important differences.
For instance, in the first row, the error of $\dot C_{\mc I \nu}(t)$ within Ehrenfest is much larger than  $\C^{\mc L}_{\mc I \nu}(t)$. 
One way to understand this is because $\mc X_{\mu \mc I}=0$ and thus, unlike for $\dot\C_{\mc I \nu}(t)$, no elements of $\C^{\mc L}(t)$ depend directly on $\C_{\mc I\nu}(t)$ [see \cref{eq:CLapp}].
We have found in previous work that the correlation functions $\C_{\mc I\nu}(t)$ tend to be more inaccurate than those initialized by a Pauli matrix.\cite{spinmap,identity}
In fact it is errors in $\C_{\mc I z}(t)$ which lead to the incorrect population prediction of the direct methods, as shown by \cref{eq:P_pm}.
One reason why $\C_{\mc I k}(t)$ is particularly difficult to get right is because it relaxes to a non-trivial value in the long-time limit whereas any correlation function initialized by a Pauli matrix is guaranteed by symmetry to relax to zero.

As we discussed above, we expect no difference between $\dot\C_{\mc I \nu}(t)$ and $\C^{\mc L}_{\mc I \nu}(t)$ calculated by spin mapping. This is consistent with our simulations, within statistical fluctuations and other numerical errors. 
We have therefore lost some flexibility in being able to replace $\dot\C_{\mc I \nu}(t)$ by more accurate values when building the kernel.
Thankfully, the accuracy of these components is already high and this is therefore not a limitation.
Looked at another way, this important relation ensures that $\dot\C_{\mc I \nu}(t)$ is as accurate as $\C^{\mc L}_{\mc I \nu}(t)$, which is one reason why the direct spin-mapping predictions are more reliable than Ehrenfest, even without using the GQME\@.
Note, however, that if it were true that all components of $\dot\C(t)$ were equal to $\C^{\mc L}(t)$, the GQME procedure would simply return the original spin-mapping correlation functions \cite{Kelly2016master}
and we would not be able to use the GQME to improve the result.

Overall, we find that the accuracy of the different components of the derivatives is either comparable or increased by replacing $\dot \C_{k\nu}(t)$ with $\C_{k\nu}^{\mc L}(t)$ within spin mapping and the most dramatic improvement occurs in the $(z,z)$ element.
This trend is consistent with the clear improvements obtained by calculating the dynamics of GQME/spin-mapping rather than direct spin mapping.
Interestingly, however, 
we notice that the error in the Ehrenfest prediction of $\C_{yz}^{\mc L}(t)$ is larger than the error in $\dot\C_{yz}(t)$. This indicates that $\C^{\mc L}(t)$ is not guaranteed to be more accurate than $\dot\C(t)$ (at least within Ehrenfest theory)
and that therefore the GQME procedure may not actually improve the results of the direct simulation.
It is because of this that it is possible to observe unphysical negative populations in the GQME/Ehrenfest predictions.
If this behaviour can be shown to be universal, one might try to improve the reliability of the GQME/Ehrenfest approach by replacing $\C^{\mc L}_{yz}(t)$ by $\dot{\C}_{yz}(t)$ (and maybe other selected elements as well) in the construction of the kernel.
However, we do not expect that this would outperform GQME/spin-mapping in general.

\subsection{Accuracy of memory kernels}\label{subsec:ker}
\begin{figure*}
\centering
\includegraphics[width=6in]{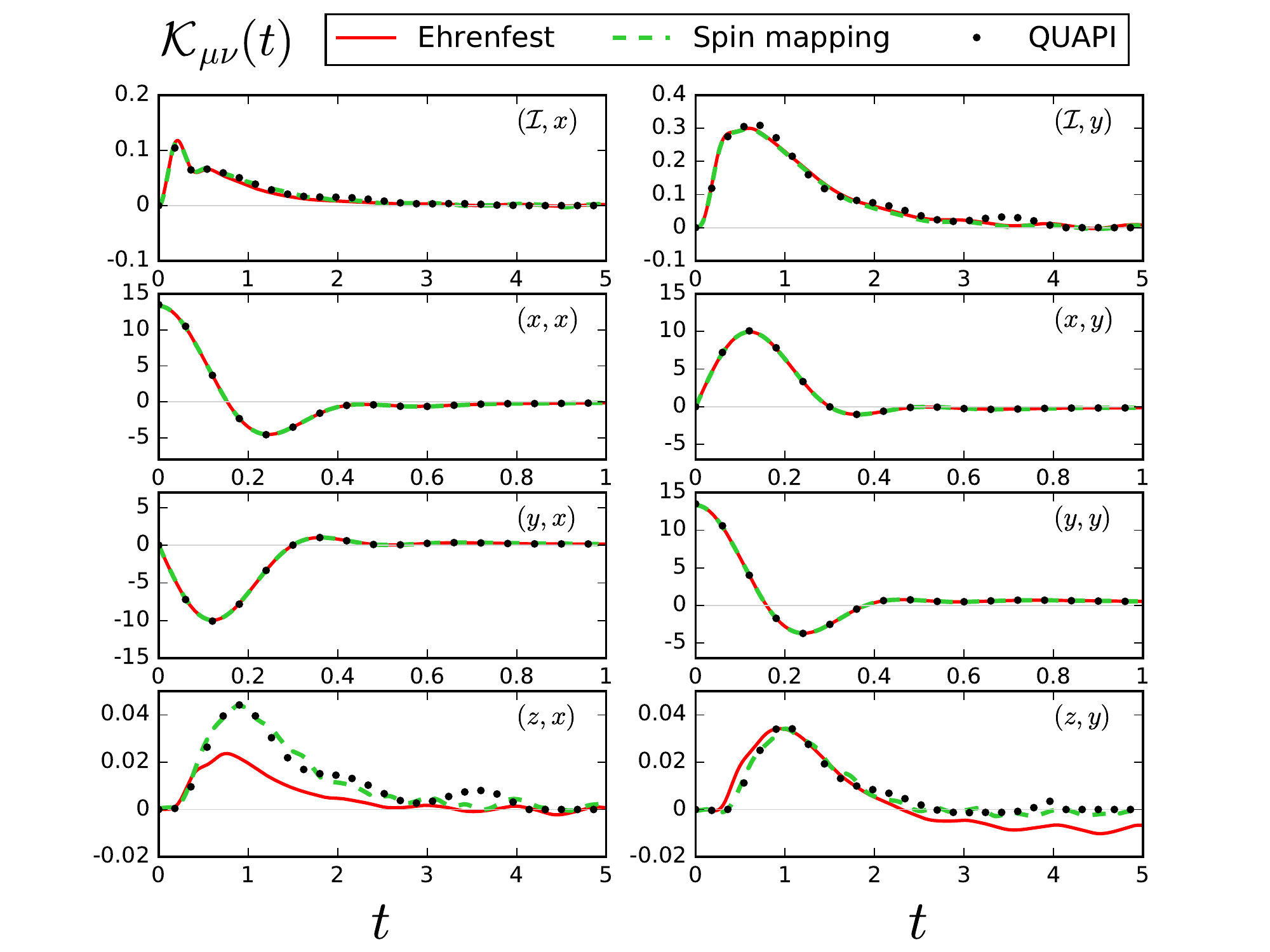}\caption{Non-zero components of the memory kernel $\mc K_{\mu\nu}(t)$ as a function of time, calculated for a representative system with $\varepsilon=5$, from spin mapping (red solid lines) and Ehrenfest (green dashed lines). The black dots denote numerically exact benchmark results calculated with QUAPI\@. The indices $(\mu,\nu)$ of the components of the kernels are shown in the upper right corner of each panel.}\label{fig:kernels_SM_MFT}
\end{figure*}

In this subsection we discuss how the predictions of the long-time dynamics from \cref{subsec:LT_pop} depend on the accuracy of the memory kernels from quasiclassical methods. 
In \cref{fig:kernels_SM_MFT} we show the components of the kernel, obtained from Ehrenfest and spin-mapping simulations.
We include the numerically exact results calculated with QUAPI\@. 
It is clear that spin mapping predicts the kernels to a higher accuracy than Ehrenfest; this ultimately leads to significant improvements in the predictions of the GQME\@. 
We note that in the case of Ehrenfest, the component $\mc K_{zx}(t)$ is particularly inaccurate at short times.
Also, $\mc K_{zy}(t)$ does not decay to zero within the range plotted, as would be expected from the QUAPI benchmark. %This issue is the reason why it is not possible to define suitable cutoff times for the GQME/Ehrenfest results in \cref{fig:pop_cutoff}.
As discussed in detail in \cref{app:zero_modeK}, the long-time populations from the GQME can be expressed as a function of the components of the integrals of the kernel %$\tilde \K(0)=\int_0^{\infty}\mr d \tau\;\K(\tau)$
$\tilde\K_{\mr{cut}}$, according to \cref{eq:sigmaz_K} and the long-time limit of the relationship \cref{eq:P_pm}.
The lack of relaxation of $\K_{zy}(t)$ is the reason why the populations predicted by GQME/Ehrenfest in \cref{fig:pop_cutoff} do not converge as a function of $\tcut$.
\footnote{Note that the quasiclassical approximation of the memory kernels will ultimately relax to zero at long enough times. 
%this in the footnote
In fact, from Eq.~(18), we can calculate
$\mc K(\infty) = -\lim_{\omega\to 0+}i\omega\tilde {\mc K}(\omega) = \left[\mc I_4 -\tilde K^{(3)}(0)\right]^{-1}\mc K^{(1)}(\infty)$, where we made use of the final-value theorem [\cref{eq:FVT}], and $\mc I_4$ is the $4\times 4$ identity matrix.
It is possible to show that $\mc K^{(1)}_{\mu\nu}(\infty)=0$ for all $\mu,\nu$. 
This follows from the \textit{mixing condition} \cite{hawkins2021,ellipsoid} $\langle AB(t)\rangle\to \langle A\rangle \langle B\rangle_{\mathrm{eq}}$ applied to Eq.~(A8a).  Here, by symmetry $\langle A\rangle =0$ and hence $\mc K_{\mu\nu}(\infty)=0$ for all $\mu,\nu$.
However, this may occur on a significantly longer timescale than it should according to the exact kernels (\cref{fig:kernels_SM_MFT}), and thus does not ensure that GQME/Ehrenfest will converge within a reasonable $\tcut$.}

Finally, by replacing $\K_{zx}(t)$ and $\K_{zy}(t)$ calculated with Ehrenfest with the corresponding terms from spin mapping, we are able to reproduce an accurate value for the thermal population. 
However, although this worked well in this specific case with $\varepsilon=5$,
when we tested this on different systems, other components also seemed to matter. In general it does not appear possible to isolate the inaccuracy due to Ehrenfest to a specific subset of the components of the full memory kernel.
The reason for this lies in the relation between the components of the input correlation functions [\cref{eq:scal_Liou,eq:CVB,eq:CVBVB}] and the full memory kernel $\K(t)$, which appears to be quite involved.
Thankfully, the problem can be simplified as shown in Sec.~\ref{sec:K3}.

\subsection{Analysis of the auxiliary kernels} \label{sec:K3}
In this section we delve deeper into the reasons why GQME/spin-mapping results in higher accuracy than GQME/Ehrenfest and where in particular the errors of the latter lie. 
In order to achieve this, we will first identify the most relevant correlation functions needed to obtain the kernels. 

We show in \cref{eq:K3_elem} that the auxiliary memory kernel $\K^{(3)}(t)$ can be written purely as a combination of correlation functions involving bath operators, $\C^{V_\ba,k\pm}_{\mu\nu}(t)$ [\cref{eq:CVB}].  
Remarkably, it does not involve any elements of the direct correlation function $\C(t)$ at all.
Furthermore, \cref{eq:K1_K3} indicates that the full kernel $\K(t)$ is in principle defined only in terms of $\K^{(3)}(t)$ and its time derivative.
It then follows that we only need to study the bath correlation functions $\C^{V_\ba,k\pm}_{\mu\nu}(t)$ in order to understand the accuracy of the GQME solution.

In particular, for a system with only $V_{\ba,z}$ system--bath coupling (such as the spin--boson model), the auxiliary kernel is given by
\begin{align*}
    \K^{(3)}_{\mc I \nu}(t) &= 2\C^{V_{\ba,z}-}_{z\nu}(t)
    &
    \K^{(3)}_{x \nu}(t) &= -2\C^{V_{\ba,z}+}_{y\nu}(t)
    \\
    \K^{(3)}_{y \nu}(t) &= 2\C^{V_{\ba,z}+}_{x\nu}(t)
    &
    \K^{(3)}_{z \nu}(t) &= 2\C^{V_{\ba,z}-}_{\mc I \nu}(t).
\end{align*}

\begin{figure*}
\includegraphics[width=7in]{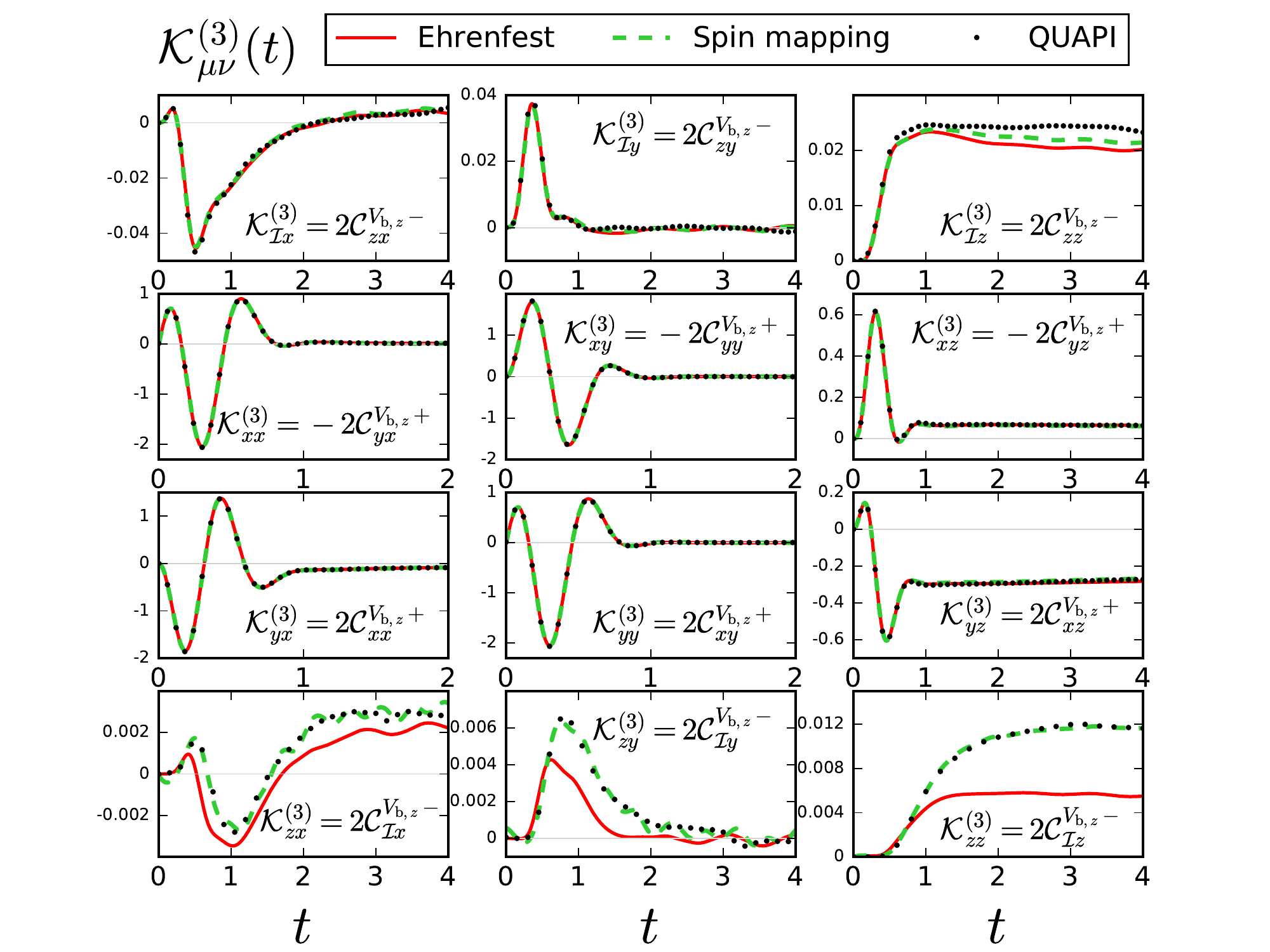}\caption{Comparison between the predictions of Ehrenfest and spin mapping for the non-zero components of the auxiliary kernel $\K^{(3)}_{\mu\nu}(t)$ for the system with $\varepsilon=5$. Benchmark results are calculated from QUAPI\@. In each panel we include the expression of the auxiliary correlation functions needed to calculate each component.}\label{fig:K3}
\end{figure*}

In \cref{fig:K3} we show the predictions of $\K^{(3)}(t)$ from Ehrenfest, spin mapping and QUAPI for the same system with $\varepsilon=5$ studied in \cref{fig:kernels_SM_MFT}. 
Interestingly, we notice that the largest relative errors occur in the three components of the last row. These elements of the kernel correspond to correlation functions $\C^{V_\ba,z-}_{\mc I k}(t)$ involving an electronic identity operator and a bath operator with a linear dependence on the momentum, as defined by \cref{eq:CVBm,eq:Vb-}.
Except for the appearance of the momentum term, these are the usual correlation functions which cause the main problems in direct simulations of Ehrenfest and are improved by spin mapping, whereas it is commonly observed that correlation functions initialized by a Pauli matrix are well described by both methods.\cite{spinmap}
This rule appears to continue to hold even when the momentum term is added.

By replacing the components of the last line of $\K^{(3)}(t)$ [i.e., $\C^{V_\ba,z-}_{\mc I \nu}(t)$] from an Ehrenfest calculation with their equivalent from spin mapping, GQME/Ehrenfest no longer predicts negative populations. 
This confirms that these three components are the key source of error in this method.
The fact that spin mapping is able to provide significantly more accuracy than Ehrenfest theory for these correlation functions in particular
is therefore the ultimate reason for the success of GQME/spin-mapping.

\section{Conclusions}
In this paper we analyzed the accuracy of the GQME, obtained by solving the memory kernels with two approximate quasiclassical methods, namely Ehrenfest mean-field theory and spin mapping. 
We compared the prediction of the equilibrium populations for asymmetric spin--boson models.
We found that the GQME/spin-mapping method consistently yields more accurate results for the long-time dynamics of the electronic populations than GQME/Ehrenfest or either of the direct quasiclassical methods. 
Although for weak asymmetry GQME/Ehrenfest can give a significant improvement in accuracy compared to direct Ehrenfest predictions, it may lead to spurious results in systems of strong asymmetry.
In particular, coupling Ehrenfest with the master equation can lead to negative electronic populations, despite the fact that populations from the direct Ehrenfest method are guaranteed to be positive. 
On the other hand, there appears to be no problem with using the GQME/spin-mapping approach even in cases where direct spin mapping predicts negative populations. This can be easily understood as the offending correlation function, $\C_{\mc I z}(t)$, is not used to construct the kernels.
The GQME procedure thus goes a long way towards fixing the negative population problem of mapping methods, although there is no guarantee that it will do so in all cases.
%\tcb{In particular, we observed a breakdown of the method in the regime of strong electronic--nuclear coupling ($\xi\gg 1$).}

In addition, we confirmed that the classical ergodic hypothesis holds for these quasiclassical nonintegrable dynamical systems, and that this  assumption can be used to predict the long-time limits of correlation functions.
We derived a useful closed relation between the integral of the memory kernel and the long-time solution of the GQME\@.
In particular, this allowed us to investigate the effect of the cutoff time and to determine whether a sensible cutoff can be defined at all.
We analyzed the accuracy of different representations of the time derivative of the correlation function (i.e., $\dot\C(t)$ and $\C^{\mc L}(t)$), which are known to lead to a trivial and nontrivial solution of the master equation, respectively. 
Although $\dot\C_{\mc I \nu}(t)=\C^{\mc L}_{\mc I \nu}(t)$ within spin mapping, other elements of these correlation functions are not identical, which means that the GQME procedure is able to change (and typically improve) the spin-mapping predictions.
However, we noticed that $\C^{\mc L}(t)$ is not systematically better than $\dot\C(t)$; this can lead to poor and even unphysical predictions for the solution of the master equation, as observed in the case of GQME/Ehrenfest.

Finally, by working in the basis of the Pauli spin matrices, we identified a small subset of auxiliary correlation functions which represent the minimal input needed to solve the GQME, all of which include a bath operator. 
Among these terms, three correlation functions involving the electronic identity operator appear to be responsible for the main error of GQME/Ehrenfest. 
Given that spin mapping is much more accurate in the predictions of these terms, the GQME/spin-mapping procedure remains reliable up to large values of $\varepsilon$.
%extensions

Nonetheless, GQME/spin-mapping is not guaranteed to return accurate results in every case.
In particular, we observed a breakdown of the method in the regime of strong electronic--nuclear coupling.
Importantly, however, it was clear in these cases that the method could not be trusted in the first place, as the long-time population failed to converge with $\tcut$.
To resolve the issue, we note that other nonadiabatic trajectory methods, such as symmetric quasiclassical windowing (SQC) \cite{Miller2016Faraday} or the mapping approach to surface hopping (MASH) \cite{MASH} may be better suited to these regimes and could be combined with the GQME in a similar way.

The formalism discussed in this work can be straightforwardly extended to more electronic states. This can be accomplished by replacing the Pauli basis with a set of generators of $\mathrm{SU}(N)$, the algebra of the total Hilbert space for a system of $N$ electronic levels. \cite{multispin} The method is therefore applicable to relevant systems in chemical physics involving more than two levels, such as the Fenna--Matthew--Olsen complex (with $N=7$ or 8). \cite{Ullah2022,FMO,Pfalzgraff2019GQME,mulvihill2021,FMOclassical} 

%applications
The GQME/spin-mapping approach could be applied to calculate nonadiabatic rate constants, which can be related to the integrals of memory kernels. \cite{xu2018} The method could also be a valid tool to study a large class of problems from nonadiabatic linear \cite{kobus2008} and nonlinear \cite{MukamelBook} spectroscopy.
%The simulation of nonlinear spectra from nonadiabatic dynamics is a powerful predictive tool in ultra-fast physics. \cite{MukamelBook}
%However, the calculation of nonlinear spectra with quasiclassical methods is in general cost-intensive.
However, in order to calculate the multi-time correlation functions required to predict nonlinear spectra, one needs to rely on partially linearized quasiclassical methods, which are more computationally expensive than linearized approaches. \cite{mannouch2022} Given that the master equation requires only short-time input to solve long-time dynamics, it may be possible to extend the present approach to become a valid tool in that framework. \cite{fetherolf2017}

A limiting factor in the implementation of the GQME discussed in this work comes from the fact that, in order to simulate the explicit Hamiltonian dynamics of the bath, a finite number of nuclear frequencies must be determined from the discretization of a continuous spectral density. This can lead to strong oscillations in time, especially in those correlations involving the explicit time evolution of the coordinates of the bath. These oscillations do not allow to fix a well defined cutoff time for the memory kernel.
For this reason, a higher number of nuclear modes was required in the simulations for building the kernels than for the direct calculation of the correlation functions.
A possible source of improvement for future work could then be to explore how different choices of the nuclear spectral density affect the relaxation of the kernels, or even to implement an implicit bath using non-Markovian friction kernels.
In particular, the reaction coordinate formulation of the spin--boson Hamiltonian \cite{Thoss2001hybrid,garg1985} is based on the identification of a single global nuclear coordinate directly coupled to the system, while the rest of the nuclear degrees of freedoms can be identified as a secondary bath. It could be computationally efficient to solve the memory kernels with a stochastic integration scheme for the dynamics of the secondary bath, as commonly used in the case of Brownian spectral densities. \cite{anto-ztrikacs2021,bellonzi2016,Lawrence2019ET}

Finally, it would be interesting to compare the accuracy of the present approach with different formulations of the GQME \cite{mulvihill2021b}, including the Tokuyama–-Mori time-convolutionless (TCL) GQME, \cite{tokuyama1981,fulinski1967,dominikus2021} and to assess whether a nonadiabatic ring-polymer formulation \cite{mapping,bossion2021} of the master equation could be an effective generalization of the method to account for nuclear quantum effects.

%\section*{Data availability}
%The data that support the findings of this study are available within the article and its supplementary material.

%\section*{Supplementary Material}
%See supplementary material for the numerical data of quasiclassical correlation functions for a representative spin--boson model with $\varepsilon=5$ and $\xi=1$.
%\tcb{A python script is also provided which evaluates the kernels and solves the GQME based on this input data.}

\section*{Acknowledgments}
The authors would like to thank Jonathan Mannouch, Johan Runeson and Joseph Lawrence for useful discussions and comments. The QUAPI results presented in this work were calculated by Jonathan Mannouch.
This project has received funding from European Union’s Horizon 2020 under MCSA Grant No.~801459 (FP-RESOMUS)
and is supported by the Cluster of
Excellence `CUI: Advanced Imaging of Matter' of the Deutsche
Forschungsgemeinschaft (DFG) -- EXC 2056 -- project ID 390715994.

\begin{appendix}
\section{Calculation of the memory kernel in terms of auxiliary correlation functions}\label{app:GQME_mapping}
In this appendix we discuss the technical details on how to construct an approximate solution of the GQME, by making use of quasiclassical approaches to calculate the auxiliary kernels $\mc K^{(1)}(t)$ and $\mc K^{(3)}(t)$ [defined in \cref{eq:Kaux}], which determine the full kernel $\K(t)$. Our methodology in this section closely follows previous work\cite{Kelly2015nonadiabatic,Montoya2016GQME,Pfalzgraff2019GQME} except that we specifically employ the Pauli basis.
To calculate the expressions in \cref{eq:Kaux_exp}, we need a quasiclassical representation of $\C^{\mc L}(t)$ [defined in \cref{eq:CL}] and its time derivative.

First, we can evaluate
%\tcb{\begin{align}%eq A3 notes Jeremy useful comment
%{\mc L}(\hat \rho_0\hat \sigma_\mu)&=\frac 12{\mathcal {L}}( \hat  \sigma_\mu\otimes \hat \rho_\ba) = \frac i 2[\hat H_\sy, \hat \sigma_\mu \otimes \hat \rho_\ba]+\nn\\
%&\quad+ \frac i 2[\hat \sigma_k\otimes \hat V_{\ba,k}, \hat \sigma_\mu \otimes \hat \rho_\ba]+ \frac i 2[\hat H_\ba, \hat \sigma_\mu \otimes \hat \rho_\ba] =  \nn \\
%&=\frac i 2[\hat H_\sy, \hat  \sigma_\mu]\otimes \hat \rho_\ba + \frac i 4\Big([\hat \sigma_k, \hat  \sigma_\mu] \otimes [\hat V_{\ba,k},\hat \rho_\ba]_+\nn\\
%&\quad+[\hat \sigma_k, \hat  \sigma_\mu]_+ \otimes [\hat V_{\ba,k},\hat \rho_\ba]\Big),
%\end{align}}
\begin{align}%eq A3 notes Jeremy %useful comment
{\mc L}\hat \rho_0\hat \sigma_\mu&=\tfrac i 2[\hat H_\sy, \hat  \sigma_\mu]\otimes \hat \rho_\ba + \tfrac i 4\Big([\hat \sigma_k, \hat  \sigma_\mu] \otimes [\hat V_{\ba,k},\hat \rho_\ba]_+\nn\\
&\quad+[\hat \sigma_k, \hat  \sigma_\mu]_+ \otimes [\hat V_{\ba,k},\hat \rho_\ba]\Big),\label{eq:Lrhosmu1}
\end{align}
where we used $[\hat H_\ba, \hat \rho_\ba]=0$. 
Next, the matrix $\mc X$, defined in \cref{eq:Xgen}, can be written as
\begin{align}\label{eq:X}
\mc X_{\mu\nu}&= \tfrac 12 \tr\left[\hat\rho_\ba\hat\sigma_\mu (i[\hat H_\sy,\hat\sigma_\nu]+i[\hat\sigma_k,\hat\sigma_\nu]\otimes \hat V_{\ba, k})\right]\nn\\
&=\tfrac 12 \tr_\sy[\hat\sigma_\mu i [\hat H_\sy,\hat\sigma_\nu]]=2\epsilon_{\mu\nu l}H_{\sy,l},
\end{align}
where we made use of the condition \cref{eq:tr_rV}. Finally, by defining the matrices $\mathcal Y^{(k)}$ and $\mathcal Z^{(k)}$, with components
\begin{subequations}
\begin{align}
\mathcal Y^{(k)}_{\mu\nu} &= \tfrac 12 \tr_\sy[\hat \sigma_\mu i [\hat \sigma_k, \hat \sigma_\nu]]=2\epsilon_{k\mu\nu}, \label{eq:Y}\\
\mathcal Z^{(k)}_{\mu\nu} &= \tfrac 12 \tr_\sy[\hat \sigma_\mu  [\hat \sigma_k, \hat\sigma_\nu]_+]=2(\delta_{\mu 0}\delta_{\nu k}+\delta_{\nu 0}\delta_{\mu k}), \label{eq:Y+}
\end{align}
we can rewrite \cref{eq:Lrhosmu1} as
\begin{align}\label{eq:Lrhosmu2}
{\mc L}\hat \rho_0\hat \sigma_\mu &= \tfrac 12 \hat\sigma_\lambda\mc X_{\lambda\mu}\otimes \hat \rho_\ba + \tfrac 14\hat\sigma_\lambda\otimes \Big( \mc Y^{(k)}_{\lambda\mu}  [\hat V_{\ba,k},\hat \rho_\ba]_+\nn\\
&\quad+\mc Z^{(k)}_{\lambda\mu}  i[\hat V_{\ba,k},\hat \rho_{\ba}]\Big),
\end{align}
\end{subequations}
and hence
\begin{align}\label{eq:CLapp}
\C^{\mathcal L}(t) &= \mc X\C(t)+ \mc Y^{(k)}\C^{V_{\ba,k}+}(t)- \mc Z^{(k)}\C^{V_{\ba,k}-}(t),
\end{align}
where 
\begin{subequations}\label{eq:CVB}
\begin{align}
\C^{V_{\ba,k}+}_{\mu\nu}(t) &= \tfrac 12\llangle\hat \sigma_\mu [\hat V_{\ba,k},\hat\rho_0]_+|\e^{\mc L t} | \hat \sigma_\nu\rrangle, \label{eq:CVBp} \\
\C^{V_{\ba,k}-}_{\mu\nu}(t) &= \tfrac 12\llangle \hat \sigma_\nu i[\hat V_{\ba,k},\hat\rho_0] |\e^{\mc L t} |\hat \sigma_\nu\rrangle \label{eq:CVBm}.
\end{align}
\end{subequations}
Likewise, 
%\tcb{given that %keep this comment
%\begin{align*}
%\dot \C^{V_{\ba,k}+}(t) &= \tfrac 12\tr\left[\hat \sigma_\mu\otimes [\hat V_{\ba,k},\hat\rho_\ba]_+  \mc L\hat \sigma_\nu(t)\right]\\
%&=\tfrac 12\tr\left[\hat \sigma_\mu\otimes [\hat V_{\ba,k},\hat\rho_\ba]_+\e^{\mc L t}i[\hat H_\sy+\hat\sigma_l\otimes\hat V_{\ba,l},\hat \sigma_\nu]\right]\\
%&=\C^{V_{\ba,k}+}_{\mu\lambda}(t)\mc X_{\lambda\nu}+\nn\\
%&\quad+\tfrac 12 \tr\left[\hat \sigma_\mu\otimes [\hat V_{\ba,k},\hat\rho_\ba]_+\e^{\mc L t}i[\hat\sigma_l,\hat \sigma_\nu]\otimes\hat V_{\ba,l}\right]\nn\\
%&=\C^{V_{\ba,k}+}_{\mu\lambda}(t)\mc X_{\lambda\nu}+\nn\\
%&\quad+\tfrac 12 \tr\left[\hat \sigma_\mu\otimes [\hat V_{\ba,k},\hat\rho_\ba]_+\e^{\mc L t}\hat\sigma_{\lambda}\mc Y^{(l)}_{\lambda\nu}\otimes\hat V_{\ba,l}\right],
%\end{align*}}
we determine the expansion of the time derivative of \cref{eq:CLapp}
\begin{align}\label{eq:CdL}
\dot\C^{\mathcal {L}}(t) =  \mathcal X \dot\C(t)&+ \mathcal Y^{(k)} \left(\C^{V_{\ba,k}+}(t)\mathcal X+\C^{V_{\ba,k}+, V_{\ba,l}}(t)\mathcal Y^{(l)} \right)\nn\\
&-\mathcal Z^{(k)} \left(\C^{V_{\ba,k}-}(t)\mathcal X+ \C^{V_{\ba,k}-, V_{\ba,l}}(t)\mathcal Y^{(l)}\right),
\end{align}
where 
\begin{subequations}\label{eq:CVBVB}
\begin{align}
\C^{V_{\ba,k}+, V_{\ba,l}}_{\mu\nu}(t) &= \tfrac 12\llangle\hat \sigma_\mu [\hat V_{\ba,k},\hat\rho_\ba]_+|\e^{\mc L t}|  \hat \sigma_\nu \hat V_{\ba,l}\rrangle,\label{eq:CVBpVB} \\
\C^{V_{\ba,k}-, V_{\ba,l}}_{\mu\nu}(t) &= \tfrac 12\llangle\hat \sigma_\mu i[\hat V_{\ba,k},\hat\rho_\ba] |\e^{\mc L t}| \hat \sigma_\nu \hat V_{\ba,l}\rrangle.\label{eq:CVBmVB}
\end{align}
\end{subequations}
One may wonder whether further improvements in accuracy could be obtained by replacing $\dot \C(t)$ with $\C^{\mc L}(t)$ in the second term of the right-hand side of \cref{eq:K1_der} and in \cref{eq:CdL}. However, these terms cancel each other when constructing $\K^{(1)}(t)$ and so it has no effect on the results. Further improvement in accuracy may, however, be possible by calculating an expression for the second derivative $\C^{\mc L^2}(t)$ to replace $\dot\C^{\mc L}(t)$. 

Together with $\C(t)$, we propagate the electronic--nuclear correlation functions in \cref{eq:CVB,eq:CVBVB} with quasiclassical methods, to construct the auxiliary kernels via \cref{eq:CLapp,eq:CdL,eq:Kaux_exp}.
Note that we can rewrite \cref{eq:Kaux_exp} as
%\tcb{ %USEFUL COMMENT
%\begin{align}
%&\mc K^{(1)}(t) = -\dot\C^{\mc L} (t)+\mc X \dot\C(t)+\C^{{\mathcal L}}(t)\mc X -\mc X\C(t)\mc X\nn \\
%&= \cancel{-\mathcal X \dot\C(t)}- \mathcal Y^{(k)} \left(\xcancel{\C^{V_{\ba,k}+}(t)\mathcal X}+\C^{V_{\ba,k+} V_{\ba,l}}(t)\mathcal Y^{(l)} \right)\nn\\
%&\quad+\mathcal Z^{(k)} \left(\xcancel{\C^{V_{\ba,k}-}(t)\mathcal X}+ \C^{V_{\ba,k}- V_{\ba,l}}(t)\mathcal Y^{(l)}\right)\nn \\
%&\quad\cancel{+\mc X\dot\C(t)}\nn \\
%&\quad+\left( \bcancel{ \mc X\C(t)}+ \xcancel{\mc Y^{(k)}\C^{V_{\ba,k}+}(t)}- \xcancel{\mc Z^{(k)}\C^{V_{\ba,k}-}(t)} \right)\mc X \bcancel{- \mc X \C(t)\mc X}
%\end{align}
%\begin{align}
%&\mc K^{(3)}(t) = - \C^{\mathcal{L}}(t) +\mc X \C(t)\nn \\
%&=- \mc X\C(t)- \mc Y^{(k)}\C^{V_{\ba,k}+}(t)+ \mc Z^{(k)}\C^{V_{\ba,k}-}(t) +\mc X \C(t)
%\end{align}}
\begin{subequations}\label{eq:K_elem}
\begin{align}
\mc K^{(1)}(t) 
&= \mc Z^{(k)}\C^{V_{\ba,k}-,V_{\ba,l}}(t)\mc Y^{(l)} - \mc Y^{(k)}\C^{V_{\ba,k}+,V_{\ba,l}}(t)\mc Y^{(k)}, \label{eq:K1_elem}\\
\mc K^{(3)}(t) &= \mc Z^{(k)}\C^{V_{\ba,k}-}(t) - \mc Y^{(k)}\C^{V_{\ba,k}+}(t),\label{eq:K3_elem}
\end{align}
\end{subequations}
which does not depend on $\C(t)$ at all. 
The explicit expressions for the matrices in \cref{eq:X,eq:Y,eq:Y+} for the spin--boson model introduced in \cref{subsec:model} are
\begin{subequations}
\begin{equation}\label{def_XYYd_mat}
    \mathcal X = \begin{pmatrix}
0 & 0 & 0 &0\\
0&0&2\varepsilon&0 \\
0 & -2\varepsilon & 0 & 2\Delta \\
0 &0&-2\Delta & 0
\end{pmatrix}, \end{equation}
\begin{equation}
\mathcal Y^{(z)} = \begin{pmatrix}
0 & 0 & 0 &0\\
0&0&2&0 \\
0 & -2 & 0 & 0 \\
0 &0&0 & 0
\end{pmatrix},
\hspace{10mm}
\mc Z^{(z)} = \begin{pmatrix}
0 & 0 & 0 &2\\
0&0&0&0 \\
0 & 0 & 0 & 0 \\
2 &0&0 & 0
\end{pmatrix},
\end{equation}
\end{subequations}
while $\mathcal Y^{(k)}$ and $\mathcal Z^{(k)}$  for $k=x,y$ are not required in this case. Finally, the classical limits of the nuclear operators appearing in the auxiliary correlation functions for the spin--boson model are
\begin{subequations}
\begin{align}
V_{\ba,k}(q,p)&= \delta_{kz}\sum_{\alpha=1}^F c_\alpha q_\alpha,\\
\{[\hat V_{\ba,k}, \hat \rho_\ba]_+\}(q,p) &= 2\rho_\ba  \delta_{kz}  \sum_{\alpha=1}^F c_\alpha q_\alpha,\\
\{i[\hat V_{\ba,k}, \hat \rho_\ba]\}(q,p) &= \beta \rho_\ba\delta_{kz} \sum_{\alpha=1}^F \frac{p_\alpha c_\alpha}{ m_\alpha}. \label{eq:Vb-}
\end{align}
\end{subequations}
\section{Relation between long-time limits and integrals over the memory kernel}\label{app:zero_modeK}
In this appendix we derive a relation between the integrals of the memory kernels and the long-time population obtained as a stationary solution of the GQME\@. \cite{cohen2013}
In \cref{subsec:cutoff} we make use of this identity to define a suitable cutoff time, $\tcut$, for the memory kernels. %provided this can be defined in the first place. 
%with a real Hamiltonian (which does not depend on $\hat\sigma_y$). 

Following Ref.~\onlinecite{Kelly2016master}, we take the Fourier--Laplace transform of the GQME [\cref{eq:GQME}] as
\begin{equation}\label{eq:GQME_FLT}
\tilde{\dot{\C}}(\omega) = -i\omega\tilde \C(\omega) -\C(0) = \tilde \C(\omega)\left(\mathcal X - \tilde\K(\omega)\right),
\end{equation}
where
\begin{equation}\label{eq:FLT}
\tilde f(\omega) = \lim_{\eta\to 0^+}\int_0^{\infty}\mr d t \; \e^{i(\omega +i\eta)t}f(t).
\end{equation}
We multiply \cref{eq:GQME_FLT} by a factor $-i\omega$ and take the $\omega\to 0$ limit of the expression, to find 
\begin{equation}\label{eq:sys_intK}
\dot{\C}(\infty)=0=\C(\infty)\left(\mathcal X - \tilde\K(0)\right),
\end{equation}
where we used the final-value theorem \cite{beerends2003}
\begin{equation}\label{eq:FVT}
\lim_{\omega \to 0}-i\omega\tilde f(\omega)=f(\infty).
\end{equation}
We can rewrite \cref{eq:sys_intK} in terms of an inhomogeneous linear system, which can be solved analytically for $\C(\infty)$.
In particular, from 
\begin{equation}
\C_{\mathcal I\mu}(\infty)\left(\mc X_{\mu\nu}-\tilde \K_{\mu\nu}(0)\right)=0
\end{equation} 
we can remove the $\nu=\mc I$ component, as
$\mc X_{\mu \mc I}=\tilde \K_{\mu \mathcal I}(0)=0$ (given that $\mc L|\hat{\mc I}\rrangle = 0$). We are then left with the set of equations
\begin{equation}\label{eq:linsys}
\C_{\mc I  j}(\infty)\left(\mc X_{jk}-\tilde \K_{jk}(0)\right) = \tilde \K_{\mc I k}(0),
\end{equation}
where we used $\C_{\mc I \mc I }(t)=1$ and $\mc X_{\mc I k}=0$.
These simultaneous equations can be solved for $\C_{\mc I  j}(\infty)$ with standard linear-algebra routines.

The solution in the case of the spin--boson model is
\begin{subequations}\label{eq:sigma_K}
\begin{align}
\C_{\mc I x}(\infty) &= \frac{\tilde{\mathcal K}_{\mc I y}(0)\tilde{\mathcal K}_{zx}(0)-\tilde{\mathcal K}_{\mathcal Ix}(0)\left[2\Delta+\tilde{\mathcal K}_{zy}(0)\right]}{\left[2\varepsilon-\tilde{\mathcal K}_{xy}(0)\right]\tilde{\mathcal K}_{zx}(0)+\tilde{\mathcal K}_{xx}(0)\left[2\Delta + \tilde{\mathcal K}_{zy}(0)\right]},\label{eq:sigmax_K}\\
\C_{\mc I z}(\infty) &= \frac{-\tilde{\mathcal K}_{\mc I y}(0)\tilde{\mathcal K}_{xx}(0)+\tilde{ K}_{\mathcal Ix}(0)\left[-2\varepsilon+\tilde{\mathcal K}_{xy}(0)\right]}{\left[2\varepsilon-\tilde{\mathcal K}_{xy}(0)\right]\tilde{\mathcal K}_{zx}(0)+\tilde{\mathcal K}_{xx}(0)\left[2\Delta + \tilde{\mathcal K}_{zy}(0)\right]},\label{eq:sigmaz_K}
\end{align}
\end{subequations}
and $\C_{\mc I y}(\infty)=0$. Note that $\C_{\mc I z}(\infty)$ determines the long-time populations [\cref{eq:P_pm}].
We observe that most of the non-zero elements of the kernel enter into the above expressions and, in general, they all affect the accuracy of the long-time predictions. 

Finally, we note that \cref{eq:linsys} can be immediately generalized to an arbitrary number of electronic states.

\end{appendix}
\FloatBarrier

%\bibliography{biblio,references} % use both reference files - but please only add new refs to biblio.bib 

\begin{thebibliography}{91}%
\makeatletter
\providecommand \@ifxundefined [1]{%
 \@ifx{#1\undefined}
}%
\providecommand \@ifnum [1]{%
 \ifnum #1\expandafter \@firstoftwo
 \else \expandafter \@secondoftwo
 \fi
}%
\providecommand \@ifx [1]{%
 \ifx #1\expandafter \@firstoftwo
 \else \expandafter \@secondoftwo
 \fi
}%
\providecommand \natexlab [1]{#1}%
\providecommand \enquote  [1]{``#1''}%
\providecommand \bibnamefont  [1]{#1}%
\providecommand \bibfnamefont [1]{#1}%
\providecommand \citenamefont [1]{#1}%
\providecommand \href@noop [0]{\@secondoftwo}%
\providecommand \href [0]{\begingroup \@sanitize@url \@href}%
\providecommand \@href[1]{\@@startlink{#1}\@@href}%
\providecommand \@@href[1]{\endgroup#1\@@endlink}%
\providecommand \@sanitize@url [0]{\catcode `\\12\catcode `\$12\catcode
  `\&12\catcode `\#12\catcode `\^12\catcode `\_12\catcode `\%12\relax}%
\providecommand \@@startlink[1]{}%
\providecommand \@@endlink[0]{}%
\providecommand \url  [0]{\begingroup\@sanitize@url \@url }%
\providecommand \@url [1]{\endgroup\@href {#1}{\urlprefix }}%
\providecommand \urlprefix  [0]{URL }%
\providecommand \Eprint [0]{\href }%
\providecommand \doibase [0]{http://dx.doi.org/}%
\providecommand \selectlanguage [0]{\@gobble}%
\providecommand \bibinfo  [0]{\@secondoftwo}%
\providecommand \bibfield  [0]{\@secondoftwo}%
\providecommand \translation [1]{[#1]}%
\providecommand \BibitemOpen [0]{}%
\providecommand \bibitemStop [0]{}%
\providecommand \bibitemNoStop [0]{.\EOS\space}%
\providecommand \EOS [0]{\spacefactor3000\relax}%
\providecommand \BibitemShut  [1]{\csname bibitem#1\endcsname}%
\let\auto@bib@innerbib\@empty
%</preamble>
\bibitem [{\citenamefont {Lee}, \citenamefont {Huo},\ and\ \citenamefont
  {Coker}(2016)}]{lee2016}%
  \BibitemOpen
  \bibfield  {author} {\bibinfo {author} {\bibfnamefont {M.~K.}\ \bibnamefont
  {Lee}}, \bibinfo {author} {\bibfnamefont {P.}~\bibnamefont {Huo}}, \ and\
  \bibinfo {author} {\bibfnamefont {D.~F.}\ \bibnamefont {Coker}},\ }\bibfield
  {title} {\enquote {\bibinfo {title} {Semiclassical path integral dynamics:
  Photosynthetic energy transfer with realistic environment interactions},}\
  }\href {\doibase 10.1146/annurev-physchem-040215-112252} {\bibfield
  {journal} {\bibinfo  {journal} {Annu. Rev. Phys. Chem.}\ }\textbf {\bibinfo
  {volume} {67}},\ \bibinfo {pages} {639--668} (\bibinfo {year}
  {2016})}\BibitemShut {NoStop}%
\bibitem [{\citenamefont {Cheng}\ and\ \citenamefont
  {Fleming}(2009)}]{cheng2009}%
  \BibitemOpen
  \bibfield  {author} {\bibinfo {author} {\bibfnamefont {Y.-C.}\ \bibnamefont
  {Cheng}}\ and\ \bibinfo {author} {\bibfnamefont {G.~R.}\ \bibnamefont
  {Fleming}},\ }\bibfield  {title} {\enquote {\bibinfo {title} {Dynamics of
  light harvesting in photosynthesis},}\ }\href {\doibase
  10.1146/annurev.physchem.040808.090259} {\bibfield  {journal} {\bibinfo
  {journal} {Annu. Rev. Phys. Chem.}\ }\textbf {\bibinfo {volume} {60}},\
  \bibinfo {pages} {241--262} (\bibinfo {year} {2009})}\BibitemShut {NoStop}%
\bibitem [{\citenamefont {Polli}\ \emph {et~al.}(2010)\citenamefont {Polli},
  \citenamefont {Alto{\`e}}, \citenamefont {Weingart}, \citenamefont
  {Spillane}, \citenamefont {Manzoni}, \citenamefont {Brida}, \citenamefont
  {Tomasello}, \citenamefont {Orlandi}, \citenamefont {Kukura}, \citenamefont
  {Mathies}, \citenamefont {Garavelli},\ and\ \citenamefont
  {Cerullo}}]{polli2010}%
  \BibitemOpen
  \bibfield  {author} {\bibinfo {author} {\bibfnamefont {D.}~\bibnamefont
  {Polli}}, \bibinfo {author} {\bibfnamefont {P.}~\bibnamefont {Alto{\`e}}},
  \bibinfo {author} {\bibfnamefont {O.}~\bibnamefont {Weingart}}, \bibinfo
  {author} {\bibfnamefont {K.~M.}\ \bibnamefont {Spillane}}, \bibinfo {author}
  {\bibfnamefont {C.}~\bibnamefont {Manzoni}}, \bibinfo {author} {\bibfnamefont
  {D.}~\bibnamefont {Brida}}, \bibinfo {author} {\bibfnamefont
  {G.}~\bibnamefont {Tomasello}}, \bibinfo {author} {\bibfnamefont
  {G.}~\bibnamefont {Orlandi}}, \bibinfo {author} {\bibfnamefont
  {P.}~\bibnamefont {Kukura}}, \bibinfo {author} {\bibfnamefont {R.~A.}\
  \bibnamefont {Mathies}}, \bibinfo {author} {\bibfnamefont {M.}~\bibnamefont
  {Garavelli}}, \ and\ \bibinfo {author} {\bibfnamefont {G.}~\bibnamefont
  {Cerullo}},\ }\bibfield  {title} {\enquote {\bibinfo {title} {Conical
  intersection dynamics of the primary photoisomerization event in vision},}\
  }\href {\doibase 10.1038/nature09346} {\bibfield  {journal} {\bibinfo
  {journal} {Nature}\ }\textbf {\bibinfo {volume} {467}},\ \bibinfo {pages}
  {440--443} (\bibinfo {year} {2010})}\BibitemShut {NoStop}%
\bibitem [{\citenamefont {Stock}\ and\ \citenamefont
  {Thoss}(2005)}]{Stock2005nonadiabatic}%
  \BibitemOpen
  \bibfield  {author} {\bibinfo {author} {\bibfnamefont {G.}~\bibnamefont
  {Stock}}\ and\ \bibinfo {author} {\bibfnamefont {M.}~\bibnamefont {Thoss}},\
  }\bibfield  {title} {\enquote {\bibinfo {title} {Classical description of
  nonadiabatic quantum dynamics},}\ }\href {\doibase 10.1002/0471739464.ch5}
  {\bibfield  {journal} {\bibinfo  {journal} {Adv. Chem. Phys.}\ }\textbf
  {\bibinfo {volume} {131}},\ \bibinfo {pages} {243--376} (\bibinfo {year}
  {2005})}\BibitemShut {NoStop}%
\bibitem [{\citenamefont {Miller}(2001)}]{Miller2001SCIVR}%
  \BibitemOpen
  \bibfield  {author} {\bibinfo {author} {\bibfnamefont {W.~H.}\ \bibnamefont
  {Miller}},\ }\bibfield  {title} {\enquote {\bibinfo {title} {The
  semiclassical initial value representation: {A} potentially practical way for
  adding quantum effects to classical molecular dynamics simulations},}\ }\href
  {\doibase 10.1021/jp003712k} {\bibfield  {journal} {\bibinfo  {journal}
  {J.~Phys. Chem.~A}\ }\textbf {\bibinfo {volume} {105}},\ \bibinfo {pages}
  {2942--2955} (\bibinfo {year} {2001})}\BibitemShut {NoStop}%
\bibitem [{\citenamefont {Kapral}(2015)}]{kapral2015}%
  \BibitemOpen
  \bibfield  {author} {\bibinfo {author} {\bibfnamefont {R.}~\bibnamefont
  {Kapral}},\ }\bibfield  {title} {\enquote {\bibinfo {title} {Quantum dynamics
  in open quantum-classical systems},}\ }\href {\doibase
  10.1088/0953-8984/27/7/073201} {\bibfield  {journal} {\bibinfo  {journal}
  {J.~Phys. Condens. Matter}\ }\textbf {\bibinfo {volume} {27}},\ \bibinfo
  {pages} {073201} (\bibinfo {year} {2015})}\BibitemShut {NoStop}%
\bibitem [{\citenamefont {Meyer}\ and\ \citenamefont
  {Miller}(1979)}]{Meyer1979nonadiabatic}%
  \BibitemOpen
  \bibfield  {author} {\bibinfo {author} {\bibfnamefont {H.-D.}\ \bibnamefont
  {Meyer}}\ and\ \bibinfo {author} {\bibfnamefont {W.~H.}\ \bibnamefont
  {Miller}},\ }\bibfield  {title} {\enquote {\bibinfo {title} {A classical
  analog for electronic degrees of freedom in nonadiabatic collision
  processes},}\ }\href {\doibase 10.1063/1.437910} {\bibfield  {journal}
  {\bibinfo  {journal} {J.~Chem. Phys.}\ }\textbf {\bibinfo {volume} {70}},\
  \bibinfo {pages} {3214--3223} (\bibinfo {year} {1979})}\BibitemShut {NoStop}%
\bibitem [{\citenamefont {Stock}\ and\ \citenamefont
  {Thoss}(1997)}]{Stock1997mapping}%
  \BibitemOpen
  \bibfield  {author} {\bibinfo {author} {\bibfnamefont {G.}~\bibnamefont
  {Stock}}\ and\ \bibinfo {author} {\bibfnamefont {M.}~\bibnamefont {Thoss}},\
  }\bibfield  {title} {\enquote {\bibinfo {title} {Semiclassical description of
  nonadiabatic quantum dynamics},}\ }\href {\doibase
  10.1103/PhysRevLett.78.578} {\bibfield  {journal} {\bibinfo  {journal} {Phys.
  Rev. Lett.}\ }\textbf {\bibinfo {volume} {78}},\ \bibinfo {pages} {578--581}
  (\bibinfo {year} {1997})}\BibitemShut {NoStop}%
\bibitem [{\citenamefont {M{\"u}ller}\ and\ \citenamefont
  {Stock}(1999)}]{Mueller1999pyrazine}%
  \BibitemOpen
  \bibfield  {author} {\bibinfo {author} {\bibfnamefont {U.}~\bibnamefont
  {M{\"u}ller}}\ and\ \bibinfo {author} {\bibfnamefont {G.}~\bibnamefont
  {Stock}},\ }\bibfield  {title} {\enquote {\bibinfo {title} {Flow of
  zero-point energy and exploration of phase space in classical simulations of
  quantum relaxation dynamics. {II.} application to nonadiabatic processes},}\
  }\href {\doibase 10.1063/1.479255} {\bibfield  {journal} {\bibinfo  {journal}
  {J.~Chem. Phys.}\ }\textbf {\bibinfo {volume} {111}},\ \bibinfo {pages} {77}
  (\bibinfo {year} {1999})}\BibitemShut {NoStop}%
\bibitem [{\citenamefont {Sun}, \citenamefont {Wang},\ and\ \citenamefont
  {Miller}(1998)}]{Sun1998mapping}%
  \BibitemOpen
  \bibfield  {author} {\bibinfo {author} {\bibfnamefont {X.}~\bibnamefont
  {Sun}}, \bibinfo {author} {\bibfnamefont {H.}~\bibnamefont {Wang}}, \ and\
  \bibinfo {author} {\bibfnamefont {W.~H.}\ \bibnamefont {Miller}},\ }\bibfield
   {title} {\enquote {\bibinfo {title} {Semiclassical theory of electronically
  nonadiabatic dynamics: Results of a linearized approximation to the initial
  value representation},}\ }\href {\doibase 10.1063/1.477389} {\bibfield
  {journal} {\bibinfo  {journal} {J.~Chem. Phys.}\ }\textbf {\bibinfo {volume}
  {109}},\ \bibinfo {pages} {7064--7074} (\bibinfo {year} {1998})}\BibitemShut
  {NoStop}%
\bibitem [{\citenamefont {Wang}\ \emph {et~al.}(1999)\citenamefont {Wang},
  \citenamefont {Song}, \citenamefont {Chandler},\ and\ \citenamefont
  {Miller}}]{Wang1999mapping}%
  \BibitemOpen
  \bibfield  {author} {\bibinfo {author} {\bibfnamefont {H.}~\bibnamefont
  {Wang}}, \bibinfo {author} {\bibfnamefont {X.}~\bibnamefont {Song}}, \bibinfo
  {author} {\bibfnamefont {D.}~\bibnamefont {Chandler}}, \ and\ \bibinfo
  {author} {\bibfnamefont {W.~H.}\ \bibnamefont {Miller}},\ }\bibfield  {title}
  {\enquote {\bibinfo {title} {Semiclassical study of electronically
  nonadiabatic dynamics in the condensed-phase: Spin-boson problem with debye
  spectral density},}\ }\href {\doibase 10.1063/1.478388} {\bibfield  {journal}
  {\bibinfo  {journal} {J.~Chem. Phys.}\ }\textbf {\bibinfo {volume} {110}},\
  \bibinfo {pages} {4828--4840} (\bibinfo {year} {1999})}\BibitemShut {NoStop}%
\bibitem [{\citenamefont {Kim}, \citenamefont {Nassimi},\ and\ \citenamefont
  {Kapral}(2008)}]{Kim2008Liouville}%
  \BibitemOpen
  \bibfield  {author} {\bibinfo {author} {\bibfnamefont {H.}~\bibnamefont
  {Kim}}, \bibinfo {author} {\bibfnamefont {A.}~\bibnamefont {Nassimi}}, \ and\
  \bibinfo {author} {\bibfnamefont {R.}~\bibnamefont {Kapral}},\ }\bibfield
  {title} {\enquote {\bibinfo {title} {{Quantum-classical Liouville dynamics in
  the mapping basis}},}\ }\href {\doibase 10.1063/1.2971041} {\bibfield
  {journal} {\bibinfo  {journal} {J.~Chem. Phys.}\ }\textbf {\bibinfo {volume}
  {129}},\ \bibinfo {pages} {084102} (\bibinfo {year} {2008})}\BibitemShut
  {NoStop}%
\bibitem [{\citenamefont {Kelly}\ \emph {et~al.}(2012)\citenamefont {Kelly},
  \citenamefont {van Zon}, \citenamefont {Schofield},\ and\ \citenamefont
  {Kapral}}]{Kelly2012mapping}%
  \BibitemOpen
  \bibfield  {author} {\bibinfo {author} {\bibfnamefont {A.}~\bibnamefont
  {Kelly}}, \bibinfo {author} {\bibfnamefont {R.}~\bibnamefont {van Zon}},
  \bibinfo {author} {\bibfnamefont {J.}~\bibnamefont {Schofield}}, \ and\
  \bibinfo {author} {\bibfnamefont {R.}~\bibnamefont {Kapral}},\ }\bibfield
  {title} {\enquote {\bibinfo {title} {{Mapping quantum-classical Liouville
  equation: Projectors and trajectories}},}\ }\href {\doibase
  10.1063/1.3685420} {\bibfield  {journal} {\bibinfo  {journal} {J.~Chem.
  Phys.}\ }\textbf {\bibinfo {volume} {136}},\ \bibinfo {pages} {084101}
  (\bibinfo {year} {2012})}\BibitemShut {NoStop}%
\bibitem [{\citenamefont {Saller}, \citenamefont {Kelly},\ and\ \citenamefont
  {Richardson}(2019)}]{identity}%
  \BibitemOpen
  \bibfield  {author} {\bibinfo {author} {\bibfnamefont {M.~A.~C.}\
  \bibnamefont {Saller}}, \bibinfo {author} {\bibfnamefont {A.}~\bibnamefont
  {Kelly}}, \ and\ \bibinfo {author} {\bibfnamefont {J.~O.}\ \bibnamefont
  {Richardson}},\ }\bibfield  {title} {\enquote {\bibinfo {title} {On the
  identity of the identity operator in nonadiabatic linearized semiclassical
  dynamics},}\ }\href {\doibase 10.1063/1.5082596} {\bibfield  {journal}
  {\bibinfo  {journal} {J. Chem. Phys.}\ }\textbf {\bibinfo {volume} {150}},\
  \bibinfo {pages} {071101} (\bibinfo {year} {2019})},\ \Eprint
  {http://arxiv.org/abs/1811.08830} {arXiv:1811.08830 [physics.chem-ph]}
  \BibitemShut {NoStop}%
\bibitem [{\citenamefont {Liu}\ \emph {et~al.}(2020)\citenamefont {Liu},
  \citenamefont {Gao}, \citenamefont {Lai}, \citenamefont {Mulvihill},\ and\
  \citenamefont {Geva}}]{Liu2020linearized}%
  \BibitemOpen
  \bibfield  {author} {\bibinfo {author} {\bibfnamefont {Y.}~\bibnamefont
  {Liu}}, \bibinfo {author} {\bibfnamefont {X.}~\bibnamefont {Gao}}, \bibinfo
  {author} {\bibfnamefont {Y.}~\bibnamefont {Lai}}, \bibinfo {author}
  {\bibfnamefont {E.}~\bibnamefont {Mulvihill}}, \ and\ \bibinfo {author}
  {\bibfnamefont {E.}~\bibnamefont {Geva}},\ }\bibfield  {title} {\enquote
  {\bibinfo {title} {Electronic dynamics through conical intersections via
  quasi-classical mapping hamiltonian methods},}\ }\href {\doibase
  10.1021/acs.jctc.0c00177} {\bibfield  {journal} {\bibinfo  {journal} {J.
  Chem. Theory Comput.}\ }\textbf {\bibinfo {volume} {16}},\ \bibinfo {pages}
  {4479--4488} (\bibinfo {year} {2020})}\BibitemShut {NoStop}%
\bibitem [{\citenamefont {Runeson}\ and\ \citenamefont
  {Richardson}(2019)}]{spinmap}%
  \BibitemOpen
  \bibfield  {author} {\bibinfo {author} {\bibfnamefont {J.~E.}\ \bibnamefont
  {Runeson}}\ and\ \bibinfo {author} {\bibfnamefont {J.~O.}\ \bibnamefont
  {Richardson}},\ }\bibfield  {title} {\enquote {\bibinfo {title} {Spin-mapping
  approach for nonadiabatic molecular dynamics},}\ }\href {\doibase
  10.1063/1.5100506} {\bibfield  {journal} {\bibinfo  {journal} {J. Chem.
  Phys.}\ }\textbf {\bibinfo {volume} {151}},\ \bibinfo {pages} {044119}
  (\bibinfo {year} {2019})},\ \Eprint {http://arxiv.org/abs/1904.08293}
  {arXiv:1904.08293 [physics.chem-ph]} \BibitemShut {NoStop}%
\bibitem [{\citenamefont {Runeson}\ and\ \citenamefont
  {Richardson}(2020)}]{multispin}%
  \BibitemOpen
  \bibfield  {author} {\bibinfo {author} {\bibfnamefont {J.~E.}\ \bibnamefont
  {Runeson}}\ and\ \bibinfo {author} {\bibfnamefont {J.~O.}\ \bibnamefont
  {Richardson}},\ }\bibfield  {title} {\enquote {\bibinfo {title} {Generalized
  spin mapping for quantum-classical dynamics},}\ }\href {\doibase
  10.1063/1.5143412} {\bibfield  {journal} {\bibinfo  {journal} {J. Chem.
  Phys.}\ }\textbf {\bibinfo {volume} {152}},\ \bibinfo {pages} {084110}
  (\bibinfo {year} {2020})},\ \Eprint {http://arxiv.org/abs/1912.10906}
  {arXiv:1912.10906 [physics.chem-ph]} \BibitemShut {NoStop}%
\bibitem [{\citenamefont {Runeson}\ \emph
  {et~al.}(2022{\natexlab{a}})\citenamefont {Runeson}, \citenamefont
  {Mannouch}, \citenamefont {Amati}, \citenamefont {Fiechter},\ and\
  \citenamefont {Richardson}}]{chimia_mapping2022}%
  \BibitemOpen
  \bibfield  {author} {\bibinfo {author} {\bibfnamefont {J.~E.}\ \bibnamefont
  {Runeson}}, \bibinfo {author} {\bibfnamefont {J.~R.}\ \bibnamefont
  {Mannouch}}, \bibinfo {author} {\bibfnamefont {G.}~\bibnamefont {Amati}},
  \bibinfo {author} {\bibfnamefont {M.~R.}\ \bibnamefont {Fiechter}}, \ and\
  \bibinfo {author} {\bibfnamefont {J.~O.}\ \bibnamefont {Richardson}},\
  }\bibfield  {title} {\enquote {\bibinfo {title} {Spin-mapping methods for
  simulating ultrafast nonadiabatic dynamics},}\ }\href {\doibase
  10.2533/chimia.2022.582} {\bibfield  {journal} {\bibinfo  {journal} {CHIMIA}\
  }\textbf {\bibinfo {volume} {76}},\ \bibinfo {pages} {582} (\bibinfo {year}
  {2022}{\natexlab{a}})}\BibitemShut {NoStop}%
\bibitem [{\citenamefont {Parandekar}\ and\ \citenamefont
  {Tully}(2006)}]{Parandekar2006Ehrenfest}%
  \BibitemOpen
  \bibfield  {author} {\bibinfo {author} {\bibfnamefont {P.~V.}\ \bibnamefont
  {Parandekar}}\ and\ \bibinfo {author} {\bibfnamefont {J.~C.}\ \bibnamefont
  {Tully}},\ }\bibfield  {title} {\enquote {\bibinfo {title} {Detailed balance
  in ehrenfest mixed quantum-classical dynamics},}\ }\href {\doibase
  10.1021/ct050213k} {\bibfield  {journal} {\bibinfo  {journal} {J. Chem.
  Theory Comput.}\ }\textbf {\bibinfo {volume} {2}},\ \bibinfo {pages}
  {229--235} (\bibinfo {year} {2006})}\BibitemShut {NoStop}%
\bibitem [{\citenamefont {Shi}\ and\ \citenamefont {Geva}(2004)}]{Shi2004GQME}%
  \BibitemOpen
  \bibfield  {author} {\bibinfo {author} {\bibfnamefont {Q.}~\bibnamefont
  {Shi}}\ and\ \bibinfo {author} {\bibfnamefont {E.}~\bibnamefont {Geva}},\
  }\bibfield  {title} {\enquote {\bibinfo {title} {A semiclassical generalized
  quantum master equation for an arbitrary system-bath coupling},}\ }\href
  {\doibase 10.1063/1.1738109} {\bibfield  {journal} {\bibinfo  {journal}
  {J.~Chem. Phys.}\ }\textbf {\bibinfo {volume} {120}},\ \bibinfo {pages}
  {10647--10658} (\bibinfo {year} {2004})}\BibitemShut {NoStop}%
\bibitem [{\citenamefont {Kelly}\ and\ \citenamefont
  {Markland}(2013)}]{Kelly2013GQME}%
  \BibitemOpen
  \bibfield  {author} {\bibinfo {author} {\bibfnamefont {A.}~\bibnamefont
  {Kelly}}\ and\ \bibinfo {author} {\bibfnamefont {T.~E.}\ \bibnamefont
  {Markland}},\ }\bibfield  {title} {\enquote {\bibinfo {title} {Efficient and
  accurate surface hopping for long time nonadiabatic quantum dynamics},}\
  }\href {\doibase 10.1063/1.4812355} {\bibfield  {journal} {\bibinfo
  {journal} {J.~Chem. Phys.}\ }\textbf {\bibinfo {volume} {139}},\ \bibinfo
  {pages} {014104} (\bibinfo {year} {2013})}\BibitemShut {NoStop}%
\bibitem [{\citenamefont {Kelly}, \citenamefont {Brackbill},\ and\
  \citenamefont {Markland}(2015)}]{Kelly2015nonadiabatic}%
  \BibitemOpen
  \bibfield  {author} {\bibinfo {author} {\bibfnamefont {A.}~\bibnamefont
  {Kelly}}, \bibinfo {author} {\bibfnamefont {N.}~\bibnamefont {Brackbill}}, \
  and\ \bibinfo {author} {\bibfnamefont {T.~E.}\ \bibnamefont {Markland}},\
  }\bibfield  {title} {\enquote {\bibinfo {title} {Accurate nonadiabatic
  quantum dynamics on the cheap: Making the most of mean field theory with
  master equations},}\ }\href {\doibase 10.1063/1.4913686} {\bibfield
  {journal} {\bibinfo  {journal} {J.~Chem. Phys.}\ }\textbf {\bibinfo {volume}
  {142}},\ \bibinfo {pages} {094110} (\bibinfo {year} {2015})}\BibitemShut
  {NoStop}%
\bibitem [{\citenamefont {Pfalzgraff}, \citenamefont {Kelly},\ and\
  \citenamefont {Markland}(2015{\natexlab{a}})}]{Pfalzgraff2015GQME}%
  \BibitemOpen
  \bibfield  {author} {\bibinfo {author} {\bibfnamefont {W.~C.}\ \bibnamefont
  {Pfalzgraff}}, \bibinfo {author} {\bibfnamefont {A.}~\bibnamefont {Kelly}}, \
  and\ \bibinfo {author} {\bibfnamefont {T.~E.}\ \bibnamefont {Markland}},\
  }\bibfield  {title} {\enquote {\bibinfo {title} {Nonadiabatic dynamics in
  atomistic environments: Harnessing quantum-classical theory with generalized
  quantum master equations},}\ }\href {\doibase 10.1021/acs.jpclett.5b02131}
  {\bibfield  {journal} {\bibinfo  {journal} {J. Phys. Chem. Lett.}\ }\textbf
  {\bibinfo {volume} {6}},\ \bibinfo {pages} {4743--4748} (\bibinfo {year}
  {2015}{\natexlab{a}})}\BibitemShut {NoStop}%
\bibitem [{\citenamefont {Kelly}\ \emph {et~al.}(2016)\citenamefont {Kelly},
  \citenamefont {Montoya-Castillo}, \citenamefont {Wang},\ and\ \citenamefont
  {Markland}}]{Kelly2016master}%
  \BibitemOpen
  \bibfield  {author} {\bibinfo {author} {\bibfnamefont {A.}~\bibnamefont
  {Kelly}}, \bibinfo {author} {\bibfnamefont {A.}~\bibnamefont
  {Montoya-Castillo}}, \bibinfo {author} {\bibfnamefont {L.}~\bibnamefont
  {Wang}}, \ and\ \bibinfo {author} {\bibfnamefont {T.~E.}\ \bibnamefont
  {Markland}},\ }\bibfield  {title} {\enquote {\bibinfo {title} {Generalized
  quantum master equations in and out of equilibrium: When can one win?}}\
  }\href {\doibase 10.1063/1.4948612} {\bibfield  {journal} {\bibinfo
  {journal} {J.~Chem. Phys.}\ }\textbf {\bibinfo {volume} {144}},\ \bibinfo
  {pages} {184105} (\bibinfo {year} {2016})}\BibitemShut {NoStop}%
\bibitem [{\citenamefont {Montoya-Castillo}\ and\ \citenamefont
  {Reichman}(2016)}]{Montoya2016GQME}%
  \BibitemOpen
  \bibfield  {author} {\bibinfo {author} {\bibfnamefont {A.}~\bibnamefont
  {Montoya-Castillo}}\ and\ \bibinfo {author} {\bibfnamefont {D.~R.}\
  \bibnamefont {Reichman}},\ }\bibfield  {title} {\enquote {\bibinfo {title}
  {{Approximate but accurate quantum dynamics from the Mori formalism: I.
  Nonequilibrium dynamics}},}\ }\href {\doibase 10.1063/1.4948408} {\bibfield
  {journal} {\bibinfo  {journal} {J.~Chem.~Phys.}\ }\textbf {\bibinfo {volume}
  {144}},\ \bibinfo {pages} {184104} (\bibinfo {year} {2016})}\BibitemShut
  {NoStop}%
\bibitem [{\citenamefont {Mulvihill}\ \emph
  {et~al.}(2019{\natexlab{a}})\citenamefont {Mulvihill}, \citenamefont {Gao},
  \citenamefont {Liu}, \citenamefont {Schubert}, \citenamefont {Dunietz},\ and\
  \citenamefont {Geva}}]{Mulvihill2019LSCGQME}%
  \BibitemOpen
  \bibfield  {author} {\bibinfo {author} {\bibfnamefont {E.}~\bibnamefont
  {Mulvihill}}, \bibinfo {author} {\bibfnamefont {X.}~\bibnamefont {Gao}},
  \bibinfo {author} {\bibfnamefont {Y.}~\bibnamefont {Liu}}, \bibinfo {author}
  {\bibfnamefont {A.}~\bibnamefont {Schubert}}, \bibinfo {author}
  {\bibfnamefont {B.~D.}\ \bibnamefont {Dunietz}}, \ and\ \bibinfo {author}
  {\bibfnamefont {E.}~\bibnamefont {Geva}},\ }\bibfield  {title} {\enquote
  {\bibinfo {title} {{Combining the mapping Hamiltonian linearized
  semiclassical approach with the generalized quantum master equation to
  simulate electronically nonadiabatic molecular dynamics}},}\ }\href {\doibase
  10.1063/1.5110891} {\bibfield  {journal} {\bibinfo  {journal} {J.~Chem.
  Phys.}\ }\textbf {\bibinfo {volume} {151}},\ \bibinfo {pages} {074103}
  (\bibinfo {year} {2019}{\natexlab{a}})}\BibitemShut {NoStop}%
\bibitem [{\citenamefont {Mulvihill}\ and\ \citenamefont
  {Geva}(2021)}]{mulvihill2021b}%
  \BibitemOpen
  \bibfield  {author} {\bibinfo {author} {\bibfnamefont {E.}~\bibnamefont
  {Mulvihill}}\ and\ \bibinfo {author} {\bibfnamefont {E.}~\bibnamefont
  {Geva}},\ }\bibfield  {title} {\enquote {\bibinfo {title} {A road map to
  various pathways for calculating the memory kernel of the generalized quantum
  master equation},}\ }\href {\doibase 10.1021/acs.jpcb.1c05719} {\bibfield
  {journal} {\bibinfo  {journal} {J. Phys. Chem. B}\ }\textbf {\bibinfo
  {volume} {125}},\ \bibinfo {pages} {9834--9852} (\bibinfo {year}
  {2021})}\BibitemShut {NoStop}%
\bibitem [{\citenamefont {Mulvihill}\ and\ \citenamefont
  {Geva}(2022)}]{mulvihill2022}%
  \BibitemOpen
  \bibfield  {author} {\bibinfo {author} {\bibfnamefont {E.~A.}\ \bibnamefont
  {Mulvihill}}\ and\ \bibinfo {author} {\bibfnamefont {E.}~\bibnamefont
  {Geva}},\ }\bibfield  {title} {\enquote {\bibinfo {title} {Simulating the
  dynamics of electronic observables via reduced-dimensionality generalized
  quantum master equations},}\ }\href {\doibase 10.1063/5.0078040} {\bibfield
  {journal} {\bibinfo  {journal} {J.~Chem. Phys.}\ }\textbf {\bibinfo {volume}
  {156}},\ \bibinfo {pages} {044119} (\bibinfo {year} {2022})}\BibitemShut
  {NoStop}%
\bibitem [{\citenamefont {Ng}\ and\ \citenamefont {Rabani}(2022)}]{ng2022}%
  \BibitemOpen
  \bibfield  {author} {\bibinfo {author} {\bibfnamefont {N.}~\bibnamefont
  {Ng}}\ and\ \bibinfo {author} {\bibfnamefont {E.}~\bibnamefont {Rabani}},\
  }\bibfield  {title} {\enquote {\bibinfo {title} {Long-time memory effects in
  a localizable central spin problem},}\ }\href {\doibase
  10.1088/1367-2630/ac4735} {\bibfield  {journal} {\bibinfo  {journal} {New J.
  Phys.}\ }\textbf {\bibinfo {volume} {24}},\ \bibinfo {pages} {013025}
  (\bibinfo {year} {2022})}\BibitemShut {NoStop}%
\bibitem [{\citenamefont {Breuer}\ and\ \citenamefont
  {Petruccione}(2002)}]{OpenQuantum}%
  \BibitemOpen
  \bibfield  {author} {\bibinfo {author} {\bibfnamefont {H.-P.}\ \bibnamefont
  {Breuer}}\ and\ \bibinfo {author} {\bibfnamefont {F.}~\bibnamefont
  {Petruccione}},\ }\href@noop {} {\emph {\bibinfo {title} {The Theory of Open
  Quantum Systems}}}\ (\bibinfo  {publisher} {Oxford University Press},\
  \bibinfo {address} {Oxford},\ \bibinfo {year} {2002})\BibitemShut {NoStop}%
\bibitem [{\citenamefont {Shi}\ and\ \citenamefont {Geva}(2003)}]{shi2003}%
  \BibitemOpen
  \bibfield  {author} {\bibinfo {author} {\bibfnamefont {Q.}~\bibnamefont
  {Shi}}\ and\ \bibinfo {author} {\bibfnamefont {E.}~\bibnamefont {Geva}},\
  }\bibfield  {title} {\enquote {\bibinfo {title} {A new approach to
  calculating the memory kernel of the generalized quantum master equation for
  an arbitrary system–bath coupling},}\ }\href {\doibase 10.1063/1.1624830}
  {\bibfield  {journal} {\bibinfo  {journal} {J.~Chem. Phys.}\ }\textbf
  {\bibinfo {volume} {119}},\ \bibinfo {pages} {12063--12076} (\bibinfo {year}
  {2003})}\BibitemShut {NoStop}%
\bibitem [{\citenamefont {Erpenbeck}\ and\ \citenamefont
  {Thoss}(2019)}]{erpenbeck2019}%
  \BibitemOpen
  \bibfield  {author} {\bibinfo {author} {\bibfnamefont {A.}~\bibnamefont
  {Erpenbeck}}\ and\ \bibinfo {author} {\bibfnamefont {M.}~\bibnamefont
  {Thoss}},\ }\bibfield  {title} {\enquote {\bibinfo {title} {Hierarchical
  quantum master equation approach to vibronic reaction dynamics at metal
  surfaces},}\ }\href {\doibase 10.1063/1.5128206} {\bibfield  {journal}
  {\bibinfo  {journal} {J.~Chem. Phys.}\ }\textbf {\bibinfo {volume} {151}},\
  \bibinfo {pages} {191101} (\bibinfo {year} {2019})}\BibitemShut {NoStop}%
\bibitem [{\citenamefont {Ullah}\ and\ \citenamefont {Dral}(2021)}]{ullah2021}%
  \BibitemOpen
  \bibfield  {author} {\bibinfo {author} {\bibfnamefont {A.}~\bibnamefont
  {Ullah}}\ and\ \bibinfo {author} {\bibfnamefont {P.~O.}\ \bibnamefont
  {Dral}},\ }\bibfield  {title} {\enquote {\bibinfo {title} {Speeding up
  quantum dissipative dynamics of open systems with kernel methods},}\ }\href
  {\doibase 10.1088/1367-2630/ac3261} {\bibfield  {journal} {\bibinfo
  {journal} {New J. Phys.}\ }\textbf {\bibinfo {volume} {23}},\ \bibinfo
  {pages} {113019} (\bibinfo {year} {2021})}\BibitemShut {NoStop}%
\bibitem [{\citenamefont {Pfalzgraff}\ \emph {et~al.}(2019)\citenamefont
  {Pfalzgraff}, \citenamefont {Montoya-Castillo}, \citenamefont {Kelly},\ and\
  \citenamefont {Markland}}]{Pfalzgraff2019GQME}%
  \BibitemOpen
  \bibfield  {author} {\bibinfo {author} {\bibfnamefont {W.~C.}\ \bibnamefont
  {Pfalzgraff}}, \bibinfo {author} {\bibfnamefont {A.}~\bibnamefont
  {Montoya-Castillo}}, \bibinfo {author} {\bibfnamefont {A.}~\bibnamefont
  {Kelly}}, \ and\ \bibinfo {author} {\bibfnamefont {T.~E.}\ \bibnamefont
  {Markland}},\ }\bibfield  {title} {\enquote {\bibinfo {title} {Efficient
  construction of generalized master equation memory kernels for multi-state
  systems from nonadiabatic quantum-classical dynamics},}\ }\href {\doibase
  10.1063/1.5095715} {\bibfield  {journal} {\bibinfo  {journal} {J.~Chem.
  Phys.}\ }\textbf {\bibinfo {volume} {150}},\ \bibinfo {pages} {244109}
  (\bibinfo {year} {2019})}\BibitemShut {NoStop}%
\bibitem [{\citenamefont {Mulvihill}\ \emph {et~al.}(2021)\citenamefont
  {Mulvihill}, \citenamefont {Lenn}, \citenamefont {Gao}, \citenamefont
  {Schubert}, \citenamefont {Dunietz},\ and\ \citenamefont
  {Geva}}]{mulvihill2021}%
  \BibitemOpen
  \bibfield  {author} {\bibinfo {author} {\bibfnamefont {E.}~\bibnamefont
  {Mulvihill}}, \bibinfo {author} {\bibfnamefont {K.~M.}\ \bibnamefont {Lenn}},
  \bibinfo {author} {\bibfnamefont {X.}~\bibnamefont {Gao}}, \bibinfo {author}
  {\bibfnamefont {A.}~\bibnamefont {Schubert}}, \bibinfo {author}
  {\bibfnamefont {B.~D.}\ \bibnamefont {Dunietz}}, \ and\ \bibinfo {author}
  {\bibfnamefont {E.}~\bibnamefont {Geva}},\ }\bibfield  {title} {\enquote
  {\bibinfo {title} {{Simulating energy transfer dynamics in the
  Fenna--Matthews--Olson complex via the modified generalized quantum master
  equation}},}\ }\href {\doibase 10.1063/5.0051101} {\bibfield  {journal}
  {\bibinfo  {journal} {J.~Chem. Phys.}\ }\textbf {\bibinfo {volume} {154}},\
  \bibinfo {pages} {204109} (\bibinfo {year} {2021})}\BibitemShut {NoStop}%
\bibitem [{\citenamefont {Gao}\ \emph {et~al.}(2020)\citenamefont {Gao},
  \citenamefont {Saller}, \citenamefont {Liu}, \citenamefont {Kelly},
  \citenamefont {Richardson},\ and\ \citenamefont {Geva}}]{linearized}%
  \BibitemOpen
  \bibfield  {author} {\bibinfo {author} {\bibfnamefont {X.}~\bibnamefont
  {Gao}}, \bibinfo {author} {\bibfnamefont {M.~A.~C.}\ \bibnamefont {Saller}},
  \bibinfo {author} {\bibfnamefont {Y.}~\bibnamefont {Liu}}, \bibinfo {author}
  {\bibfnamefont {A.}~\bibnamefont {Kelly}}, \bibinfo {author} {\bibfnamefont
  {J.~O.}\ \bibnamefont {Richardson}}, \ and\ \bibinfo {author} {\bibfnamefont
  {E.}~\bibnamefont {Geva}},\ }\bibfield  {title} {\enquote {\bibinfo {title}
  {Benchmarking quasiclassical mapping hamiltonian methods for simulating
  electronically nonadiabatic molecular dynamics},}\ }\href {\doibase
  10.1021/acs.jctc.9b01267} {\bibfield  {journal} {\bibinfo  {journal}
  {J.~Chem.~Theory Comput.}\ }\textbf {\bibinfo {volume} {16}},\ \bibinfo
  {pages} {2883--2895} (\bibinfo {year} {2020})}\BibitemShut {NoStop}%
\bibitem [{\citenamefont {Runeson}\ \emph
  {et~al.}(2022{\natexlab{b}})\citenamefont {Runeson}, \citenamefont
  {Lawrence}, \citenamefont {Mannouch},\ and\ \citenamefont
  {Richardson}}]{FMOclassical}%
  \BibitemOpen
  \bibfield  {author} {\bibinfo {author} {\bibfnamefont {J.~E.}\ \bibnamefont
  {Runeson}}, \bibinfo {author} {\bibfnamefont {J.~E.}\ \bibnamefont
  {Lawrence}}, \bibinfo {author} {\bibfnamefont {J.~R.}\ \bibnamefont
  {Mannouch}}, \ and\ \bibinfo {author} {\bibfnamefont {J.~O.}\ \bibnamefont
  {Richardson}},\ }\bibfield  {title} {\enquote {\bibinfo {title} {Explaining
  the efficiency of photosynthesis: quantum uncertainty or classical
  vibrations?}}\ }\href {\doibase 10.1021/acs.jpclett.2c00538} {\bibfield
  {journal} {\bibinfo  {journal} {J.~Phys. Chem. Lett.}\ }\textbf {\bibinfo
  {volume} {13}},\ \bibinfo {pages} {3392--3399} (\bibinfo {year}
  {2022}{\natexlab{b}})}\BibitemShut {NoStop}%
\bibitem [{\citenamefont {Nakajima}(1958)}]{nakajima1958}%
  \BibitemOpen
  \bibfield  {author} {\bibinfo {author} {\bibfnamefont {S.}~\bibnamefont
  {Nakajima}},\ }\bibfield  {title} {\enquote {\bibinfo {title} {{On Quantum
  Theory of Transport Phenomena: Steady Diffusion}},}\ }\href {\doibase
  10.1143/PTP.20.948} {\bibfield  {journal} {\bibinfo  {journal} {Prog. Theor.
  Phys.}\ }\textbf {\bibinfo {volume} {20}},\ \bibinfo {pages} {948--959}
  (\bibinfo {year} {1958})}\BibitemShut {NoStop}%
\bibitem [{\citenamefont {Zwanzig}(1960)}]{zwanzig1960}%
  \BibitemOpen
  \bibfield  {author} {\bibinfo {author} {\bibfnamefont {R.}~\bibnamefont
  {Zwanzig}},\ }\bibfield  {title} {\enquote {\bibinfo {title} {Ensemble method
  in the theory of irreversibility},}\ }\href {\doibase 10.1063/1.1731409}
  {\bibfield  {journal} {\bibinfo  {journal} {J.~Chem. Phys.}\ }\textbf
  {\bibinfo {volume} {33}},\ \bibinfo {pages} {1338--1341} (\bibinfo {year}
  {1960})}\BibitemShut {NoStop}%
\bibitem [{\citenamefont {Mori}(1965)}]{mori1965}%
  \BibitemOpen
  \bibfield  {author} {\bibinfo {author} {\bibfnamefont {H.}~\bibnamefont
  {Mori}},\ }\bibfield  {title} {\enquote {\bibinfo {title} {{Transport,
  Collective Motion, and Brownian Motion}},}\ }\href {\doibase
  10.1143/PTP.33.423} {\bibfield  {journal} {\bibinfo  {journal} {Prog. Theor.
  Phys.}\ }\textbf {\bibinfo {volume} {33}},\ \bibinfo {pages} {423--455}
  (\bibinfo {year} {1965})}\BibitemShut {NoStop}%
\bibitem [{\citenamefont {Gyamfi}(2020)}]{gyamfi2020}%
  \BibitemOpen
  \bibfield  {author} {\bibinfo {author} {\bibfnamefont {J.~A.}\ \bibnamefont
  {Gyamfi}},\ }\bibfield  {title} {\enquote {\bibinfo {title} {{Fundamentals of
  quantum mechanics in Liouville space}},}\ }\href {\doibase
  10.1088/1361-6404/ab9fdd} {\bibfield  {journal} {\bibinfo  {journal} {Eur. J.
  Phys.}\ }\textbf {\bibinfo {volume} {41}},\ \bibinfo {pages} {063002}
  (\bibinfo {year} {2020})}\BibitemShut {NoStop}%
\bibitem [{\citenamefont {Press}\ \emph {et~al.}(2007)\citenamefont {Press},
  \citenamefont {Teukolsky}, \citenamefont {Vetterling},\ and\ \citenamefont
  {Flannery}}]{NumRep}%
  \BibitemOpen
  \bibfield  {author} {\bibinfo {author} {\bibfnamefont {W.~H.}\ \bibnamefont
  {Press}}, \bibinfo {author} {\bibfnamefont {S.~A.}\ \bibnamefont
  {Teukolsky}}, \bibinfo {author} {\bibfnamefont {W.~T.}\ \bibnamefont
  {Vetterling}}, \ and\ \bibinfo {author} {\bibfnamefont {B.~P.}\ \bibnamefont
  {Flannery}},\ }\href@noop {} {\emph {\bibinfo {title} {Numerical Recipes:
  {The} Art of Scientific Computing}}},\ \bibinfo {edition} {3rd}\ ed.\
  (\bibinfo  {publisher} {Cambridge University Press},\ \bibinfo {address}
  {Cambridge},\ \bibinfo {year} {2007})\BibitemShut {NoStop}%
\bibitem [{\citenamefont {Runeson}(2022)}]{runesonPhD}%
  \BibitemOpen
  \bibfield  {author} {\bibinfo {author} {\bibfnamefont {J.~E.}\ \bibnamefont
  {Runeson}},\ }\emph {\bibinfo {title} {Spin-mapping approaches for mixed
  quantum-classical dynamics}},\ \href {\doibase 10.3929/ethz-b-000543913}
  {Ph.D. thesis},\ \bibinfo  {school} {ETH Zurich} (\bibinfo {year} {2022}),\
  \bibinfo {note} {\url{https://doi.org/10.3929/ethz-b-000543913}}\BibitemShut
  {NoStop}%
\bibitem [{\citenamefont {Stratonovich}(1957)}]{stratonovich1957}%
  \BibitemOpen
  \bibfield  {author} {\bibinfo {author} {\bibfnamefont {R.~L.}\ \bibnamefont
  {Stratonovich}},\ }\bibfield  {title} {\enquote {\bibinfo {title} {On
  distributions in representation space},}\ }\href
  {http://jetp.ras.ru/cgi-bin/e/index/e/4/6/p891?a=list} {\bibfield  {journal}
  {\bibinfo  {journal} {Sov. Phys. JETP}\ }\textbf {\bibinfo {volume} {4}},\
  \bibinfo {pages} {891--898} (\bibinfo {year} {1957})}\BibitemShut {NoStop}%
\bibitem [{\citenamefont {Weigert}\ and\ \citenamefont
  {Müller}(1995)}]{weigert1995}%
  \BibitemOpen
  \bibfield  {author} {\bibinfo {author} {\bibfnamefont {S.}~\bibnamefont
  {Weigert}}\ and\ \bibinfo {author} {\bibfnamefont {G.}~\bibnamefont
  {Müller}},\ }\bibfield  {title} {\enquote {\bibinfo {title} {Quantum
  integrability and action operators in spin dynamics},}\ }\href {\doibase
  https://doi.org/10.1016/0960-0779(95)00021-U} {\bibfield  {journal} {\bibinfo
   {journal} {Chaos Solit. Fractals}\ }\textbf {\bibinfo {volume} {5}},\
  \bibinfo {pages} {1419--1438} (\bibinfo {year} {1995})}\BibitemShut {NoStop}%
\bibitem [{\citenamefont {Hele}\ and\ \citenamefont
  {Ananth}(2016)}]{Hele2016Faraday}%
  \BibitemOpen
  \bibfield  {author} {\bibinfo {author} {\bibfnamefont {T.~J.~H.}\
  \bibnamefont {Hele}}\ and\ \bibinfo {author} {\bibfnamefont {N.}~\bibnamefont
  {Ananth}},\ }\bibfield  {title} {\enquote {\bibinfo {title} {Deriving the
  exact nonadiabatic quantum propagator in the mapping variable
  representation},}\ }\href {\doibase 10.1039/C6FD00106H} {\bibfield  {journal}
  {\bibinfo  {journal} {Faraday Discuss.}\ }\textbf {\bibinfo {volume} {195}},\
  \bibinfo {pages} {269--289} (\bibinfo {year} {2016})}\BibitemShut {NoStop}%
\bibitem [{\citenamefont {Sato}, \citenamefont {Kelly},\ and\ \citenamefont
  {Rubio}(2018)}]{Sato2018CFBT}%
  \BibitemOpen
  \bibfield  {author} {\bibinfo {author} {\bibfnamefont {S.~A.}\ \bibnamefont
  {Sato}}, \bibinfo {author} {\bibfnamefont {A.}~\bibnamefont {Kelly}}, \ and\
  \bibinfo {author} {\bibfnamefont {A.}~\bibnamefont {Rubio}},\ }\bibfield
  {title} {\enquote {\bibinfo {title} {Coupled forward-backward trajectory
  approach for nonequilibrium electron-ion dynamics},}\ }\href {\doibase
  10.1103/PhysRevB.97.134308} {\bibfield  {journal} {\bibinfo  {journal} {Phys.
  Rev. B}\ }\textbf {\bibinfo {volume} {97}},\ \bibinfo {pages} {134308}
  (\bibinfo {year} {2018})}\BibitemShut {NoStop}%
\bibitem [{\citenamefont {Leggett}\ \emph {et~al.}(1987)\citenamefont
  {Leggett}, \citenamefont {Chakravarty}, \citenamefont {Dorsey}, \citenamefont
  {Fisher}, \citenamefont {Garg},\ and\ \citenamefont
  {Zwerger}}]{Leggett1987spinboson}%
  \BibitemOpen
  \bibfield  {author} {\bibinfo {author} {\bibfnamefont {A.~J.}\ \bibnamefont
  {Leggett}}, \bibinfo {author} {\bibfnamefont {S.}~\bibnamefont
  {Chakravarty}}, \bibinfo {author} {\bibfnamefont {A.~T.}\ \bibnamefont
  {Dorsey}}, \bibinfo {author} {\bibfnamefont {M.~P.~A.}\ \bibnamefont
  {Fisher}}, \bibinfo {author} {\bibfnamefont {A.}~\bibnamefont {Garg}}, \ and\
  \bibinfo {author} {\bibfnamefont {W.}~\bibnamefont {Zwerger}},\ }\bibfield
  {title} {\enquote {\bibinfo {title} {Dynamics of the dissipative two-state
  system},}\ }\href {\doibase 10.1103/RevModPhys.59.1} {\bibfield  {journal}
  {\bibinfo  {journal} {Rev. Mod. Phys.}\ }\textbf {\bibinfo {volume} {59}},\
  \bibinfo {pages} {1} (\bibinfo {year} {1987})}\BibitemShut {NoStop}%
\bibitem [{\citenamefont {Craig}\ and\ \citenamefont
  {Manolopoulos}(2005)}]{craig2005}%
  \BibitemOpen
  \bibfield  {author} {\bibinfo {author} {\bibfnamefont {I.~R.}\ \bibnamefont
  {Craig}}\ and\ \bibinfo {author} {\bibfnamefont {D.~E.}\ \bibnamefont
  {Manolopoulos}},\ }\bibfield  {title} {\enquote {\bibinfo {title} {Chemical
  reaction rates from ring polymer molecular dynamics},}\ }\href {\doibase
  10.1063/1.1850093} {\bibfield  {journal} {\bibinfo  {journal} {J.~Chem.
  Phys.}\ }\textbf {\bibinfo {volume} {122}},\ \bibinfo {pages} {084106}
  (\bibinfo {year} {2005})}\BibitemShut {NoStop}%
\bibitem [{\citenamefont {Habershon}\ and\ \citenamefont
  {Manolopoulos}(2009)}]{Habershon2009water}%
  \BibitemOpen
  \bibfield  {author} {\bibinfo {author} {\bibfnamefont {S.}~\bibnamefont
  {Habershon}}\ and\ \bibinfo {author} {\bibfnamefont {D.~E.}\ \bibnamefont
  {Manolopoulos}},\ }\bibfield  {title} {\enquote {\bibinfo {title} {Zero point
  energy leakage in condensed phase dynamics: An assessment of quantum
  simulation methods for liquid water},}\ }\href {\doibase 10.1063/1.3276109}
  {\bibfield  {journal} {\bibinfo  {journal} {J.~Chem. Phys.}\ }\textbf
  {\bibinfo {volume} {131}},\ \bibinfo {pages} {244518} (\bibinfo {year}
  {2009})}\BibitemShut {NoStop}%
\bibitem [{\citenamefont {Hawkins}(2021)}]{hawkins2021}%
  \BibitemOpen
  \bibfield  {author} {\bibinfo {author} {\bibfnamefont {J.}~\bibnamefont
  {Hawkins}},\ }\href {https://books.google.ch/books?id=Yd0XEAAAQBAJ} {\emph
  {\bibinfo {title} {Ergodic Dynamics: From Basic Theory to Applications}}},\
  Graduate Texts in Mathematics\ (\bibinfo  {publisher} {Springer Nature
  Switzerland},\ \bibinfo {address} {Cham},\ \bibinfo {year}
  {2021})\BibitemShut {NoStop}%
\bibitem [{\citenamefont {Evans}, \citenamefont {Searles},\ and\ \citenamefont
  {Williams}(2009)}]{evans2009}%
  \BibitemOpen
  \bibfield  {author} {\bibinfo {author} {\bibfnamefont {D.~J.}\ \bibnamefont
  {Evans}}, \bibinfo {author} {\bibfnamefont {D.~J.}\ \bibnamefont {Searles}},
  \ and\ \bibinfo {author} {\bibfnamefont {S.~R.}\ \bibnamefont {Williams}},\
  }\bibfield  {title} {\enquote {\bibinfo {title} {Dissipation and the
  relaxation to equilibrium},}\ }\href {\doibase
  10.1088/1742-5468/2009/07/p07029} {\bibfield  {journal} {\bibinfo  {journal}
  {J. Stat. Mech. Theory Exp.}\ }\textbf {\bibinfo {volume} {2009}},\ \bibinfo
  {pages} {P07029} (\bibinfo {year} {2009})}\BibitemShut {NoStop}%
\bibitem [{\citenamefont {Mauri}, \citenamefont {Car},\ and\ \citenamefont
  {Tosatti}(1993)}]{mauri1993}%
  \BibitemOpen
  \bibfield  {author} {\bibinfo {author} {\bibfnamefont {F.}~\bibnamefont
  {Mauri}}, \bibinfo {author} {\bibfnamefont {R.}~\bibnamefont {Car}}, \ and\
  \bibinfo {author} {\bibfnamefont {E.}~\bibnamefont {Tosatti}},\ }\bibfield
  {title} {\enquote {\bibinfo {title} {Canonical statistical averages of
  coupled quantum-classical systems},}\ }\href {\doibase
  10.1209/0295-5075/24/6/001} {\bibfield  {journal} {\bibinfo  {journal}
  {Europhys. Lett.}\ }\textbf {\bibinfo {volume} {24}},\ \bibinfo {pages}
  {431--436} (\bibinfo {year} {1993})}\BibitemShut {NoStop}%
\bibitem [{\citenamefont {Thoss}, \citenamefont {Wang},\ and\ \citenamefont
  {Miller}(2001)}]{Thoss2001hybrid}%
  \BibitemOpen
  \bibfield  {author} {\bibinfo {author} {\bibfnamefont {M.}~\bibnamefont
  {Thoss}}, \bibinfo {author} {\bibfnamefont {H.}~\bibnamefont {Wang}}, \ and\
  \bibinfo {author} {\bibfnamefont {W.~H.}\ \bibnamefont {Miller}},\ }\bibfield
   {title} {\enquote {\bibinfo {title} {Self-consistent hybrid approach for
  complex systems: Application to the spin-boson model with {D}ebye spectral
  density},}\ }\href {\doibase 10.1063/1.1385562} {\bibfield  {journal}
  {\bibinfo  {journal} {J.~Chem. Phys.}\ }\textbf {\bibinfo {volume} {115}},\
  \bibinfo {pages} {2991} (\bibinfo {year} {2001})}\BibitemShut {NoStop}%
\bibitem [{\citenamefont {Wang}\ and\ \citenamefont {Thoss}(2017)}]{wang2017}%
  \BibitemOpen
  \bibfield  {author} {\bibinfo {author} {\bibfnamefont {H.}~\bibnamefont
  {Wang}}\ and\ \bibinfo {author} {\bibfnamefont {M.}~\bibnamefont {Thoss}},\
  }\bibfield  {title} {\enquote {\bibinfo {title} {{A multilayer
  multiconfiguration time-dependent Hartree simulation of the
  reaction-coordinate spin-boson model employing an interaction picture}},}\
  }\href {\doibase 10.1063/1.4978901} {\bibfield  {journal} {\bibinfo
  {journal} {J.~Chem. Phys.}\ }\textbf {\bibinfo {volume} {146}},\ \bibinfo
  {pages} {124112} (\bibinfo {year} {2017})}\BibitemShut {NoStop}%
\bibitem [{\citenamefont {{G. Amati, and J. E. Runeson}}\ and\ \citenamefont
  {Richardson}(2022)}]{ellipsoid}%
  \BibitemOpen
  \bibfield  {author} {\bibinfo {author} {\bibnamefont {{G. Amati, and J. E.
  Runeson}}}\ and\ \bibinfo {author} {\bibfnamefont {J.~O.}\ \bibnamefont
  {Richardson}},\ }\href@noop {} {\enquote {\bibinfo {title} {Detailed balance
  in mixed quantum--classical dynamics: Ellipsoid mapping},}\ } (\bibinfo
  {year} {2022}),\ \bibinfo {note} {to appear}\BibitemShut {NoStop}%
\bibitem [{\citenamefont {M{\"u}ller}\ and\ \citenamefont
  {Stock}(1998)}]{Mueller1998mapping}%
  \BibitemOpen
  \bibfield  {author} {\bibinfo {author} {\bibfnamefont {U.}~\bibnamefont
  {M{\"u}ller}}\ and\ \bibinfo {author} {\bibfnamefont {G.}~\bibnamefont
  {Stock}},\ }\bibfield  {title} {\enquote {\bibinfo {title} {Consistent
  treatment of quantum-mechanical and classical degrees of freedom in mixed
  quantum-classical simulations},}\ }\href {\doibase 10.1063/1.476184}
  {\bibfield  {journal} {\bibinfo  {journal} {J.~Chem. Phys.}\ }\textbf
  {\bibinfo {volume} {108}},\ \bibinfo {pages} {7516--7526} (\bibinfo {year}
  {1998})}\BibitemShut {NoStop}%
\bibitem [{\citenamefont {Saller}, \citenamefont {Kelly},\ and\ \citenamefont
  {Richardson}(2020)}]{FMO}%
  \BibitemOpen
  \bibfield  {author} {\bibinfo {author} {\bibfnamefont {M.~A.~C.}\
  \bibnamefont {Saller}}, \bibinfo {author} {\bibfnamefont {A.}~\bibnamefont
  {Kelly}}, \ and\ \bibinfo {author} {\bibfnamefont {J.~O.}\ \bibnamefont
  {Richardson}},\ }\bibfield  {title} {\enquote {\bibinfo {title} {Improved
  population operators for multi-state nonadiabatic dynamics with the mixed
  quantum-classical mapping approach},}\ }\href {\doibase 10.1039/C9FD00050J}
  {\bibfield  {journal} {\bibinfo  {journal} {Faraday Discuss.}\ }\textbf
  {\bibinfo {volume} {221}},\ \bibinfo {pages} {150--167} (\bibinfo {year}
  {2020})},\ \Eprint {http://arxiv.org/abs/1904.11847} {arXiv:1904.11847
  [physics.chem-ph]} \BibitemShut {NoStop}%
\bibitem [{\citenamefont {Bellonzi}, \citenamefont {Jain},\ and\ \citenamefont
  {Subotnik}(2016)}]{bellonzi2016}%
  \BibitemOpen
  \bibfield  {author} {\bibinfo {author} {\bibfnamefont {N.}~\bibnamefont
  {Bellonzi}}, \bibinfo {author} {\bibfnamefont {A.}~\bibnamefont {Jain}}, \
  and\ \bibinfo {author} {\bibfnamefont {J.~E.}\ \bibnamefont {Subotnik}},\
  }\bibfield  {title} {\enquote {\bibinfo {title} {An assessment of mean-field
  mixed semiclassical approaches: Equilibrium populations and algorithm
  stability},}\ }\href {\doibase 10.1063/1.4946810} {\bibfield  {journal}
  {\bibinfo  {journal} {J.~Chem. Phys.}\ }\textbf {\bibinfo {volume} {144}},\
  \bibinfo {pages} {154110} (\bibinfo {year} {2016})}\BibitemShut {NoStop}%
\bibitem [{\citenamefont {Mulvihill}\ \emph
  {et~al.}(2019{\natexlab{b}})\citenamefont {Mulvihill}, \citenamefont
  {Schubert}, \citenamefont {Sun}, \citenamefont {Dunietz},\ and\ \citenamefont
  {Geva}}]{mulvihill2019_no_sb}%
  \BibitemOpen
  \bibfield  {author} {\bibinfo {author} {\bibfnamefont {E.}~\bibnamefont
  {Mulvihill}}, \bibinfo {author} {\bibfnamefont {A.}~\bibnamefont {Schubert}},
  \bibinfo {author} {\bibfnamefont {X.}~\bibnamefont {Sun}}, \bibinfo {author}
  {\bibfnamefont {B.~D.}\ \bibnamefont {Dunietz}}, \ and\ \bibinfo {author}
  {\bibfnamefont {E.}~\bibnamefont {Geva}},\ }\bibfield  {title} {\enquote
  {\bibinfo {title} {A modified approach for simulating electronically
  nonadiabatic dynamics via the generalized quantum master equation},}\ }\href
  {\doibase 10.1063/1.5055756} {\bibfield  {journal} {\bibinfo  {journal}
  {J.~Chem. Phys.}\ }\textbf {\bibinfo {volume} {150}},\ \bibinfo {pages}
  {034101} (\bibinfo {year} {2019}{\natexlab{b}})}\BibitemShut {NoStop}%
\bibitem [{\citenamefont {Makri}(1995)}]{makri1995}%
  \BibitemOpen
  \bibfield  {author} {\bibinfo {author} {\bibfnamefont {N.}~\bibnamefont
  {Makri}},\ }\bibfield  {title} {\enquote {\bibinfo {title} {Numerical path
  integral techniques for long time dynamics of quantum dissipative systems},}\
  }\href {\doibase 10.1063/1.531046} {\bibfield  {journal} {\bibinfo  {journal}
  {J.~Math. Phys.}\ }\textbf {\bibinfo {volume} {36}},\ \bibinfo {pages}
  {2430--2457} (\bibinfo {year} {1995})}\BibitemShut {NoStop}%
\bibitem [{\citenamefont {Manzano}(2020)}]{manzano2020}%
  \BibitemOpen
  \bibfield  {author} {\bibinfo {author} {\bibfnamefont {D.}~\bibnamefont
  {Manzano}},\ }\bibfield  {title} {\enquote {\bibinfo {title} {{A short
  introduction to the Lindblad master equation}},}\ }\href {\doibase
  10.1063/1.5115323} {\bibfield  {journal} {\bibinfo  {journal} {AIP Advances}\
  }\textbf {\bibinfo {volume} {10}},\ \bibinfo {pages} {025106} (\bibinfo
  {year} {2020})}\BibitemShut {NoStop}%
\bibitem [{\citenamefont {Rivas}\ \emph {et~al.}(2010)\citenamefont {Rivas},
  \citenamefont {Plato}, \citenamefont {Huelga},\ and\ \citenamefont
  {Plenio}}]{rivas2010}%
  \BibitemOpen
  \bibfield  {author} {\bibinfo {author} {\bibfnamefont {{\'{A}}.}~\bibnamefont
  {Rivas}}, \bibinfo {author} {\bibfnamefont {A.~D.~K.}\ \bibnamefont {Plato}},
  \bibinfo {author} {\bibfnamefont {S.~F.}\ \bibnamefont {Huelga}}, \ and\
  \bibinfo {author} {\bibfnamefont {M.~B.}\ \bibnamefont {Plenio}},\ }\bibfield
   {title} {\enquote {\bibinfo {title} {{Markovian master equations: A critical
  study}},}\ }\href {\doibase 10.1088/1367-2630/12/11/113032} {\bibfield
  {journal} {\bibinfo  {journal} {New J. Phys.}\ }\textbf {\bibinfo {volume}
  {12}},\ \bibinfo {pages} {113032} (\bibinfo {year} {2010})}\BibitemShut
  {NoStop}%
\bibitem [{\citenamefont {Kidon}, \citenamefont {Wilner},\ and\ \citenamefont
  {Rabani}(2015)}]{kidon2015}%
  \BibitemOpen
  \bibfield  {author} {\bibinfo {author} {\bibfnamefont {L.}~\bibnamefont
  {Kidon}}, \bibinfo {author} {\bibfnamefont {E.~Y.}\ \bibnamefont {Wilner}}, \
  and\ \bibinfo {author} {\bibfnamefont {E.}~\bibnamefont {Rabani}},\
  }\bibfield  {title} {\enquote {\bibinfo {title} {Exact calculation of the
  time convolutionless master equation generator: Application to the
  nonequilibrium resonant level model},}\ }\href {\doibase 10.1063/1.4937396}
  {\bibfield  {journal} {\bibinfo  {journal} {J.~Chem. Phys.}\ }\textbf
  {\bibinfo {volume} {143}},\ \bibinfo {pages} {234110} (\bibinfo {year}
  {2015})}\BibitemShut {NoStop}%
\bibitem [{\citenamefont {Pfalzgraff}, \citenamefont {Kelly},\ and\
  \citenamefont {Markland}(2015{\natexlab{b}})}]{pfalzgraff2015}%
  \BibitemOpen
  \bibfield  {author} {\bibinfo {author} {\bibfnamefont {W.~C.}\ \bibnamefont
  {Pfalzgraff}}, \bibinfo {author} {\bibfnamefont {A.}~\bibnamefont {Kelly}}, \
  and\ \bibinfo {author} {\bibfnamefont {T.~E.}\ \bibnamefont {Markland}},\
  }\bibfield  {title} {\enquote {\bibinfo {title} {Nonadiabatic dynamics in
  atomistic environments: Harnessing quantum-classical theory with generalized
  quantum master equations},}\ }\href {\doibase 10.1021/acs.jpclett.5b02131}
  {\bibfield  {journal} {\bibinfo  {journal} {J.~Phys. Chem. Lett.}\ }\textbf
  {\bibinfo {volume} {6}},\ \bibinfo {pages} {4743--4748} (\bibinfo {year}
  {2015}{\natexlab{b}})}\BibitemShut {NoStop}%
\bibitem [{\citenamefont {Mannouch}\ and\ \citenamefont
  {Richardson}(2020{\natexlab{a}})}]{spinPLDM1}%
  \BibitemOpen
  \bibfield  {author} {\bibinfo {author} {\bibfnamefont {J.~R.}\ \bibnamefont
  {Mannouch}}\ and\ \bibinfo {author} {\bibfnamefont {J.~O.}\ \bibnamefont
  {Richardson}},\ }\bibfield  {title} {\enquote {\bibinfo {title} {{A partially
  linearized spin-mapping approach for nonadiabatic dynamics. I. Derivation of
  the theory}},}\ }\href {\doibase 10.1063/5.0031168} {\bibfield  {journal}
  {\bibinfo  {journal} {J. Chem. Phys.}\ }\textbf {\bibinfo {volume} {153}},\
  \bibinfo {pages} {194109} (\bibinfo {year} {2020}{\natexlab{a}})},\ \Eprint
  {http://arxiv.org/abs/2007.05047} {arXiv:2007.05047} \BibitemShut {NoStop}%
\bibitem [{\citenamefont {Mannouch}\ and\ \citenamefont
  {Richardson}(2020{\natexlab{b}})}]{spinPLDM2}%
  \BibitemOpen
  \bibfield  {author} {\bibinfo {author} {\bibfnamefont {J.~R.}\ \bibnamefont
  {Mannouch}}\ and\ \bibinfo {author} {\bibfnamefont {J.~O.}\ \bibnamefont
  {Richardson}},\ }\bibfield  {title} {\enquote {\bibinfo {title} {{A partially
  linearized spin-mapping approach for nonadiabatic dynamics. II. Analysis and
  comparison with related approaches}},}\ }\href {\doibase 10.1063/5.0031173}
  {\bibfield  {journal} {\bibinfo  {journal} {J. Chem. Phys.}\ }\textbf
  {\bibinfo {volume} {153}},\ \bibinfo {pages} {194110} (\bibinfo {year}
  {2020}{\natexlab{b}})},\ \Eprint {http://arxiv.org/abs/2007.05048}
  {arXiv:2007.05048} \BibitemShut {NoStop}%
\bibitem [{\citenamefont {Kapral}\ and\ \citenamefont
  {Ciccotti}(1999)}]{Kapral1999QCLE}%
  \BibitemOpen
  \bibfield  {author} {\bibinfo {author} {\bibfnamefont {R.}~\bibnamefont
  {Kapral}}\ and\ \bibinfo {author} {\bibfnamefont {G.}~\bibnamefont
  {Ciccotti}},\ }\bibfield  {title} {\enquote {\bibinfo {title} {Mixed
  quantum-classical dynamics},}\ }\href@noop {} {\bibfield  {journal} {\bibinfo
   {journal} {J. Chem. Phys.}\ }\textbf {\bibinfo {volume} {110}},\ \bibinfo
  {pages} {8919--8929} (\bibinfo {year} {1999})}\BibitemShut {NoStop}%
\bibitem [{\citenamefont {Leaf}(1968)}]{leaf1968}%
  \BibitemOpen
  \bibfield  {author} {\bibinfo {author} {\bibfnamefont {B.}~\bibnamefont
  {Leaf}},\ }\bibfield  {title} {\enquote {\bibinfo {title} {Weyl
  transformation and the classical limit of quantum mechanics},}\ }\href
  {\doibase 10.1063/1.1664478} {\bibfield  {journal} {\bibinfo  {journal}
  {J.~Math. Phys.}\ }\textbf {\bibinfo {volume} {9}},\ \bibinfo {pages}
  {65--72} (\bibinfo {year} {1968})}\BibitemShut {NoStop}%
\bibitem [{\citenamefont {Zwanzig}(2001)}]{Zwanzig}%
  \BibitemOpen
  \bibfield  {author} {\bibinfo {author} {\bibfnamefont {R.}~\bibnamefont
  {Zwanzig}},\ }\href@noop {} {\emph {\bibinfo {title} {Nonequilibrium
  Statistical Mechanics}}}\ (\bibinfo  {publisher} {Oxford University Press},\
  \bibinfo {address} {New York},\ \bibinfo {year} {2001})\BibitemShut {NoStop}%
\bibitem [{\citenamefont {Sun}\ and\ \citenamefont
  {Miller}(1997)}]{Sun1997mapping}%
  \BibitemOpen
  \bibfield  {author} {\bibinfo {author} {\bibfnamefont {X.}~\bibnamefont
  {Sun}}\ and\ \bibinfo {author} {\bibfnamefont {W.~H.}\ \bibnamefont
  {Miller}},\ }\bibfield  {title} {\enquote {\bibinfo {title} {Semiclassical
  initial value representation for electronically nonadiabatic molecular
  dynamics},}\ }\href {\doibase 10.1063/1.473624} {\bibfield  {journal}
  {\bibinfo  {journal} {J.~Chem. Phys.}\ }\textbf {\bibinfo {volume} {106}},\
  \bibinfo {pages} {6346--6353} (\bibinfo {year} {1997})}\BibitemShut {NoStop}%
\bibitem [{\citenamefont {Bonella}\ and\ \citenamefont
  {Coker}(2003)}]{Bonella2003mapping}%
  \BibitemOpen
  \bibfield  {author} {\bibinfo {author} {\bibfnamefont {S.}~\bibnamefont
  {Bonella}}\ and\ \bibinfo {author} {\bibfnamefont {D.~F.}\ \bibnamefont
  {Coker}},\ }\bibfield  {title} {\enquote {\bibinfo {title} {{Semiclassical
  implementation of the mapping Hamiltonian approach for nonadiabatic dynamics
  using focused initial distribution sampling}},}\ }\href {\doibase
  10.1063/1.1542883} {\bibfield  {journal} {\bibinfo  {journal} {J.~Chem.
  Phys.}\ }\textbf {\bibinfo {volume} {118}},\ \bibinfo {pages} {4370--4385}
  (\bibinfo {year} {2003})}\BibitemShut {NoStop}%
\bibitem [{Note1()}]{Note1}%
  \BibitemOpen
  \bibinfo {note} {Note that the quasiclassical approximation of the memory
  kernels will ultimately relax to zero at long enough times. In fact, from
  Eq.~(18), we can calculate $\protect \mathcal K(\infty ) = -\protect \qopname
  \relax m{lim}_{\omega \to 0+}i\omega \protect \tilde {\protect \mathcal
  K}(\omega ) = \left [\protect \mathcal I_4 -\protect \tilde K^{(3)}(0)\right
  ]^{-1}\protect \mathcal K^{(1)}(\infty )$, where we made use of the
  final-value theorem [\protect \cref {eq:FVT}], and $\protect \mathcal I_4$ is
  the $4\times 4$ identity matrix. It is possible to show that $\protect
  \mathcal K^{(1)}_{\mu \nu }(\infty )=0$ for all $\mu ,\nu $. This follows
  from the \protect \textit {mixing condition} \cite {hawkins2021,ellipsoid}
  $\langle AB(t)\rangle \to \langle A\rangle \langle B\rangle _{\protect
  \mathrm {eq}}$ applied to Eq.~(A8a). Here, by symmetry $\langle A\rangle =0$
  and hence $\protect \mathcal K_{\mu \nu }(\infty )=0$ for all $\mu ,\nu $.
  However, this may occur on a significantly longer timescale than it should
  according to the exact kernels (\protect \cref {fig:kernels_SM_MFT}), and
  thus does not ensure that GQME/Ehrenfest will converge within a reasonable
  $t_{\protect \mathrm {cut}}$.}\BibitemShut {Stop}%
\bibitem [{\citenamefont {Miller}\ and\ \citenamefont
  {Cotton}(2016)}]{Miller2016Faraday}%
  \BibitemOpen
  \bibfield  {author} {\bibinfo {author} {\bibfnamefont {W.~H.}\ \bibnamefont
  {Miller}}\ and\ \bibinfo {author} {\bibfnamefont {S.~J.}\ \bibnamefont
  {Cotton}},\ }\bibfield  {title} {\enquote {\bibinfo {title} {Classical
  molecular dynamics simulation of electronically non-adiabatic processes},}\
  }\href {\doibase 10.1039/C6FD00181E} {\bibfield  {journal} {\bibinfo
  {journal} {Faraday Discuss.}\ }\textbf {\bibinfo {volume} {195}},\ \bibinfo
  {pages} {9--30} (\bibinfo {year} {2016})}\BibitemShut {NoStop}%
\bibitem [{\citenamefont {Mannouch}\ and\ \citenamefont {Richardson}()}]{MASH}%
  \BibitemOpen
  \bibfield  {author} {\bibinfo {author} {\bibfnamefont {J.~R.}\ \bibnamefont
  {Mannouch}}\ and\ \bibinfo {author} {\bibfnamefont {J.~O.}\ \bibnamefont
  {Richardson}},\ }\href@noop {} {\enquote {\bibinfo {title} {A mapping
  approach to surface hopping},}\ }\bibinfo {note} {In preparation}\BibitemShut
  {NoStop}%
\bibitem [{\citenamefont {Ullah}\ and\ \citenamefont {Dral}(2022)}]{Ullah2022}%
  \BibitemOpen
  \bibfield  {author} {\bibinfo {author} {\bibfnamefont {A.}~\bibnamefont
  {Ullah}}\ and\ \bibinfo {author} {\bibfnamefont {P.~O.}\ \bibnamefont
  {Dral}},\ }\bibfield  {title} {\enquote {\bibinfo {title} {Predicting the
  future of excitation energy transfer in light-harvesting complex with
  artificial intelligence-based quantum dynamics},}\ }\href {\doibase
  10.1038/s41467-022-29621-w} {\bibfield  {journal} {\bibinfo  {journal} {Nat.
  Commun.}\ }\textbf {\bibinfo {volume} {13}},\ \bibinfo {pages} {1930}
  (\bibinfo {year} {2022})}\BibitemShut {NoStop}%
\bibitem [{\citenamefont {Xu}\ \emph {et~al.}(2018)\citenamefont {Xu},
  \citenamefont {Yan}, \citenamefont {Liu},\ and\ \citenamefont
  {Shi}}]{xu2018}%
  \BibitemOpen
  \bibfield  {author} {\bibinfo {author} {\bibfnamefont {M.}~\bibnamefont
  {Xu}}, \bibinfo {author} {\bibfnamefont {Y.}~\bibnamefont {Yan}}, \bibinfo
  {author} {\bibfnamefont {Y.}~\bibnamefont {Liu}}, \ and\ \bibinfo {author}
  {\bibfnamefont {Q.}~\bibnamefont {Shi}},\ }\bibfield  {title} {\enquote
  {\bibinfo {title} {{Convergence of high order memory kernels in the
  Nakajima-Zwanzig generalized master equation and rate constants: Case study
  of the spin-boson model}},}\ }\href {\doibase 10.1063/1.5022761} {\bibfield
  {journal} {\bibinfo  {journal} {J.~Chem. Phys.}\ }\textbf {\bibinfo {volume}
  {148}},\ \bibinfo {pages} {164101} (\bibinfo {year} {2018})}\BibitemShut
  {NoStop}%
\bibitem [{\citenamefont {Kobus}\ \emph {et~al.}(2008)\citenamefont {Kobus},
  \citenamefont {Gorbunov}, \citenamefont {Nguyen},\ and\ \citenamefont
  {Stock}}]{kobus2008}%
  \BibitemOpen
  \bibfield  {author} {\bibinfo {author} {\bibfnamefont {M.}~\bibnamefont
  {Kobus}}, \bibinfo {author} {\bibfnamefont {R.~D.}\ \bibnamefont {Gorbunov}},
  \bibinfo {author} {\bibfnamefont {P.~H.}\ \bibnamefont {Nguyen}}, \ and\
  \bibinfo {author} {\bibfnamefont {G.}~\bibnamefont {Stock}},\ }\bibfield
  {title} {\enquote {\bibinfo {title} {Nonadiabatic vibrational dynamics and
  spectroscopy of peptides: A quantum-classical description},}\ }\href
  {\doibase https://doi.org/10.1016/j.chemphys.2007.10.034} {\bibfield
  {journal} {\bibinfo  {journal} {Chem. Phys.}\ }\textbf {\bibinfo {volume}
  {347}},\ \bibinfo {pages} {208--217} (\bibinfo {year} {2008})}\BibitemShut
  {NoStop}%
\bibitem [{\citenamefont {Mukamel}(1995)}]{MukamelBook}%
  \BibitemOpen
  \bibfield  {author} {\bibinfo {author} {\bibfnamefont {S.}~\bibnamefont
  {Mukamel}},\ }\href@noop {} {\emph {\bibinfo {title} {Principles of Nonlinear
  Optical Spectroscopy}}}\ (\bibinfo  {publisher} {Oxford University Press},\
  \bibinfo {address} {Oxford},\ \bibinfo {year} {1995})\BibitemShut {NoStop}%
\bibitem [{\citenamefont {Mannouch}\ and\ \citenamefont
  {Richardson}(2022)}]{mannouch2022}%
  \BibitemOpen
  \bibfield  {author} {\bibinfo {author} {\bibfnamefont {J.~R.}\ \bibnamefont
  {Mannouch}}\ and\ \bibinfo {author} {\bibfnamefont {J.~O.}\ \bibnamefont
  {Richardson}},\ }\bibfield  {title} {\enquote {\bibinfo {title} {A partially
  linearized spin-mapping approach for simulating nonlinear optical spectra},}\
  }\href {\doibase 10.1063/5.0077744} {\bibfield  {journal} {\bibinfo
  {journal} {J.~Chem. Phys.}\ }\textbf {\bibinfo {volume} {156}},\ \bibinfo
  {pages} {024108} (\bibinfo {year} {2022})}\BibitemShut {NoStop}%
\bibitem [{\citenamefont {Fetherolf}\ and\ \citenamefont
  {Berkelbach}(2017)}]{fetherolf2017}%
  \BibitemOpen
  \bibfield  {author} {\bibinfo {author} {\bibfnamefont {J.~H.}\ \bibnamefont
  {Fetherolf}}\ and\ \bibinfo {author} {\bibfnamefont {T.~C.}\ \bibnamefont
  {Berkelbach}},\ }\bibfield  {title} {\enquote {\bibinfo {title} {Linear and
  nonlinear spectroscopy from quantum master equations},}\ }\href {\doibase
  10.1063/1.5006824} {\bibfield  {journal} {\bibinfo  {journal} {J.~Chem.
  Phys.}\ }\textbf {\bibinfo {volume} {147}},\ \bibinfo {pages} {244109}
  (\bibinfo {year} {2017})}\BibitemShut {NoStop}%
\bibitem [{\citenamefont {Garg}, \citenamefont {Onuchic},\ and\ \citenamefont
  {Ambegaokar}(1985)}]{garg1985}%
  \BibitemOpen
  \bibfield  {author} {\bibinfo {author} {\bibfnamefont {A.}~\bibnamefont
  {Garg}}, \bibinfo {author} {\bibfnamefont {J.~N.}\ \bibnamefont {Onuchic}}, \
  and\ \bibinfo {author} {\bibfnamefont {V.}~\bibnamefont {Ambegaokar}},\
  }\bibfield  {title} {\enquote {\bibinfo {title} {Effect of friction on
  electron transfer in biomolecules},}\ }\href {\doibase 10.1063/1.449017}
  {\bibfield  {journal} {\bibinfo  {journal} {J.~Chem. Phys.}\ }\textbf
  {\bibinfo {volume} {83}},\ \bibinfo {pages} {4491--4503} (\bibinfo {year}
  {1985})}\BibitemShut {NoStop}%
\bibitem [{\citenamefont {Anto-Sztrikacs}\ and\ \citenamefont
  {Segal}(2021)}]{anto-ztrikacs2021}%
  \BibitemOpen
  \bibfield  {author} {\bibinfo {author} {\bibfnamefont {N.}~\bibnamefont
  {Anto-Sztrikacs}}\ and\ \bibinfo {author} {\bibfnamefont {D.}~\bibnamefont
  {Segal}},\ }\bibfield  {title} {\enquote {\bibinfo {title} {{Capturing
  non-Markovian dynamics with the reaction coordinate method}},}\ }\href
  {\doibase 10.1103/PhysRevA.104.052617} {\bibfield  {journal} {\bibinfo
  {journal} {Phys. Rev. A}\ }\textbf {\bibinfo {volume} {104}},\ \bibinfo
  {pages} {052617} (\bibinfo {year} {2021})}\BibitemShut {NoStop}%
\bibitem [{\citenamefont {Lawrence}\ \emph {et~al.}(2019)\citenamefont
  {Lawrence}, \citenamefont {Fletcher}, \citenamefont {Lindoy},\ and\
  \citenamefont {Manolopoulos}}]{Lawrence2019ET}%
  \BibitemOpen
  \bibfield  {author} {\bibinfo {author} {\bibfnamefont {J.~E.}\ \bibnamefont
  {Lawrence}}, \bibinfo {author} {\bibfnamefont {T.}~\bibnamefont {Fletcher}},
  \bibinfo {author} {\bibfnamefont {L.~P.}\ \bibnamefont {Lindoy}}, \ and\
  \bibinfo {author} {\bibfnamefont {D.~E.}\ \bibnamefont {Manolopoulos}},\
  }\bibfield  {title} {\enquote {\bibinfo {title} {On the calculation of
  quantum mechanical electron transfer rates},}\ }\href {\doibase
  10.1063/1.5116800} {\bibfield  {journal} {\bibinfo  {journal} {J.~Chem.
  Phys.}\ }\textbf {\bibinfo {volume} {151}},\ \bibinfo {pages} {114119}
  (\bibinfo {year} {2019})}\BibitemShut {NoStop}%
\bibitem [{\citenamefont {Tokuyama}(1981)}]{tokuyama1981}%
  \BibitemOpen
  \bibfield  {author} {\bibinfo {author} {\bibfnamefont {M.}~\bibnamefont
  {Tokuyama}},\ }\bibfield  {title} {\enquote {\bibinfo {title}
  {{Statistical-dynamical theory of nonlinear stochastic processes: II.
  Time-convolutionless projector method in nonequilibrium open systems}},}\
  }\href {\doibase https://doi.org/10.1016/0378-4371(81)90041-8} {\bibfield
  {journal} {\bibinfo  {journal} {Phys. A: Stat. Mech. Appl.}\ }\textbf
  {\bibinfo {volume} {109}},\ \bibinfo {pages} {128--160} (\bibinfo {year}
  {1981})}\BibitemShut {NoStop}%
\bibitem [{\citenamefont {Fuliński}(1967)}]{fulinski1967}%
  \BibitemOpen
  \bibfield  {author} {\bibinfo {author} {\bibfnamefont {A.}~\bibnamefont
  {Fuliński}},\ }\bibfield  {title} {\enquote {\bibinfo {title} {On the
  “memory” properties of generalized master equations},}\ }\href {\doibase
  https://doi.org/10.1016/0375-9601(67)90198-3} {\bibfield  {journal} {\bibinfo
   {journal} {Phys. Lett. A}\ }\textbf {\bibinfo {volume} {24}},\ \bibinfo
  {pages} {63--64} (\bibinfo {year} {1967})}\BibitemShut {NoStop}%
\bibitem [{\citenamefont {Brian}\ and\ \citenamefont
  {Sun}(2021)}]{dominikus2021}%
  \BibitemOpen
  \bibfield  {author} {\bibinfo {author} {\bibfnamefont {D.}~\bibnamefont
  {Brian}}\ and\ \bibinfo {author} {\bibfnamefont {X.}~\bibnamefont {Sun}},\
  }\bibfield  {title} {\enquote {\bibinfo {title} {Generalized quantum master
  equation: A tutorial review and recent advances},}\ }\href {\doibase
  10.1063/1674-0068/cjcp2109157} {\bibfield  {journal} {\bibinfo  {journal}
  {Chinese J. Chem. Phys.}\ }\textbf {\bibinfo {volume} {34}},\ \bibinfo
  {pages} {497--524} (\bibinfo {year} {2021})}\BibitemShut {NoStop}%
\bibitem [{\citenamefont {Richardson}\ and\ \citenamefont
  {Thoss}(2013)}]{mapping}%
  \BibitemOpen
  \bibfield  {author} {\bibinfo {author} {\bibfnamefont {J.~O.}\ \bibnamefont
  {Richardson}}\ and\ \bibinfo {author} {\bibfnamefont {M.}~\bibnamefont
  {Thoss}},\ }\bibfield  {title} {\enquote {\bibinfo {title} {Communication:
  {N}onadiabatic ring-polymer molecular dynamics},}\ }\href {\doibase
  10.1063/1.4816124} {\bibfield  {journal} {\bibinfo  {journal} {J.~Chem.
  Phys.}\ }\textbf {\bibinfo {volume} {139}},\ \bibinfo {pages} {031102}
  (\bibinfo {year} {2013})}\BibitemShut {NoStop}%
\bibitem [{\citenamefont {Bossion}, \citenamefont {Chowdhury},\ and\
  \citenamefont {Huo}(2021)}]{bossion2021}%
  \BibitemOpen
  \bibfield  {author} {\bibinfo {author} {\bibfnamefont {D.}~\bibnamefont
  {Bossion}}, \bibinfo {author} {\bibfnamefont {S.~N.}\ \bibnamefont
  {Chowdhury}}, \ and\ \bibinfo {author} {\bibfnamefont {P.}~\bibnamefont
  {Huo}},\ }\bibfield  {title} {\enquote {\bibinfo {title} {Non-adiabatic ring
  polymer molecular dynamics with spin mapping variables},}\ }\href {\doibase
  10.1063/5.0051456} {\bibfield  {journal} {\bibinfo  {journal} {J.~Chem.
  Phys.}\ }\textbf {\bibinfo {volume} {154}},\ \bibinfo {pages} {184106}
  (\bibinfo {year} {2021})}\BibitemShut {NoStop}%
\bibitem [{\citenamefont {Cohen}, \citenamefont {Wilner},\ and\ \citenamefont
  {Rabani}(2013)}]{cohen2013}%
  \BibitemOpen
  \bibfield  {author} {\bibinfo {author} {\bibfnamefont {G.}~\bibnamefont
  {Cohen}}, \bibinfo {author} {\bibfnamefont {E.~Y.}\ \bibnamefont {Wilner}}, \
  and\ \bibinfo {author} {\bibfnamefont {E.}~\bibnamefont {Rabani}},\
  }\bibfield  {title} {\enquote {\bibinfo {title} {Generalized projected
  dynamics for non-system observables of non-equilibrium quantum impurity
  models},}\ }\href {\doibase 10.1088/1367-2630/15/7/073018} {\bibfield
  {journal} {\bibinfo  {journal} {New J. Phys.}\ }\textbf {\bibinfo {volume}
  {15}},\ \bibinfo {pages} {073018} (\bibinfo {year} {2013})}\BibitemShut
  {NoStop}%
\bibitem [{\citenamefont {Beerends}\ \emph {et~al.}(2003)\citenamefont
  {Beerends}, \citenamefont {Morsche}, \citenamefont {van~den Berg},\ and\
  \citenamefont {van~de Vrie}}]{beerends2003}%
  \BibitemOpen
  \bibfield  {author} {\bibinfo {author} {\bibfnamefont {R.}~\bibnamefont
  {Beerends}}, \bibinfo {author} {\bibfnamefont {H.}~\bibnamefont {Morsche}},
  \bibinfo {author} {\bibfnamefont {J.}~\bibnamefont {van~den Berg}}, \ and\
  \bibinfo {author} {\bibfnamefont {E.}~\bibnamefont {van~de Vrie}},\ }\href
  {https://books.google.de/books?id=frT5\_rfyO4IC} {\emph {\bibinfo {title}
  {{Fourier and Laplace Transforms}}}}\ (\bibinfo  {publisher} {Cambridge
  University Press},\ \bibinfo {year} {2003})\BibitemShut {NoStop}%
\end{thebibliography}
%\bibliographystyle{ieeetr} 

\end{document}